\documentclass[preprint2,iopj]{emulateapj}
\usepackage{threeparttable,rotating,amsmath}
\usepackage{natbib}

\newcommand{\fermi}{\textit{Fermi }}

\begin{document}

\title{Gamma-Ray Observational Properties of TeV-Detected Blazars}

\author{
G.~D.~\c{S}ent\"{u}rk\altaffilmark{1},  
M.~Errando\altaffilmark{2,3},
M.~B\"ottcher\altaffilmark{4,5}, and
R.~Mukherjee\altaffilmark{2}
}

\altaffiltext{1}{Physics Department, Columbia University, 550 West 120th Street, New York, NY 10027, USA; gds2110@columbia.edu}
\altaffiltext{2}{Department of Physics \& Astronomy, Barnard College, Columbia University, 3009 Broadway, New York, NY 10027, USA}
\altaffiltext{3}{Columbia Astrophysics Laboratory, Columbia University, 550 West 120th Street, New York, NY 10027, USA}
\altaffiltext{4}{Astrophysical Institute, Department of Physics and Astronomy, Ohio University, Athens, OH, USA}
\altaffiltext{5}{Centre for Space Research, North-West University Potchefstroom, Potchefstroom 2531, South Africa}

\begin{abstract}
The synergy between the {\it Fermi}-LAT and ground-based Cherenkov telescope arrays gives us the opportunity for the first time to characterize the high-energy emission from blazars over 5 decades in energy, from 100\,MeV to 10\,TeV. 
In this study, we perform a {\it Fermi}-LAT spectral analysis for TeV-detected blazars and combine it with archival TeV data. We examine the observational properties in the $\gamma$-ray band of our sample of TeV-detected blazars and compare the results with X-ray and GeV-selected populations. 
The spectral energy distributions (SEDs) that result from combining {\it Fermi}-LAT and ground-based spectra are studied in detail. 
Simple parameterizations such as a power-law function do not always reproduce the high-energy SEDs, where spectral features that could indicate intrinsic absorption are observed.

\end{abstract}

\keywords{galaxies: active -- galaxies: nuclei -- gamma rays: general}

\section{Introduction}
Active Galactic Nuclei (AGNs) are extreme objects with observed luminosity outshining their host galaxy. 
These sources are believed to be powered by accretion onto a central supermassive black hole, commonly display relativistic jets, and exhibit non-thermal continuum emission extending from the radio band to X and $\gamma$ rays. 
Blazars constitute a subclass of AGNs, with jet axes oriented close to the observer's line of sight. 
Relativistic beaming gives rise to distinctive observational features in blazars, such as strongly anisotropic radiation, superluminal motion, high polarization and rapid variability~\citep{urry95}. 
Blazars are divided into two subclasses, flat spectrum radio quasars (FSRQs) and BL Lacertae objects (BL Lacs). 
FSRQs are observationally characterized by  broad  spectral lines in the optical band, which are weak or not present in BL Lacs. 
The spectral energy distribution (SED) of blazars exhibits a two-component structure, with a low-energy component peaking between  infrared (IR) and X-ray energies, and a high-energy one between X and $\gamma$ rays. 
The low energy component is believed to be dominated by synchrotron emission from relativistic electrons in the jet~\citep{kembhavi}. 
The peak frequency of the synchrotron component of the SED ($\nu_\mathrm{syn}$) is used to sub-classify BL Lacs into low (LBLs, $\nu_\mathrm{syn} < 10^{14}$\,Hz), intermediate (IBL, $\nu_\mathrm{syn} \sim 10^{14}-10^{15}$\,Hz) and high-frequency-peaked BL Lacs (HBL, $\nu_\mathrm{syn} > 10^{15}$\,Hz).
\par 
The high-energy component of the blazar SED has been historically less studied, due to the later development of hard X-ray and $\gamma$-ray detectors compared to those of longer frequency bands.
The Synchrotron self-Compton (SSC) model is the simplest scenario that explains the high-energy emission of blazars, by inverse-Compton (IC) up-scattering of soft synchrotron photons off the same electrons that have undergone synchrotron cooling~\citep{maraschi92}. 
Throughout the text, we refer to the high-energy component of the blazar SED as ``IC component".
An additional IC target photon field external to the jet is often invoked \citep[see, e.g.,][]{ver_wCom_flare}. 
This mechanism is referred to as external-Compton (EC), and several possible sources for the external photon field have been set forth~\citep{markus_rev}. 
Other models suggest a significant contribution from hadronic processes to the high-energy output~\citep{proton_3c279}. 
\begin{figure*}[t]
\centering
\includegraphics[width=130mm]{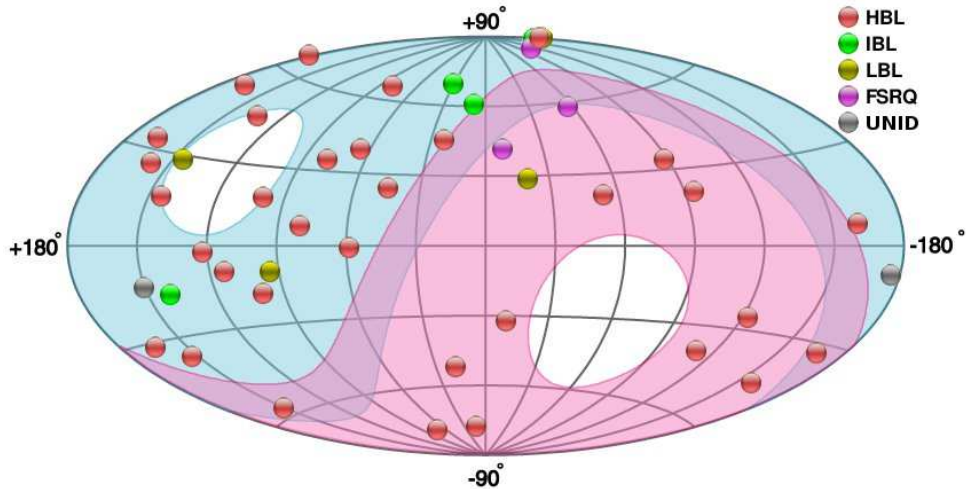}
\caption{\footnotesize Skymap of TeV blazars in galactic coordinates, as of 2012 January, generated using TeVCat (http://tevcat.uchicago.edu/). Blue and pink shaded areas represent VERITAS/MAGIC and HESS visibilities, respectively. A total of 46 sources consisting of 33 HBLs, 4 IBLs, 4 LBLs, 3 FSRQs, and 2 sources that were formerly classified as AGN of unknown type (UNID), namely IC~310 and VER~J0521+211, are shown. VER~J0521+211 is now identified as a BL Lac~\citep{manel_fermi2011} and recent studies suggest that the high-energy radiation from IC~310 originates from a blazar-like emission mechanism~\citep{kadler2012}.}
\label{fig:tev_skymap}
\end{figure*} 
\par
A good spectral characterization of the high-energy peak of the blazar spectrum (keV - TeV band) is essential to discriminate between the aforementioned models. During the EGRET era covering 1991-2001~\citep{egret}, only five blazars were known at TeV energies: Mrk~421, Mrk~501, 1ES~1959+650, PKS~2155-304 and 1ES~2344+514; thanks to the first generation of ground-based Imaging Atmospheric Cherenkov Telescopes (IACTs; Whipple~\citep{whipple}, HEGRA \citep{hegra}, Durham Mark 6~\citep{durham}, and Telescope Array~\citep{tel_array}). Only three of these sources were detected by EGRET (Mrk~501 was only marginally detected and 1ES~2344+514 not seen at all). By the time {\it Fermi} started operations (2008 August), the number of known TeV blazars had increased to 21 with the second generation of IACTs in operation (VERITAS~\citep{veritas}, MAGIC~\citep{magic}, HESS~\citep{hess}). This number has doubled since then, with most TeV blazars being also detected in the GeV range by {\it Fermi}-LAT.
\par
For the first time now, good quality spectra are available both from {\it Fermi}-LAT in the high-energy (HE, $0.1$\,GeV$<E<100$\,GeV) $\gamma$-ray band and IACTs in the very high energy (VHE, $E>100$\,GeV) $\gamma$-ray band for more than two dozen sources. 
The combined spectral data covers up to five decades in energy, giving a detailed description of the high-energy peak of the blazar SED. 
Recent studies have explored this newly available data sample, focusing on the GeV properties of TeV-selected blazars~\citep{steve_tev}, or deriving jet parameters assuming leptonic emission models~\citep{zhang2011}. 
These studies are similar to earlier studies carried out on a limited sample of TeV-detected blazars~\citep[e.g.,][]{wagner_synoptic}.
\par
In this paper we study the GeV-TeV observational properties of the high-energy emission in blazars that are detected in the TeV band. 
Section 2 describes the population of TeV blazars, giving census information, investigating luminosity, redshift and photon index distributions among different blazar types. 
In Section 3, we study TeV blazars that appear in the \fermi data and outline their GeV properties with respect to the rest of the \fermi blazars. 
Section 4 defines our sample and focuses on general observational TeV properties of our objects.
In section 5, we give a detailed description of the \fermi analysis that we performed on our TeV blazar sample. 
Finally, Section 6 discusses various observational characteristics of the studied sources based on their GeV-TeV spectral shapes, such as the peak frequency of the IC component, absorption-like spectral features and variability.
Throughout the text, the symbol $\sigma$ is used to designate the standard deviation, as a measure of statistical significance.

\section{TeV Blazars}
Mkn 421 was the first blazar and extragalactic object to be discovered as a VHE $\gamma$-ray emitter, detected with the Whipple telescope in 1992~\citep{punch92}. 
Since then, different candidate selection methods have been applied to radio, X-ray or HE data with the aim of finding new ``TeV" blazars, i.e. detected in the VHE regime, \citep{costamante02,perez03,behera09}, leading to the discovery of most of the known TeV blazars.
To date, 44 blazars and 2 AGNs of unknown type have been detected in the VHE range\footnote{\label{tevcat}http://tevcat.uchicago.edu/}, with a census consisting of 33 HBLs, 4 IBLs, 4 LBLs and 3 FSRQs (see Figure~\ref{fig:tev_skymap}). 
In this work, we have studied the blazars that have a published TeV spectrum as of 2011 February (referred to as the ``sample" in the remainder of the text).
\begin{figure}[h!]
\centering
\includegraphics[width=80mm]{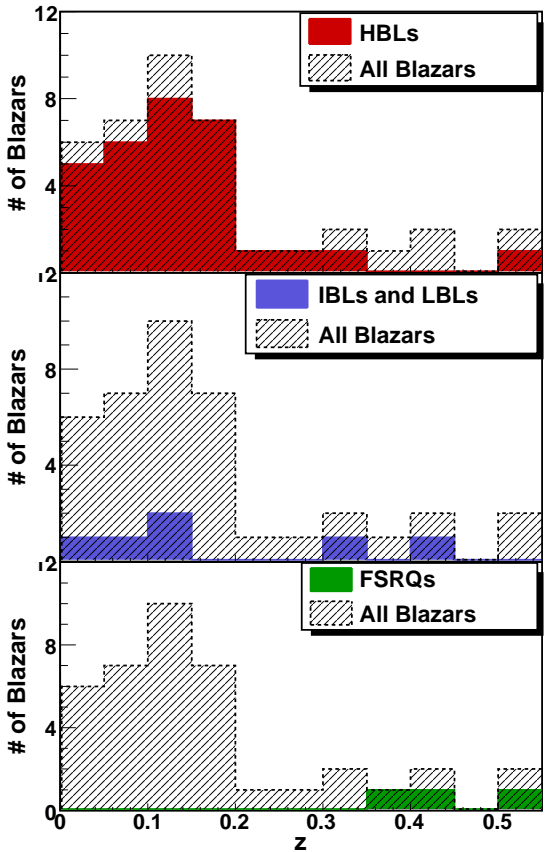}
\caption{{\footnotesize Redshift ($z$) distribution of all TeV blazars as of 2012 January. Top, middle and bottom panels also show HBLs, IBLs and LBLs, and FSRQs, respectively. Most of the blazars with known redshifts (30 out of 39) are located at $z<0.2$. FSRQs are the most distant objects in the population. The farthest object is the FSRQ 3C\,279, with a redshift of 0.536. The rest of the population does not have a secure redshift.}}
\label{fig:z_dist}
\end{figure}
\begin{figure}[h]
\centering
\includegraphics[width=80mm]{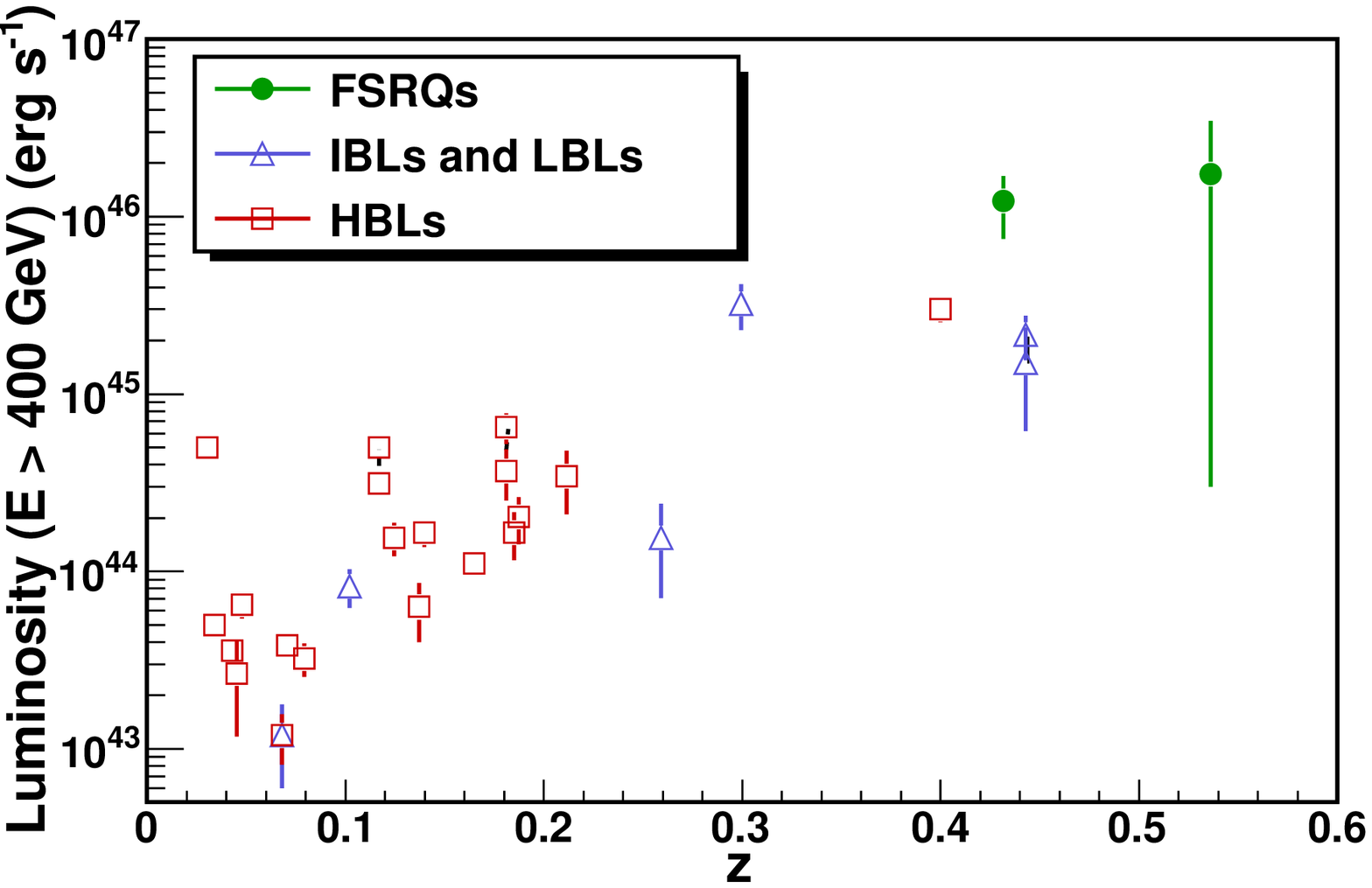}
\includegraphics[width=80mm]{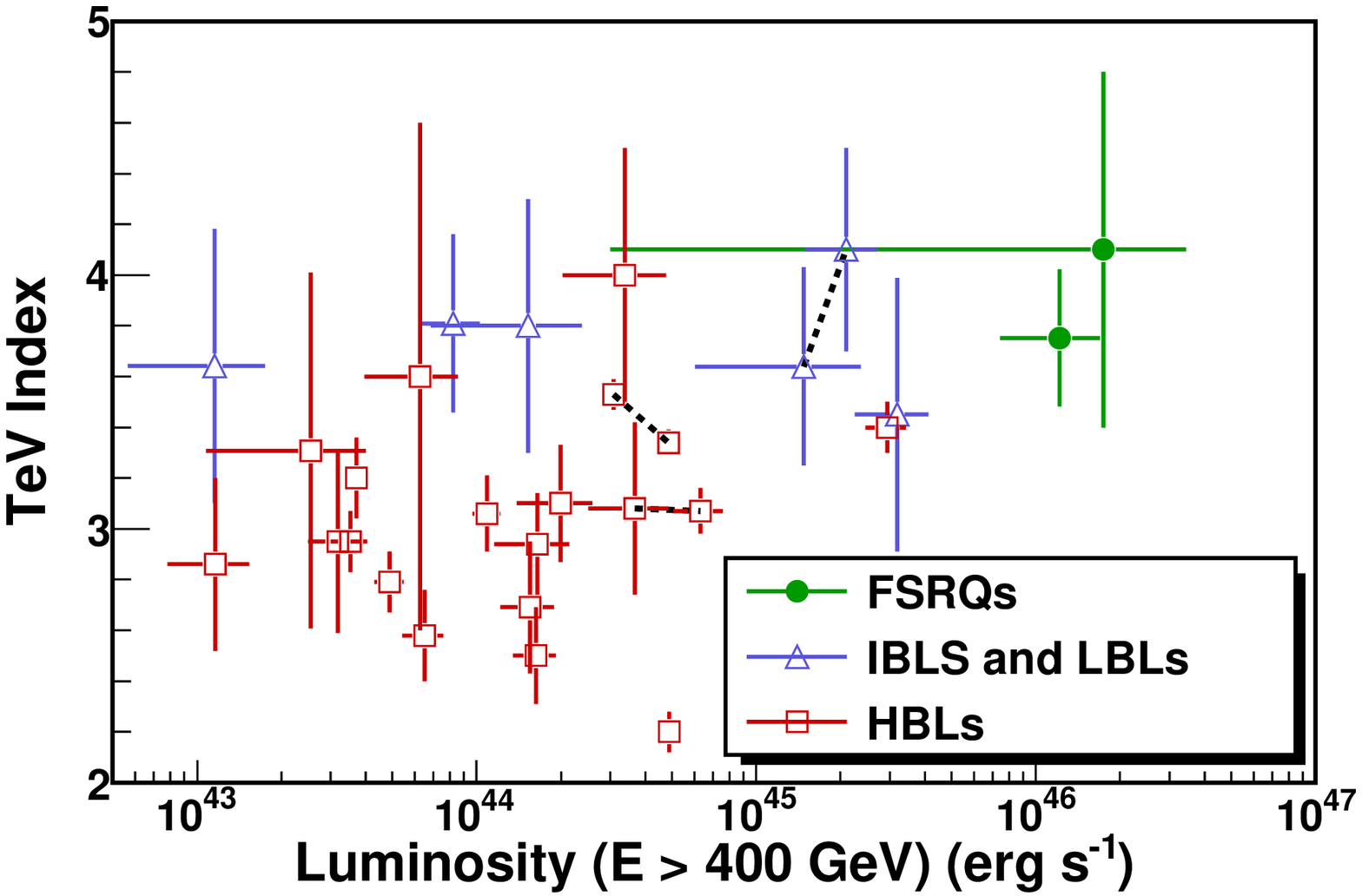}
\caption{{\footnotesize Top: apparent luminosity (not EBL-corrected) vs. redshift plot of TeV
blazars. As expected, high-redshift objects are less likely to be seen, and in
fact most of them were detected in flare states. Bottom: TeV index vs.
apparent luminosity distribution of the same sample. For fluxes up to $\sim
10^{45}\mathrm{erg}\,\mathrm{s^{-1}}$, the TeV index distribution is fairly
homogenous. At higher luminosities, mostly soft spectrum sources are detected.
FSRQs are the most luminous TeV emitters but one should note that their data come
from flares only. Squares represent HBLs, triangles IBLs and LBLs, and circles FSRQs. Points representing different flux states of the same source are connected with dashed black lines (bottom).}}
\label{fig:tev_lum}
\end{figure}

\par
The redshift ($z$) of TeV blazars in our sample ranges from 0.031 (HBL Mrk\,421) to 0.536 (FSRQ 3C\,279), and nearly one fourth of the population does not have a secure redshift. 
This lack is due to the fact that optical emission lines are typically weak or absent in BL Lac objects, rendering direct redshift measurements difficult.
The majority of known-redshift TeV blazars are located at $z<0.2$, mostly due to the absorption from the extragalactic background light (EBL). 
Figure~\ref{fig:z_dist} illustrates the redshift distribution of all TeV blazars. 
The TeV FSRQs are the most distant objects in the population with redshift ranging from 0.36 to 0.536. 

Using archival data, we calculated the apparent isotropic luminosity of the blazars in our sample (see Section~\ref{sec:sample}) for $E>400\,GeV$, with the following formula:
\begin{equation} 
L=F\times4\pi D_{L}^{2}/(1+z)^{2-\Gamma}
\label{eq:luminosity}
\end{equation}
where $F$ is the energy flux for energies above $E>400$ GeV, $D_{L}$ is the luminosity distance with Hubble constant $H_{o}=71\,\mathrm{km}\,\mathrm{s}^{-1} \mathrm{Mpc}^{-1}$ and the cosmological constant $\Omega_\mathrm{\lambda}=0.730$, $z$ is the redshift, and $\Gamma$ is the observed photon index for each blazar.
Figure~\ref{fig:tev_lum} top panel shows the luminosity versus redshift correlation for the sample. 
Sources at high redshifts tend to be scarce and much more luminous. 
The reason why we see only luminous sources at high redshifts is that the less luminous ones are too weak to be detected. 
On the other hand, the reason why we do not detect them in low redshifts is that we integrate over a much smaller volume and thus are less likely to see high-luminosity sources that should be scarce compared to low-luminosity ones. 
Note that if the luminosity is corrected for EBL absorption, which is stronger at TeV energies and high redshifts, the correlation will be steeper.
Figure~\ref{fig:tev_lum} bottom panel shows the TeV photon index versus luminosity correlation of the same sample. 
For luminosities up to $\sim 10^{45}\mathrm{erg}\,\mathrm{s^{-1}}$, the photon index distribution is fairly homogenous.

\begin{figure}[h!]
\centering
\includegraphics[width=80mm]{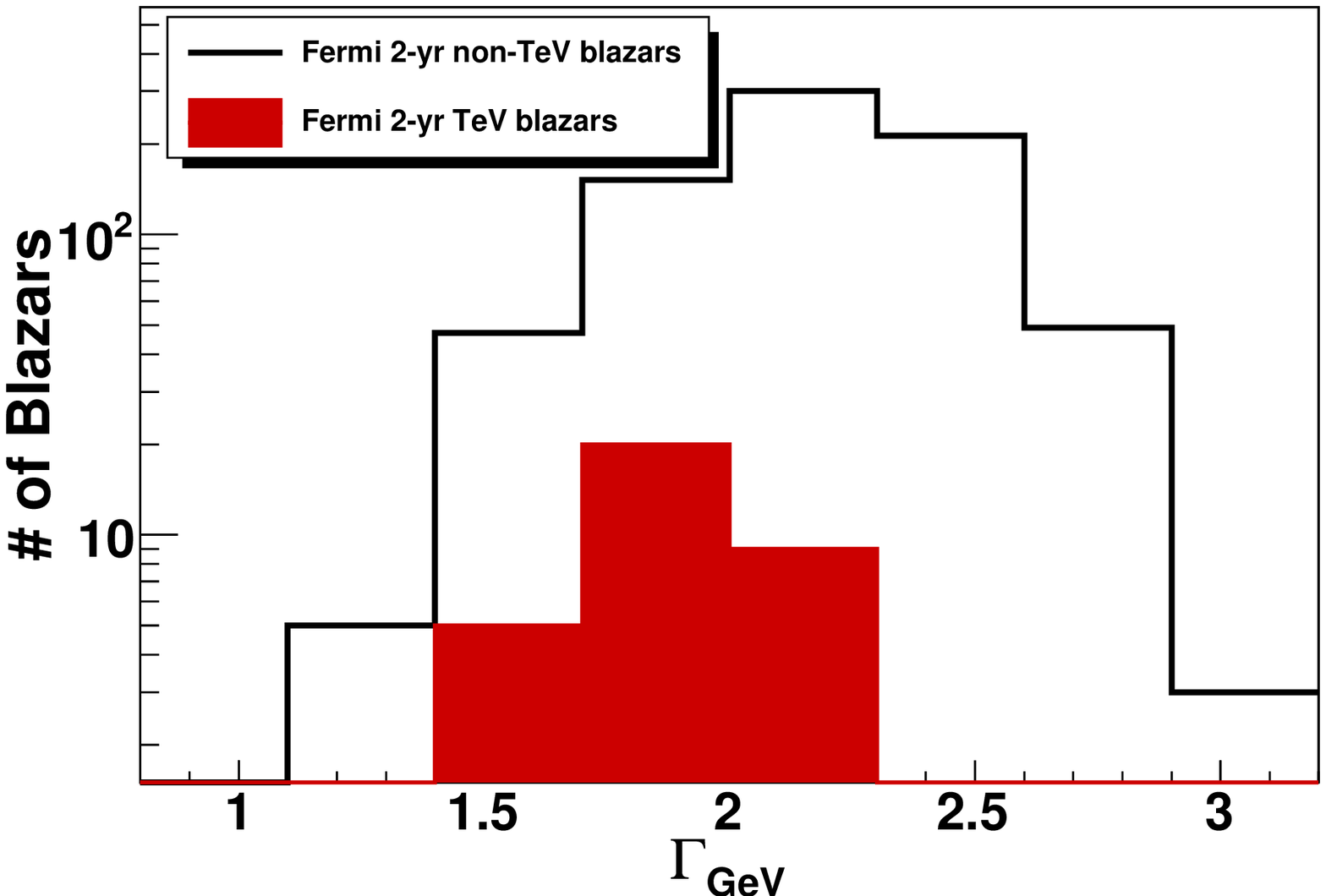}
\includegraphics[width=80mm]{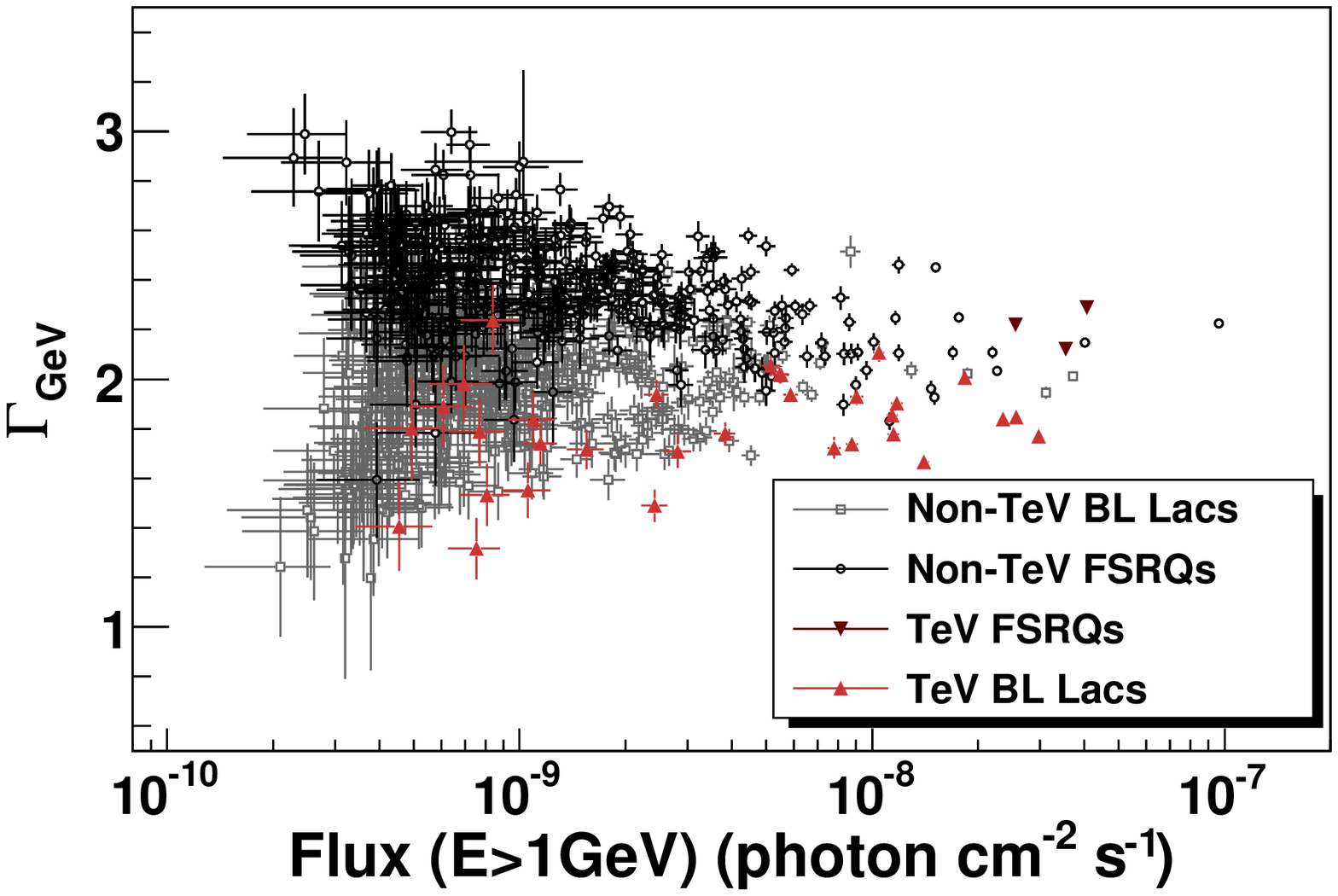}
\caption{{\footnotesize Top: GeV photon index ($\Gamma_\mathrm{GeV}$)
distribution for non-TeV blazars (black solid line) and TeV blazars (red shaded
area) from the 2FGL catalog~\citep{2fgl}. TeV emitters tend to have hard
spectra in the GeV band. Bottom: $\Gamma_\mathrm{GeV}$ vs. integral flux for
$E>1\mathrm{GeV}$ for the same sample, with gray open squares for non-TeV
BL Lacs, black open circles for non-TeV FSRQs, red triangles for TeV BL Lacs and
dark red inverse triangles for TeV FSRQs. TeV emitters are mostly located along
the bottom of the distribution. TeV FSRQs are at the edge of the group where the luminosity is relatively higher and the index softer. }}
\label{fig:gev_indices}
\end{figure}
\begin{figure*}[t]
\centering
\includegraphics[width=80mm]{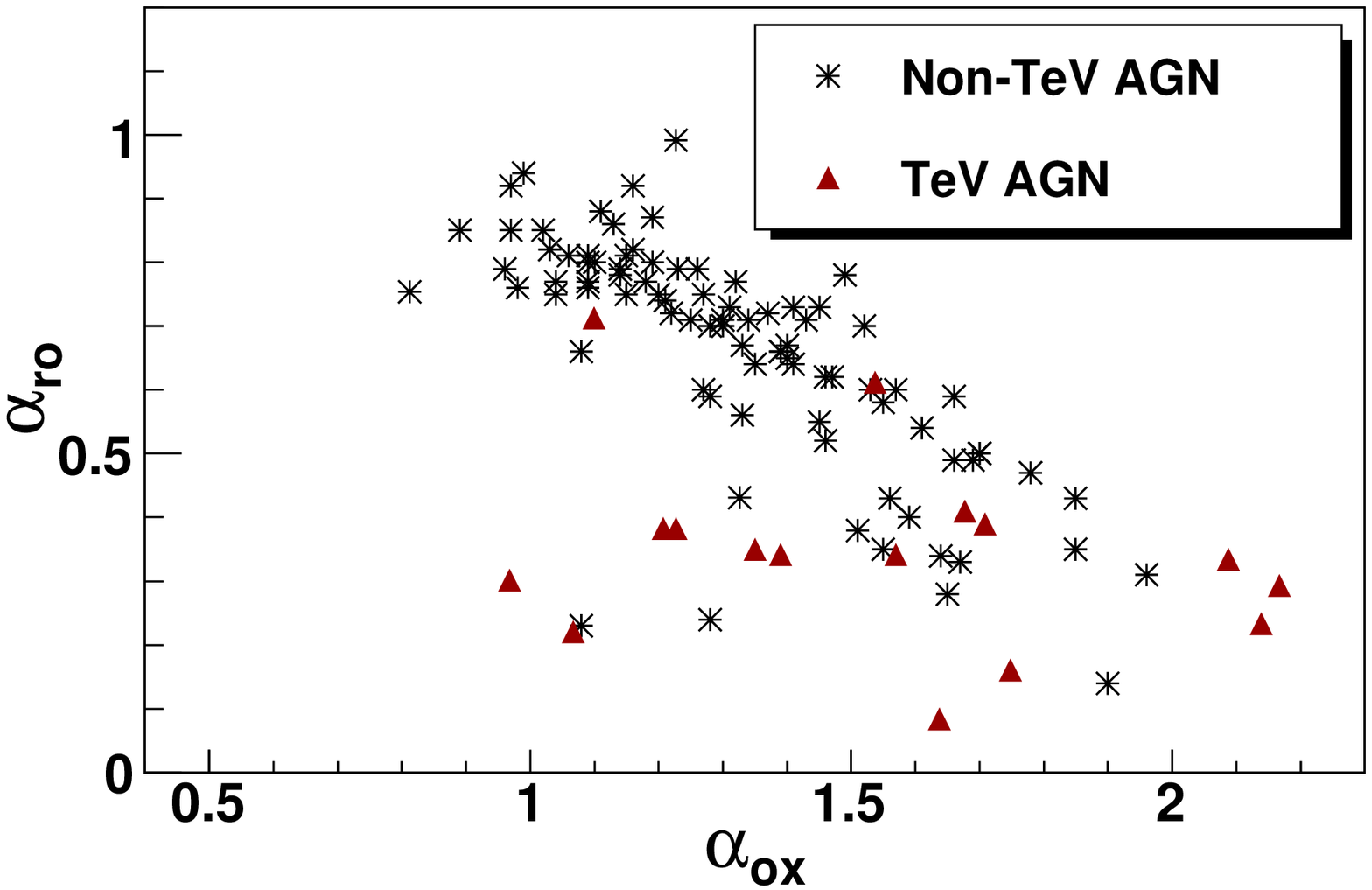}
\includegraphics[width=80mm]{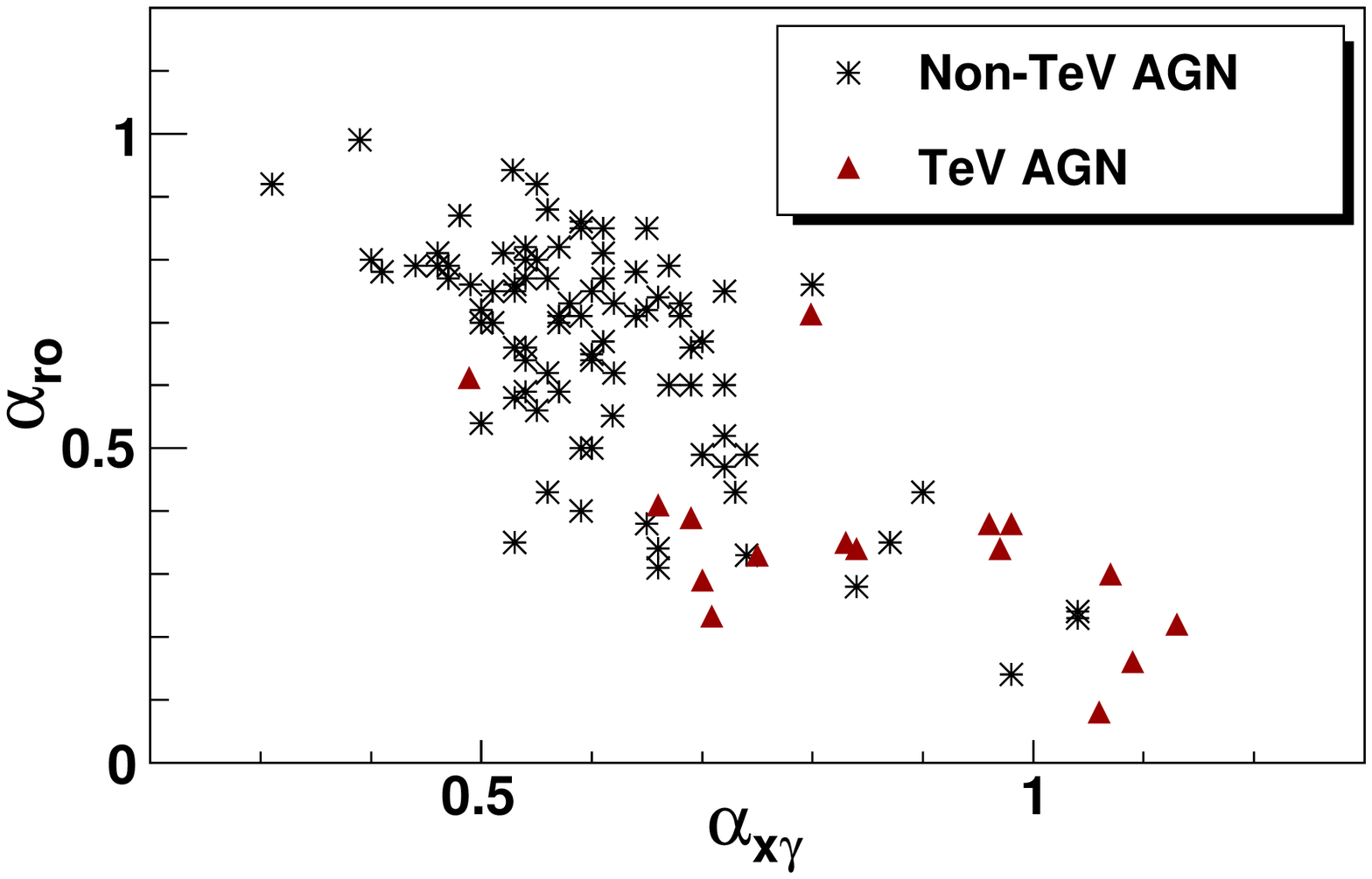}
\caption{{\footnotesize Scatter plots of $\alpha_\mathrm{ro}$ (5\,GHz -- 5000\,\AA), $\alpha_\mathrm{ox}$ (5000\,\AA -- 1\,keV) and $\alpha_\mathrm{x\gamma}$ (1\,keV -- 100\,MeV) spectral indices for the {\it Fermi}-LAT bright AGN sample (LBAS). Data are taken from~\citep{steve_LBAS}. Note the correlation between $\alpha_\mathrm{ro}$ and $\alpha_\mathrm{x\gamma}$. Red triangles represent TeV-detected AGNs, and asterisks non-TeV-detected ones. TeV-emitters seem to be well isolated in $\alpha_\mathrm{ro}$ - $\alpha_\mathrm{x\gamma}$ parameter space.}}
\label{fig:LBAS_scatters}
\end{figure*}
\section{\fermi TeV Blazars}\label{fermi_tev_agn}
The second {\it Fermi}-LAT catalog (2FGL) contains 1873 sources, among which 1062 are AGN, with 435 BL Lacs, 370 FSRQs and 257 AGNs of unknown class~\citep{2fgl}. 
Thirty six of these AGNs are TeV emitters. 
Figure~\ref{fig:gev_indices} (top) shows the distribution of \fermi spectral indices ($\Gamma_\mathrm{GeV}$) for TeV  and non-TeV blazars in the 2FGL catalog. 
TeV-detected blazars tend to have harder GeV indices. 
As can be seen from Figure~\ref{fig:gev_indices} (bottom), another distinguishing parameter for TeV emitters within the \fermi blazar population is the integral flux for energies above 1\,GeV.
It follows that an effective method for TeV-candidate selection in the HE $\gamma$-ray band is to look for bright hard spectrum sources, and select the candidates based on their extrapolated fluxes at VHE energies. 
For all TeV blazars, $\Gamma_\mathrm{GeV}<2.3$ and for most of them $\Gamma_\mathrm{GeV}<2$, in agreement with an inverse-Compton peak frequency ($\nu_\mathrm{IC}$) located in the high-energy tail of the \fermi range or beyond.  
Figure~\ref{fig:LBAS_scatters} shows scatter plots of spectral indices of \emph{Fermi}-bright AGN~\citep{steve_LBAS} in radio, optical and X-ray bands, comparing TeV and non-TeV sources.
The TeV and non-TeV AGNs occupy separate regions in the parameter space, consistent with the results in~\citep{steve_LBAS}, considering that most TeV AGNs are HBLs.

\begin{figure}[h]
\centering
\includegraphics[width=80mm]{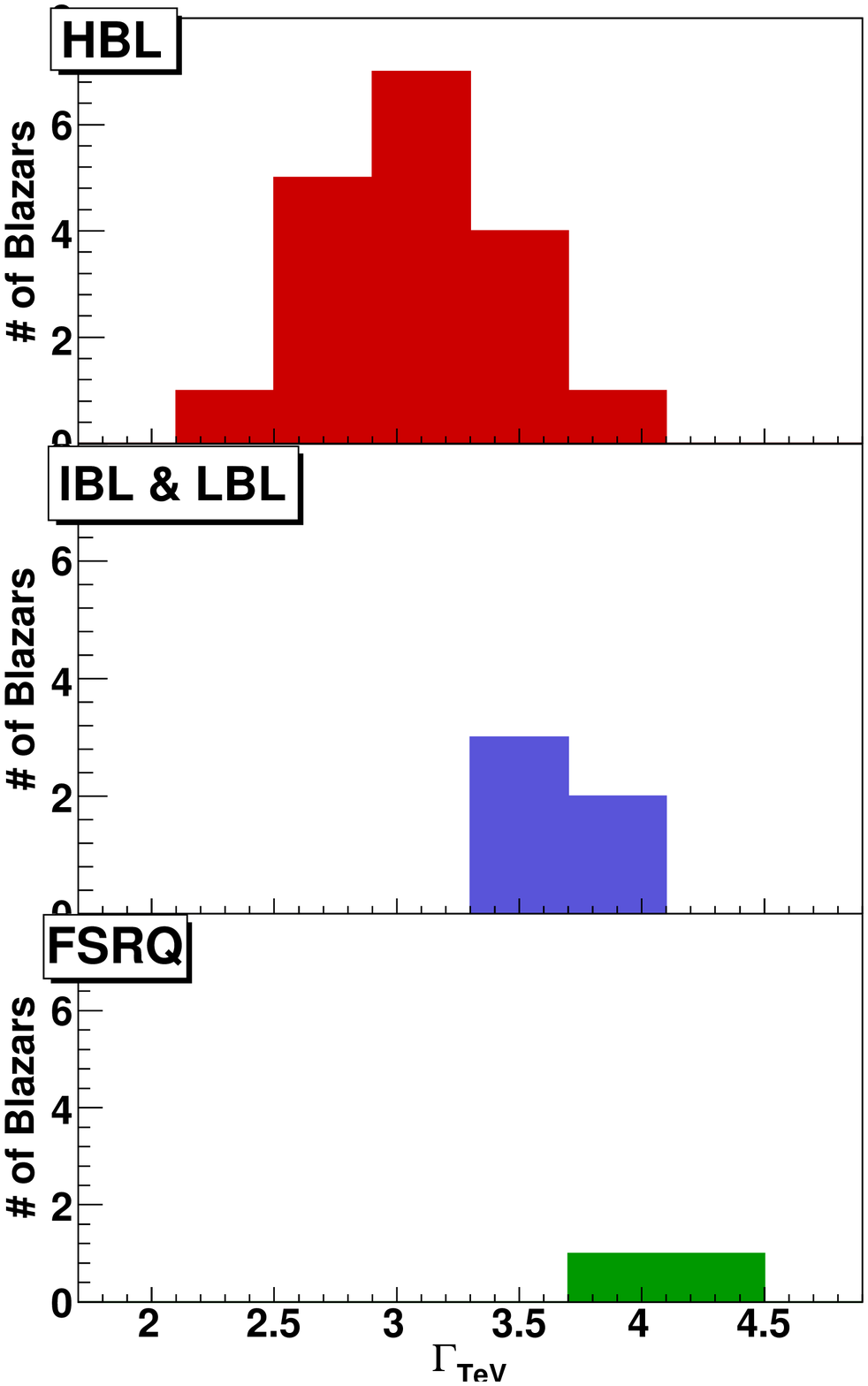}
\caption{{\footnotesize TeV photon index distributions for our sample (see Table~\ref{tab:sample}). For blazars that have multiple published results in Table~\ref{tab:sample}, the most recent one was used. Top, middle and bottom panels show HBLs, IBLs+LBLs, and FSRQs, respectively. HBLs tend to have harder spectra than the rest of the sample. FSRQs have the softest spectra. Note that the TeV indices are not EBL-corrected.}}
\label{fig:tev_indices}
\end{figure}

\section{Data Sample}\label{sec:sample}
Our blazar sample contains all blazars with a published VHE spectrum before 2011 February, including a total of 26 sources (see Table~\ref{tab:sample}): 19 HBLs, 3 IBLs, 2 LBLs and 2 FSRQs. 
TeV spectral index distributions of the whole sample are shown in Figure~\ref{fig:tev_indices}.
Three of these blazars have insecure redshifts either because the spectroscopic
measurements were inconclusive, or the calculations were made indirectly based on EBL absorption studies. 
References for the adopted redshift values in these three cases are given in Table~\ref{tab:sample}.
Seven of our targets were detected with EGRET and 23 of them are in the 2FGL catalog~\citep{2fgl}. 
The ones that are missing in the \fermi data (1ES 0229+200, 1ES 0347-121, PKS 0548-322) are very hard spectrum sources that
would be weak in the \fermi band. 
More than half of the sample have been detected multiple times in the VHE band. 
These multiple detections extending over several years and obtained mostly with different instruments suggest that spectral variability in the VHE band is a common property for VHE blazars. 
Even though no general pattern has been established for VHE variability, several sources have been observed to have a flux increase up to a few times their baseline emission~\citep{ver_wCom_flare,magic_501_flare,pks2155_flare}, occasionally accompanied by a change in spectral index~\citep{magic_501_flare} and minute-scale flux doubling times~\citep{magic_501_flare,pks2155_flare}. 
\par
The first 27 month \fermi data and archival VHE spectra published before 2011 February were used to construct combined GeV-TeV SEDs in this study. 
Only in seven cases (RGB J0710+591, 1ES 1218+304, PKS 1222+21 (4C +21.35), PKS 1424+240, PKS 2155-304, and two different measurements for 3C 66A) were the VHE data found to overlap with the \fermi era.
The remainder of the VHE data were taken before the \fermi mission. 

\begin{table*}
\begin{threeparttable}
\begin{tiny}
\begin{tabular}{|cccccccccc|}
\hline
Name 		& SED type 	& $z$ 	&$\mathrm{log_{10}}(\nu_\mathrm{syn}/1\,\mathrm{Hz})$	& \fermi var.	& \fermi state	& $\Gamma_\mathrm{TeV}$			&$I$	&$E_\mathrm{th}$ (GeV)& Reference 	\\
		& (1)		& (2)	& (3)			& (4)		& (5)		& (6)					& (7)	&(8) 			& (9)	\\
\hline
\hline
RGB J0152+017	& HBL 		& 0.080 & --	& --		& average	& $2.95\pm0.36_\mathrm{stat}\pm0.20_\mathrm{syst}$ 	& 2\%		& 300		& \citep{aharonian08} \\
\hline
3C 66A* 	& IBL		& 0.444\tnote{a}& 15.63	& 171		& MAGIC	& $3.64\pm0.39_\mathrm{stat}\pm0.25_\mathrm{syst}$ 	& 6\%	& 200			& \citep{aleksic11a} \\
		&		&	&	&  	&VERITAS	& $4.1\pm0.4_\mathrm{stat}\pm0.6_\mathrm{syst}$ 		& 8\%	& 100			& \citep{acciari09b} \\
\hline
1ES 0229+200 	& HBL 		& 0.140	&19.45	& -- 	& average 	& $2.50\pm0.19_\mathrm{stat}\pm0.10_\mathrm{syst}$ 		& 2\%	& 580			& \citep{aharonian07c} \\
\hline
1ES 0347-121 	& HBL 		& 0.188 &17.94	& --	&average 	& $3.10\pm0.23_\mathrm{stat}\pm0.10_\mathrm{syst}$ 		& 2\%	& 250			& \citep{aharonian07b} \\
\hline
PKS 0548-322	& HBL		& 0.069	&16.84	& --	&average	& $2.86\pm0.34_\mathrm{stat}\pm0.10_\mathrm{syst}$ 		& 1\%	& 200			& \citep{aharonian10} \\
\hline
RGB J0710+591	& HBL		& 0.125	&21.05	& 6 	&VERITAS	& $2.69\pm0.26_\mathrm{stat}\pm0.20_\mathrm{syst}$ 		& 3\%	& 300			& \citep{acciari10c} \\
\hline
S5 0716+714	& LBL		& 0.300	&14.46	& 266	&high		& $3.45\pm0.54_\mathrm{stat}\pm0.2_\mathrm{syst}$ 		& 9\%	& 400			& \citep{anderhub09a} \\
\hline
1ES 0806+524	& HBL		& 0.138	&16.56	& 20 	&average	& $3.6\pm1.0_\mathrm{stat}\pm0.3_\mathrm{syst}$ 		& 2\%	& 300			& \citep{acciari09a} \\
\hline
1ES 1011+496	& HBL		& 0.212	&16.74	& 16	&high		& $4.0\pm0.5_\mathrm{stat}\pm0.2_\mathrm{syst}$			& 6\%	& 200			& \citep{albert07d} \\
\hline
1ES 1101-232	& HBL		& 0.186	&16.88	& 1 	&average	& $2.94\pm0.20_\mathrm{stat}$					& 3\%	& 225			& \citep{aharonian07a} \\
\hline
Markarian 421*	& HBL		& 0.031	&18.49	& 44	&medium	& $2.20\pm0.08_\mathrm{stat}\pm0.2_\mathrm{syst}$ 			& 50--200\%	& 200		& \citep{albert07b} \\
\hline
Markarian 180	& HBL		& 0.046	&18.61	& 10	&average	& $3.3\pm0.7_\mathrm{stat}\pm0.2_\mathrm{syst}$			& 11\%	& 200			& \citep{albert06} \\
\hline
1ES 1218+304*	& HBL		& 0.182	&19.14	& 15	&average	& $3.08\pm0.34_\mathrm{stat}\pm0.2_\mathrm{syst}$ 		& 7\%	& 200			& \citep{acciari09c} \\
		&		&	&	&  		& VERITAS	& $3.07\pm0.09_\mathrm{stat}$				& 6\%	& 200			& \citep{acciari10b} \\
\hline
W Comae*	& IBL		& 0.102 &14.84	& 47	&high		& $3.81\pm0.35_\mathrm{stat}\pm0.34_\mathrm{syst}$ 		& 9\%	& 200			& \citep{acciari08} \\
\hline
PKS 1222+21 & FSRQ		& 0.432	&13.27	& 101	&MAGIC	&$3.75\pm0.27_\mathrm{stat}\pm0.2_\mathrm{syst}$ 			& 100\%	& 100			& \citep{aleksic11b} \\
\hline
3C 279		& FSRQ		& 0.536	&12.67	& 898	&high		& $4.1\pm0.7_\mathrm{stat}\pm0.2_\mathrm{syst}$			& 15\%	& 200			& \citep{3c279} \\	
\hline
PKS 1424+240	& IBL		& 0.260\tnote{b}&15.7	& 26	&VERITAS	& $3.80\pm0.5_\mathrm{stat}\pm0.3_\mathrm{syst}$ 	& 3\%	& 140			& \citep{acciari10a} \\
\hline
H 1426+428	& HBL		& 0.129	&18.55	& 7	&average	& --								& 3\%	& 1000			& \citep{horns04} \\
\hline
PG 1553+113	& HBL		& 0.4\tnote{c}	&16.49	& 44	&high		& $3.4\pm0.1_\mathrm{stat}\pm0.2_\mathrm{syst}$		& 8\%	& 200			& \citep{aleksic10} \\
\hline
Markarian 501*	& HBL		& 0.034	&16.84	& 46	&low		& $2.79\pm0.12_\mathrm{stat}$					& 20\%	& 200			& \citep{anderhub09b} \\
\hline
1ES 1959+650*	& HBL		& 0.048	&18.03	& 16	&low		& $2.58\pm0.18_\mathrm{stat}$					& 10\%	& 200			& \citep{1es_1959} \\
\hline
PKS 2005-489*	& HBL		& 0.071	&--	& 9	&average	& $3.20\pm0.16_\mathrm{stat}\pm0.10_\mathrm{syst}$ 		& 3\%	& 400			& \citep{acero10} \\
\hline
PKS 2155-304*	& HBL		& 0.117	&15.7	& 63	&HESS	& $3.34\pm0.05_\mathrm{stat}\pm0.1_\mathrm{syst}$ 			& 14\%	& 400			& \citep{aharonian09} \\
		&		&	&	& 		& low		& $3.53\pm0.06_\mathrm{stat}\pm0.10_\mathrm{syst}$ 	& 15\%	& 200			&\citep{abramowski10b} \\
\hline
BL Lacertae	& LBL		& 0.069	&14.28	& 35 	&high		& $3.64\pm0.54_\mathrm{stat}\pm0.2_\mathrm{syst}$ 		& 3\%	& 200			& \citep{albert07c} \\
\hline
1ES 2344+514*	& HBL		& 0.044	&16.4	& 10	&average	& $2.95\pm0.12_\mathrm{stat}\pm0.2_\mathrm{syst}$ 		& 11\%	& 200			& \citep{albert07a} \\
\hline
H 2356-309	& HBL		& 0.165	&17.24	& 8	&average	& $3.06\pm0.15_\mathrm{stat}\pm0.10_\mathrm{syst}$ 		& 2\%	& 240			&\citep{abramowski10a} \\
\hline
\end{tabular}
\end{tiny}
\begin{tablenotes}
\item[a]\citep{miller78,lanzetta93}
\item[b]\citep{prandini2011}
\item[c]\citep{mazin07}
\item[*] Blazars that are reported as variable in the TeV band, according to TeVCat (http://tevcat.uchicago.edu/).
\end{tablenotes}

\caption{{\footnotesize GeV-TeV properties of the VHE blazar sample taken from the literature. Columns (1), (2) and (3) show the spectral energy distribution (SED) type, redshift, and synchrotron peak frequency ($\nu_\mathrm{syn}$), respectively. \fermi variability indices (4) were taken from the 1FGL catalog~\citep{abdo2010}. \fermi states (5) are identified in this work using 27 month \fermi light curves as described in Section~\ref{fermi}. In cases where \fermi data are contemporaneous with TeV observations, the corresponding TeV instruments are listed in Column (5). TeV spectral indices (6) were taken from the references listed (9). TeV integral fluxes (7) are above the listed energy threshold (8) and in units of Crab Nebula flux. For the Crab Nebula unit conversions, spectral measurements above 350\,GeV from~\citep{ozlem_thesis} are used.}}
\label{tab:sample}
\end{threeparttable}
\end{table*}

\par
All VHE spectra were corrected for the EBL absorption using the model by~\citep{dominguez2011}. 
Other background models are also available ~\citep[e.g.,][]{finke_bg}. 
However, with a different EBL model, we do not expect any significant differences in our results up to a few TeV, given the redshift and energy range of our sample. 
See Section~\ref{tev_peaked} for a more detailed discussion on the EBL correction effects on our study.

\section{\emph{Fermi} Analysis}\label{fermi}
The fact that most of the GeV and TeV data are not contemporaneous makes it hard to interpret the combined spectra of blazars. 
Moreover, \emph{Fermi} data represent an average state over relatively long periods, whereas the VHE spectra consist of ``snapshots", mostly taken during flares. 
To account for blazar variability and the non-contemporaneous nature of the data set, for bright enough sources, the \fermi data were split into ``low" and ``high" flux states as described below. 
Thus, non-contemporaneous GeV and TeV measurements were matched in a more realistic way than directly using all the time-averaged \emph{Fermi} data.
Table~\ref{tab:sample} summarizes the \fermi flux states and VHE spectra used for each source.
\begin{figure*}[!b]
\centering
\includegraphics[width=80mm]{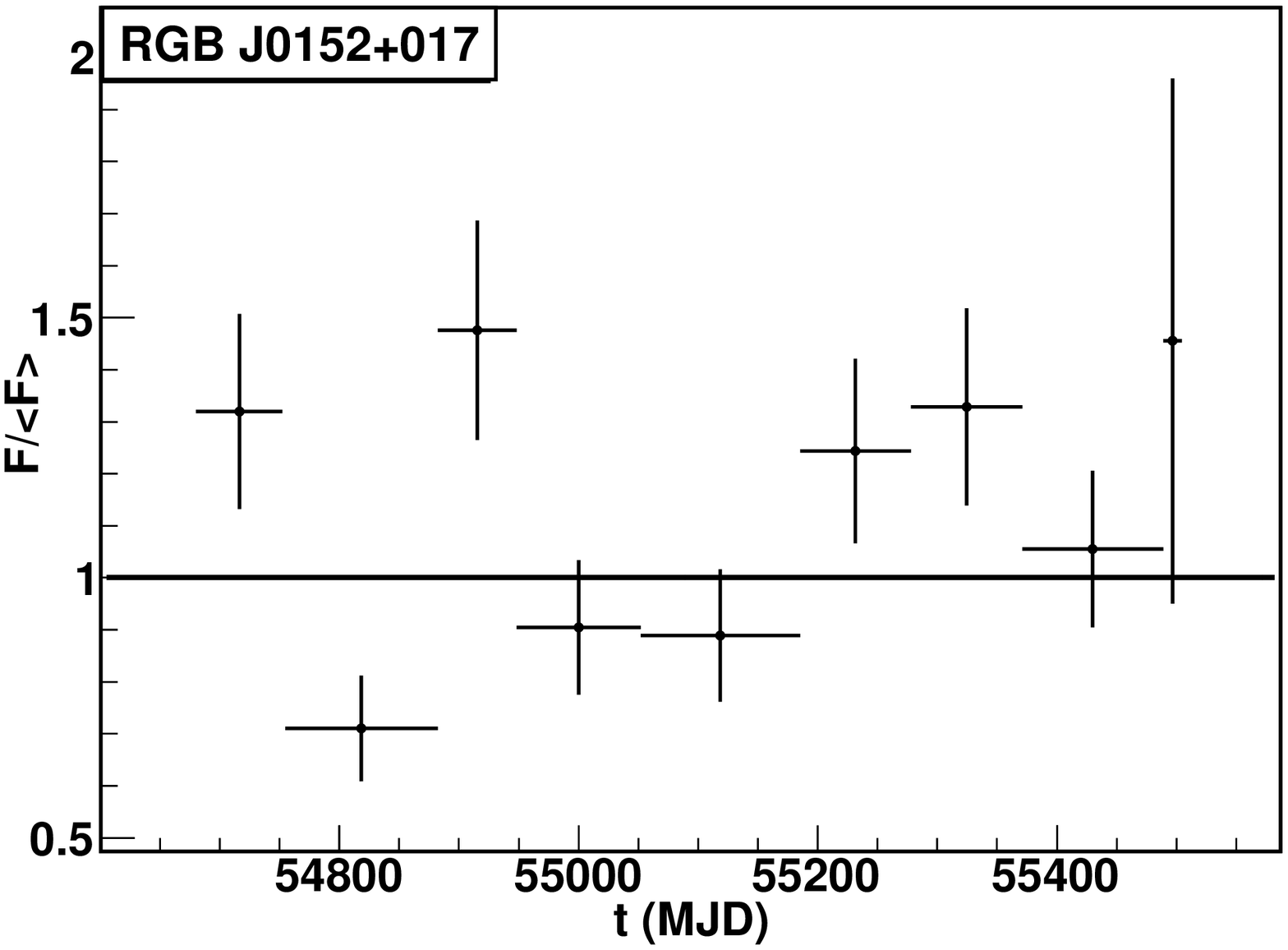}
\includegraphics[width=80mm]{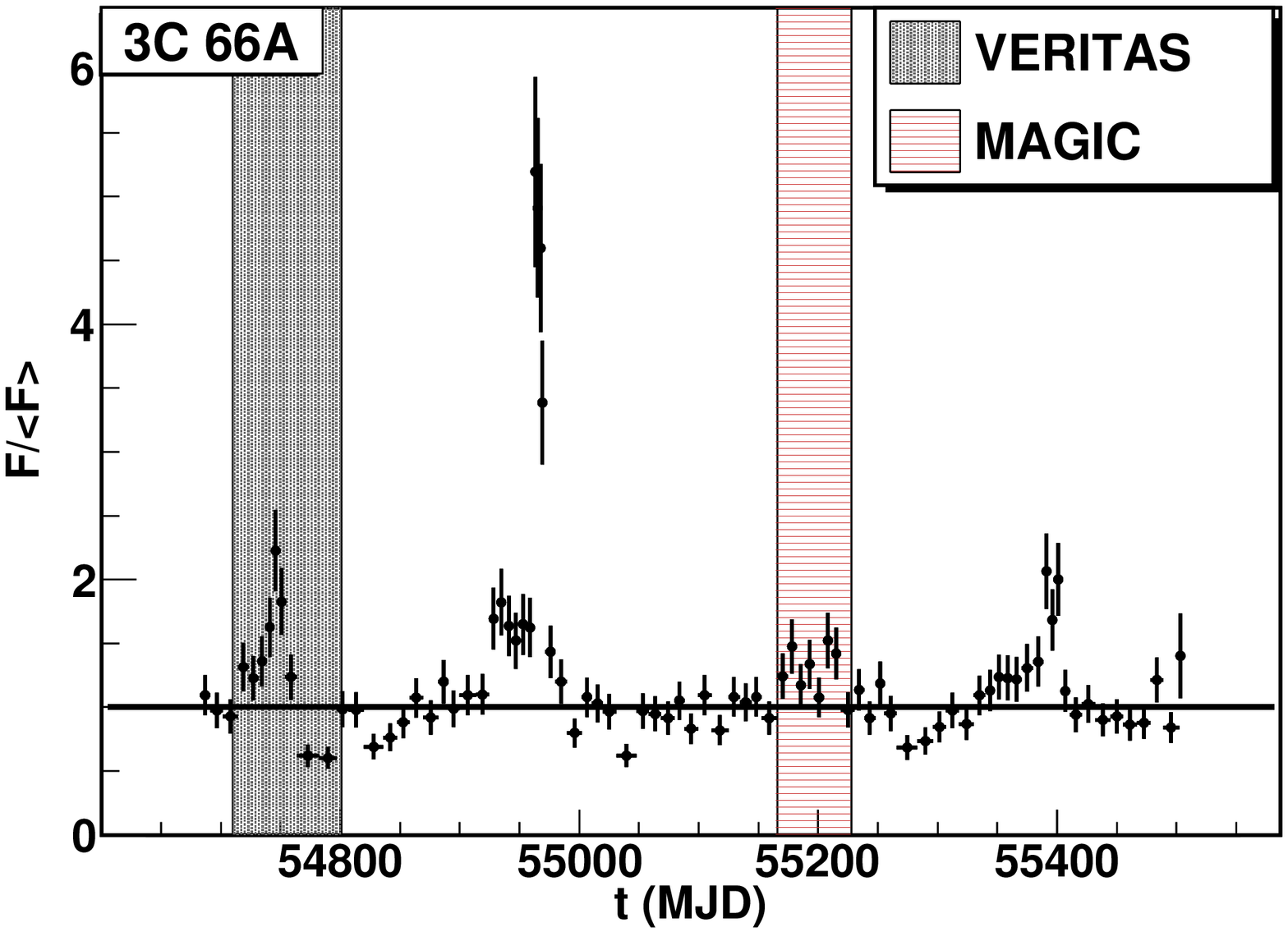}
\includegraphics[width=80mm]{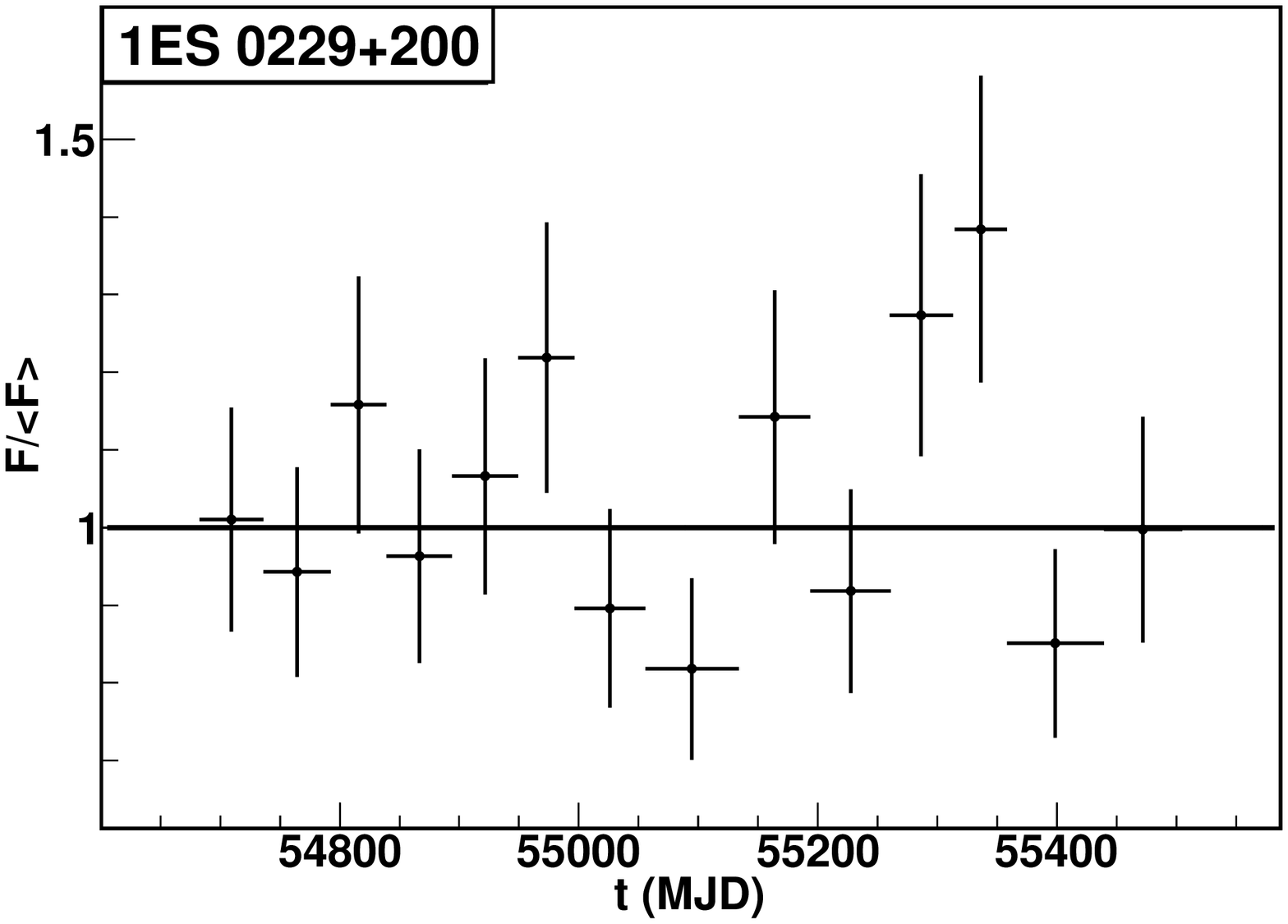}
\includegraphics[width=80mm]{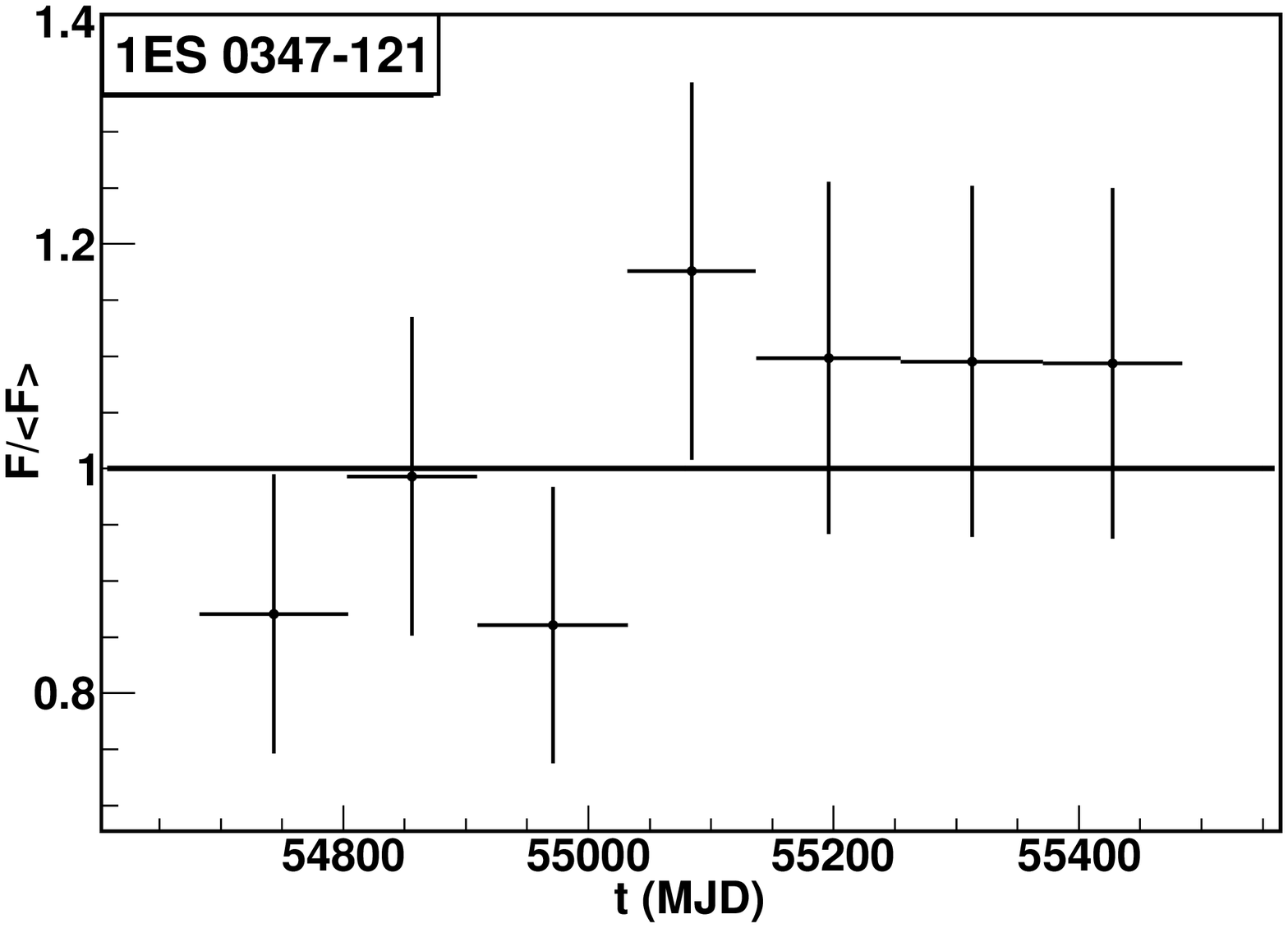}
\includegraphics[width=80mm]{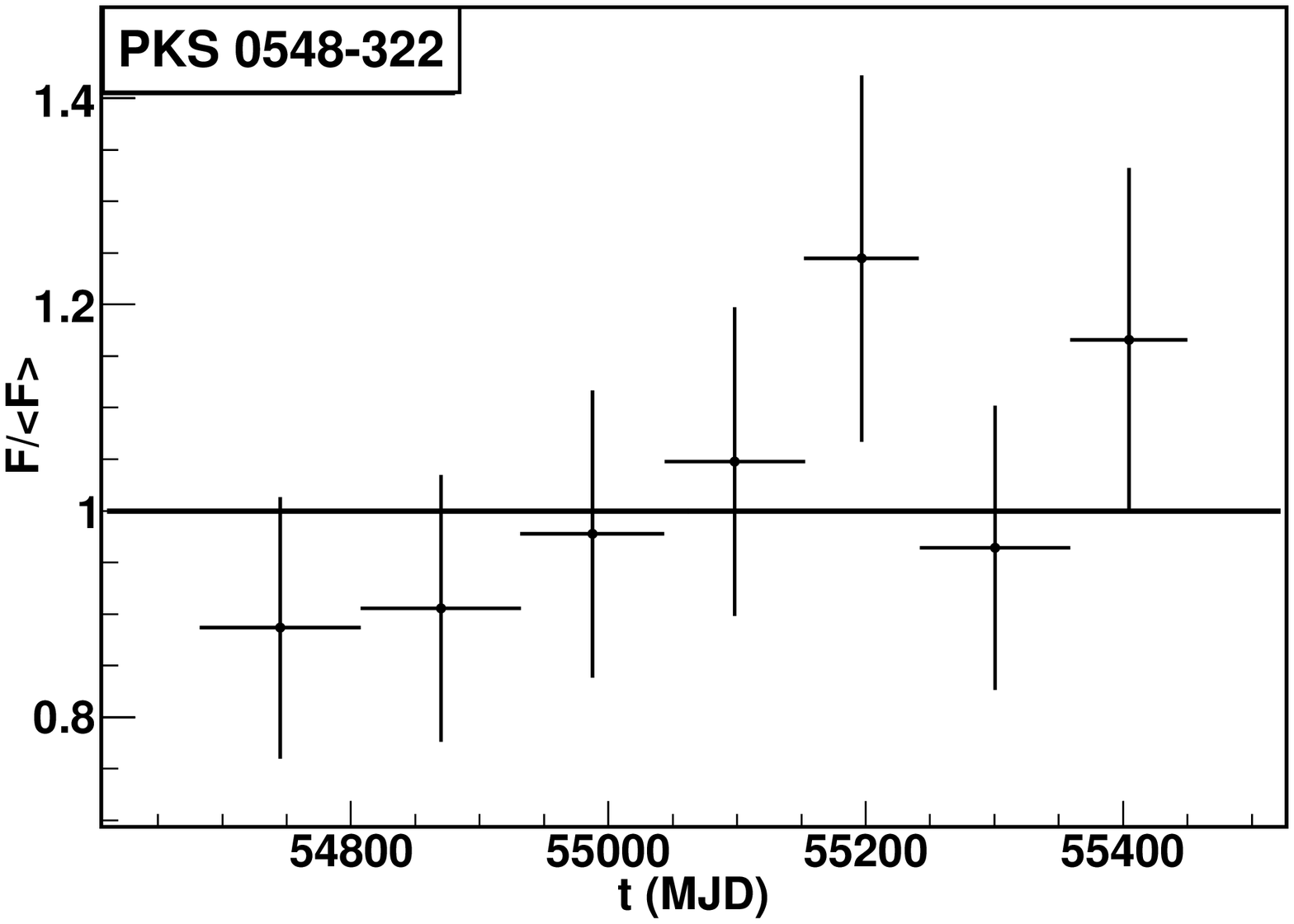}
\includegraphics[width=80mm]{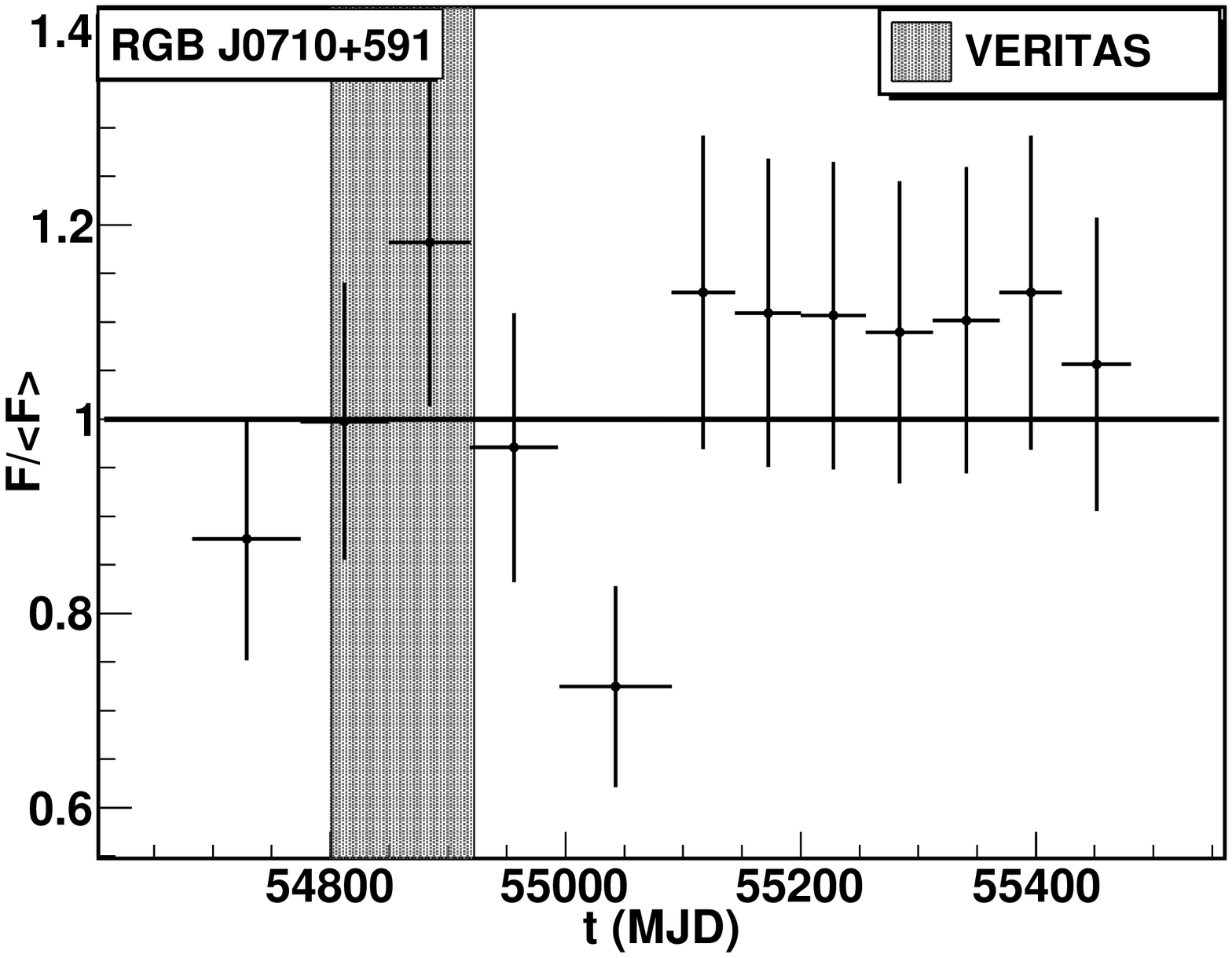}
\includegraphics[width=80mm]{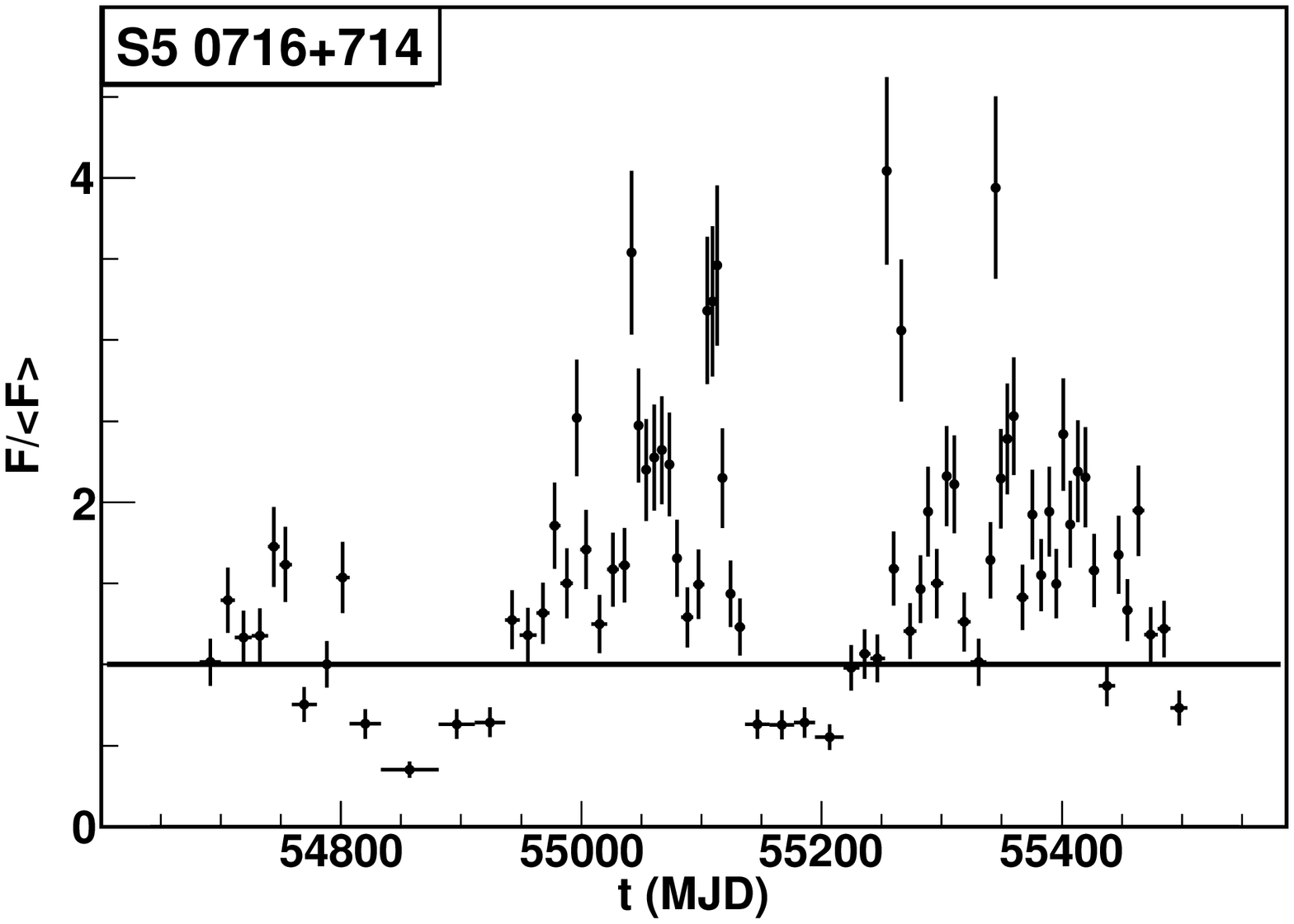}
\includegraphics[width=80mm]{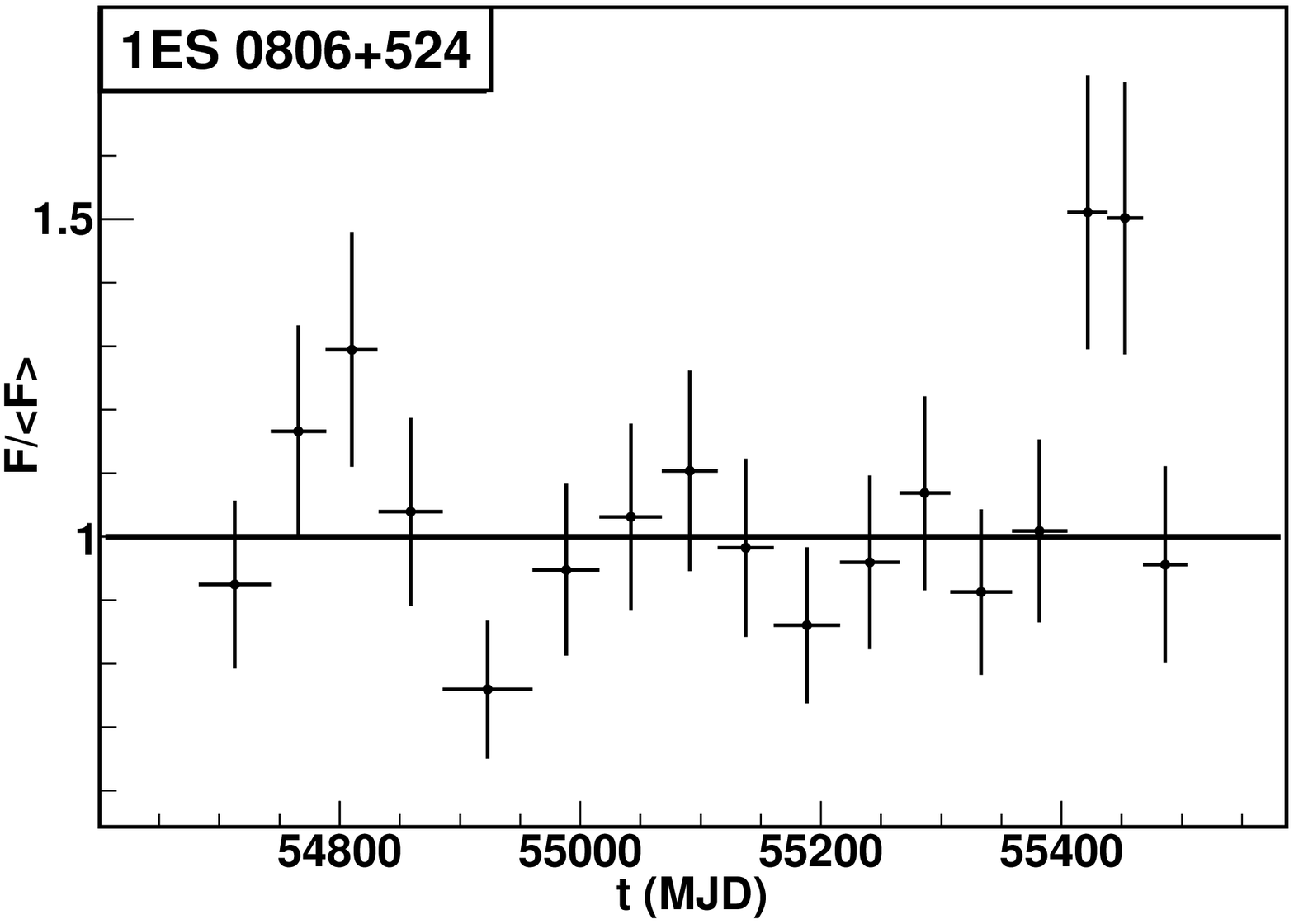}
\end{figure*}

\begin{figure*}
\centering
\includegraphics[width=80mm]{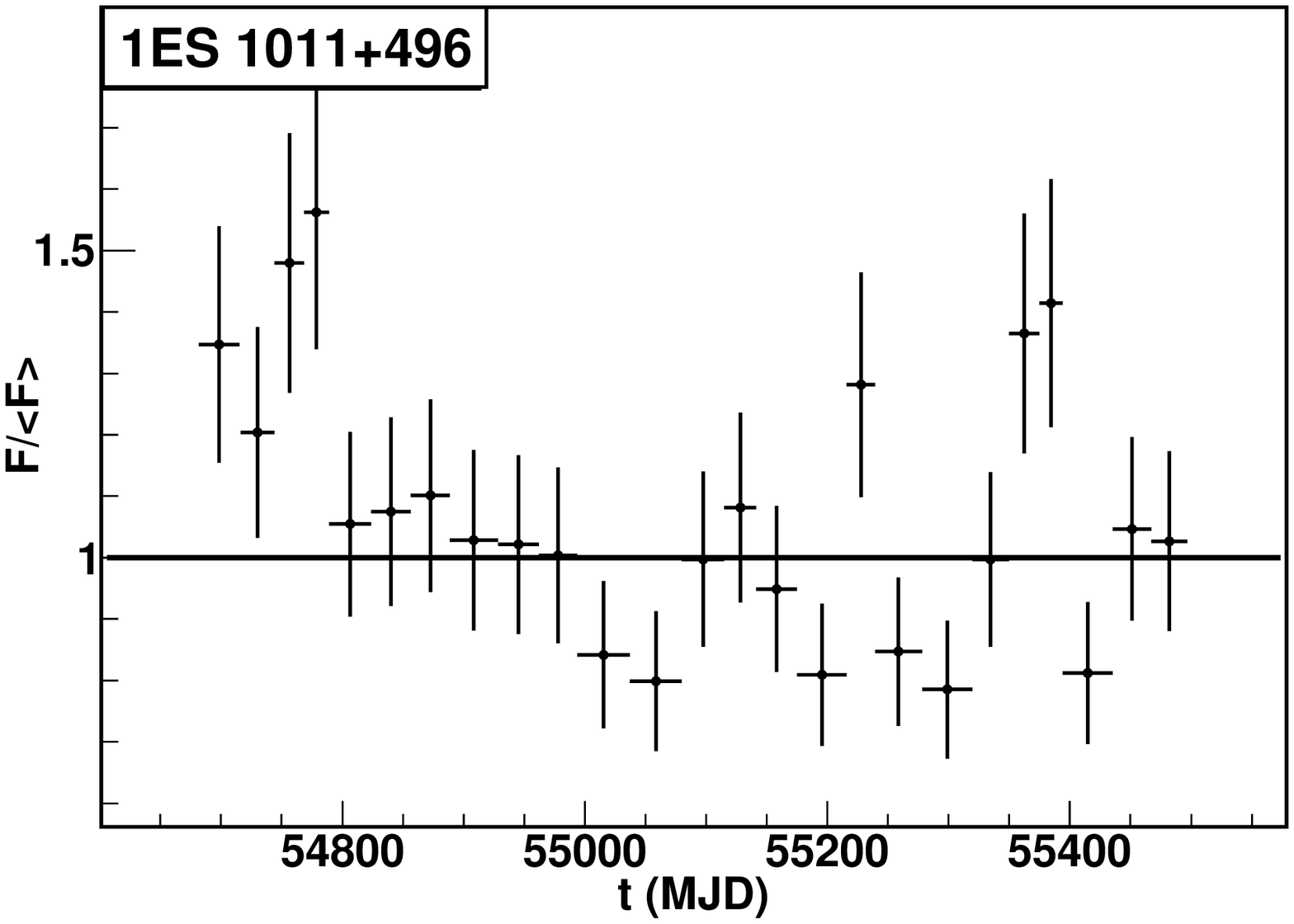}
\includegraphics[width=80mm]{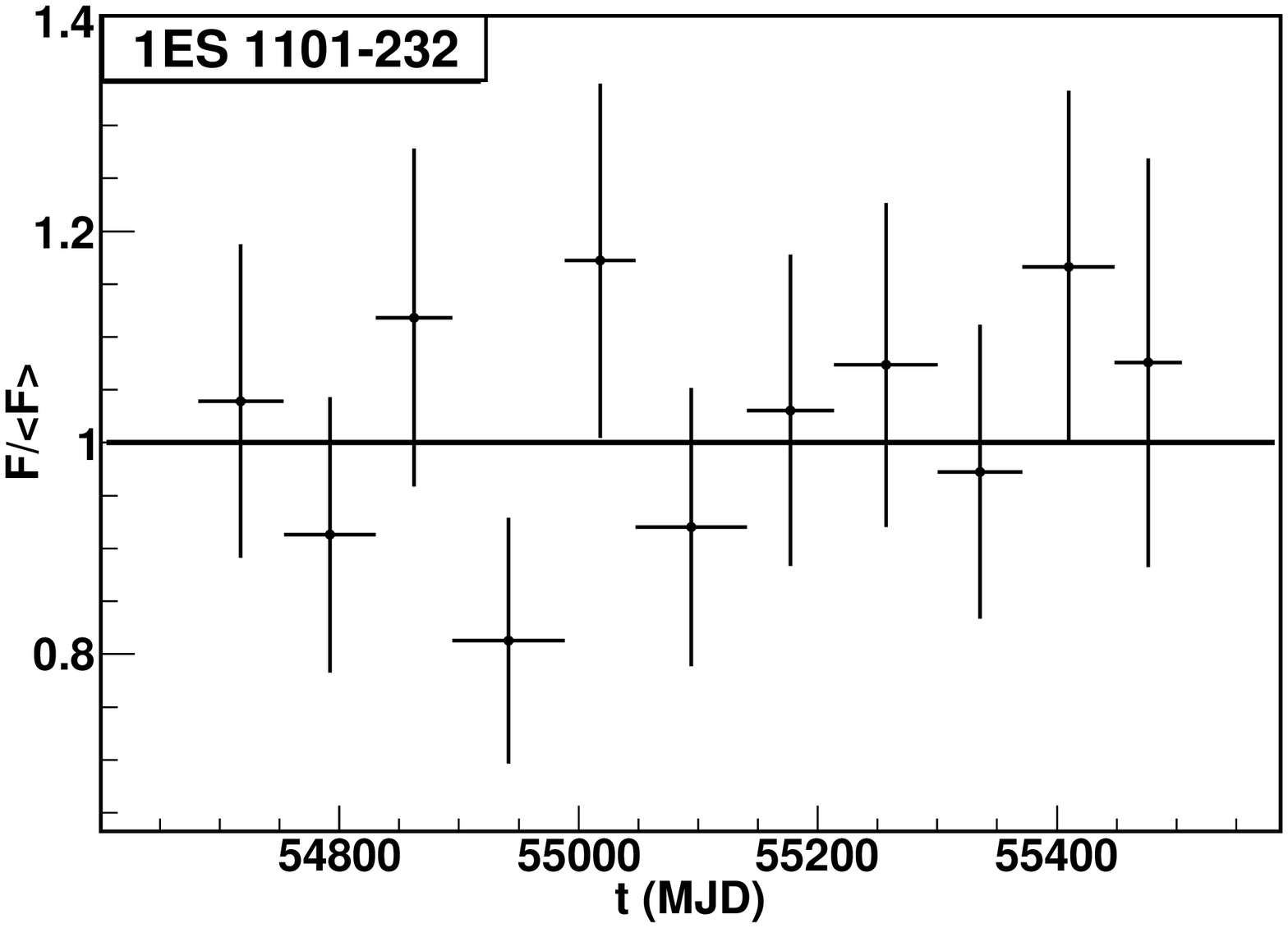}
\includegraphics[width=80mm]{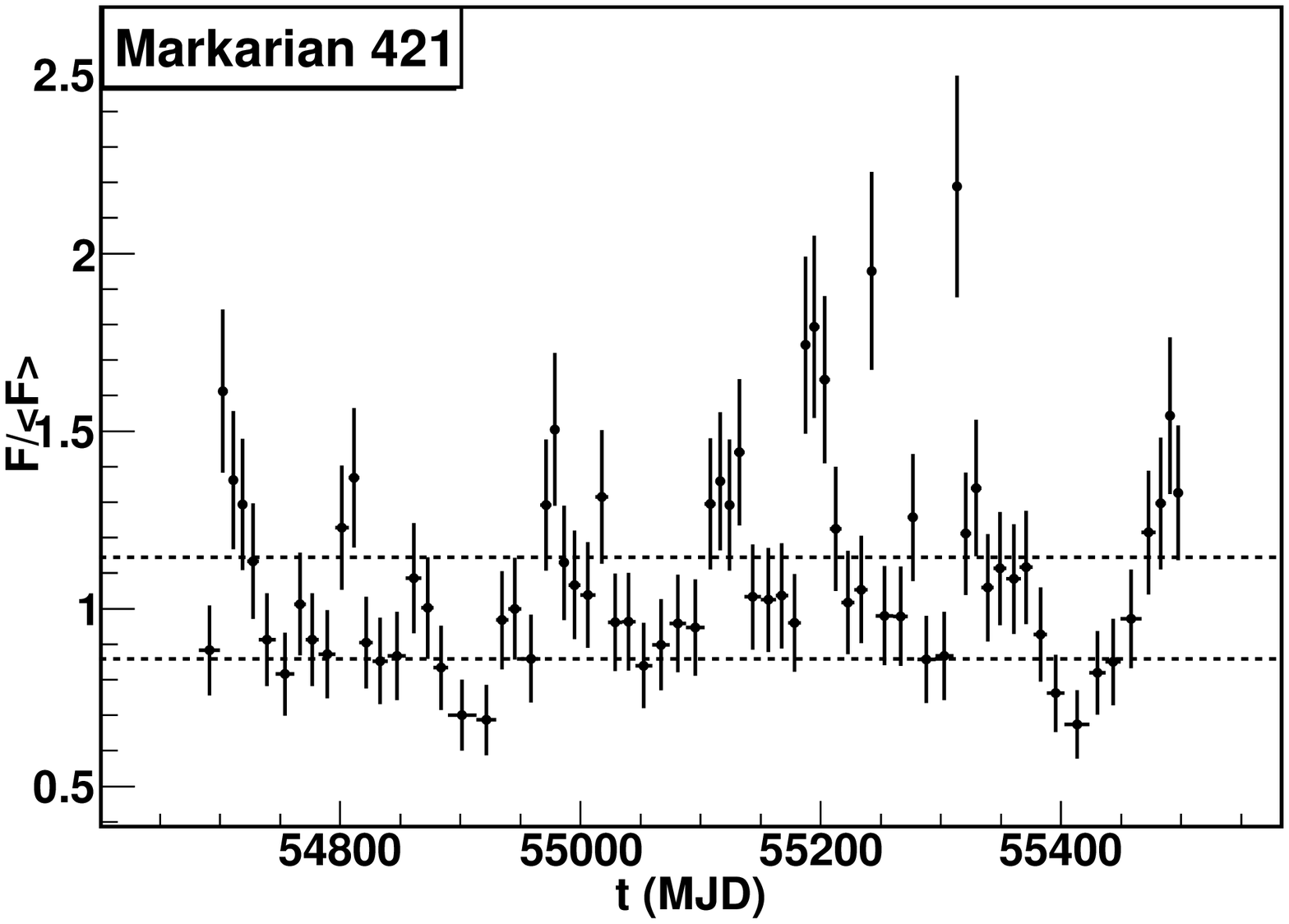}
\includegraphics[width=80mm]{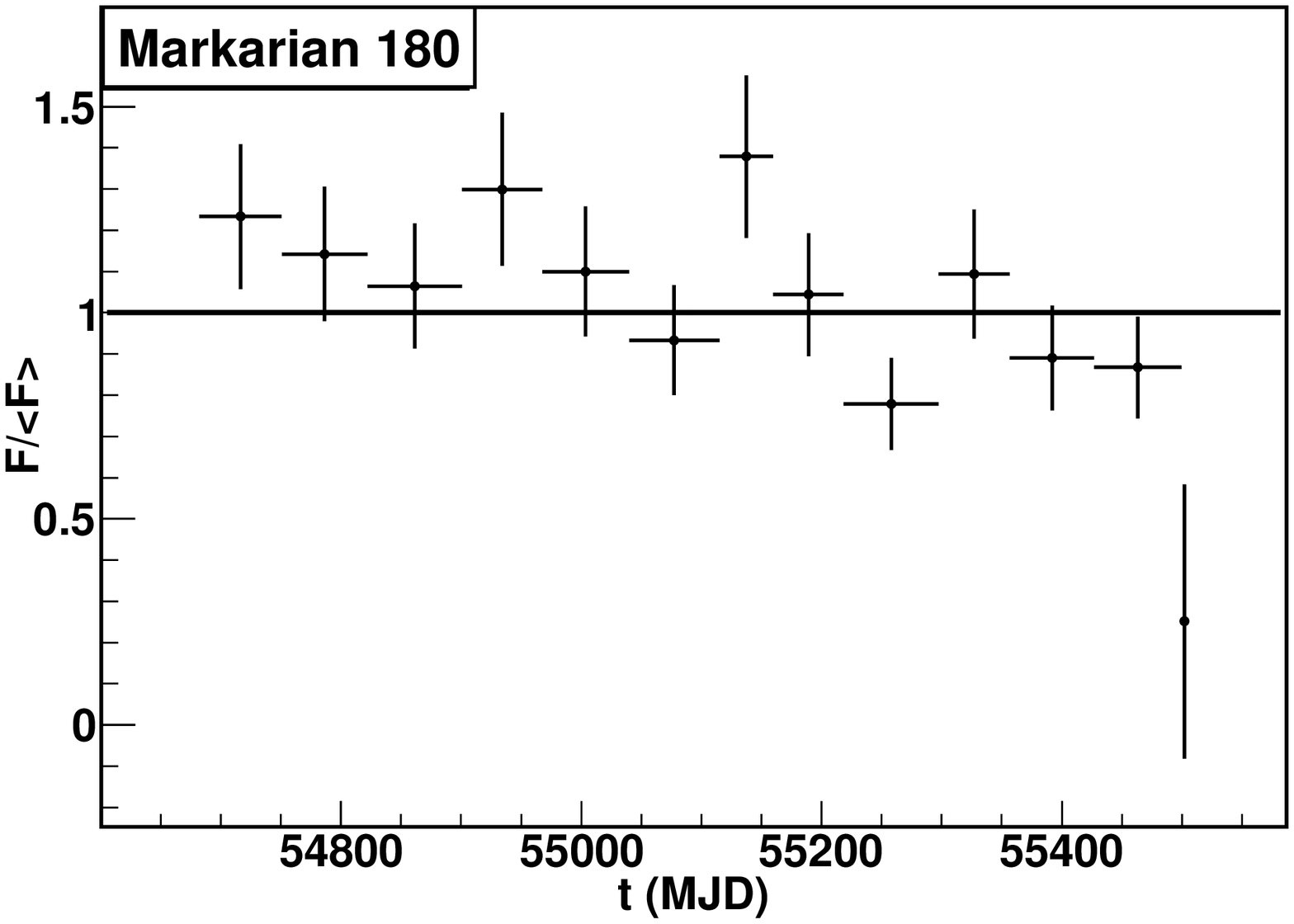}
\includegraphics[width=80mm]{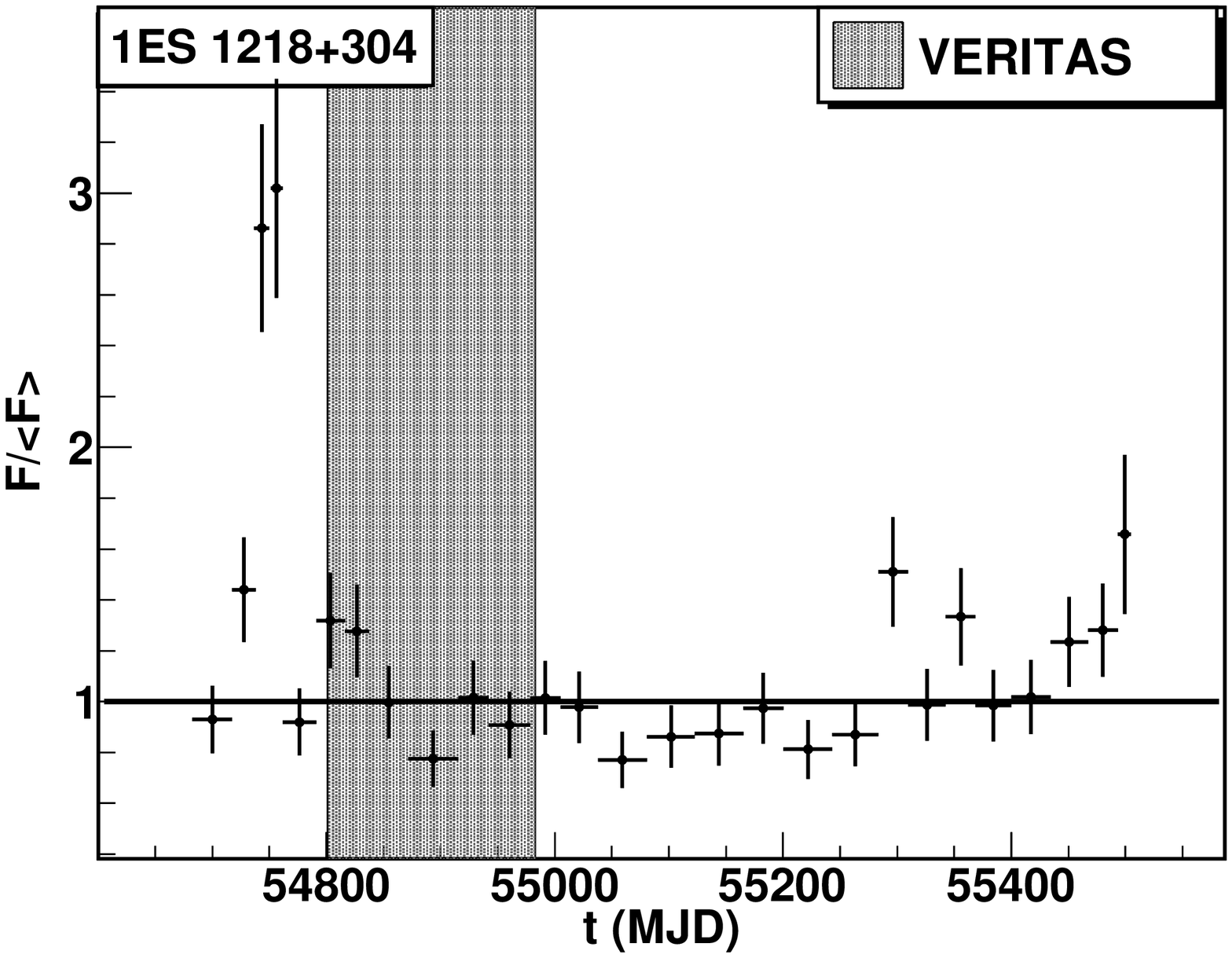}
\includegraphics[width=80mm]{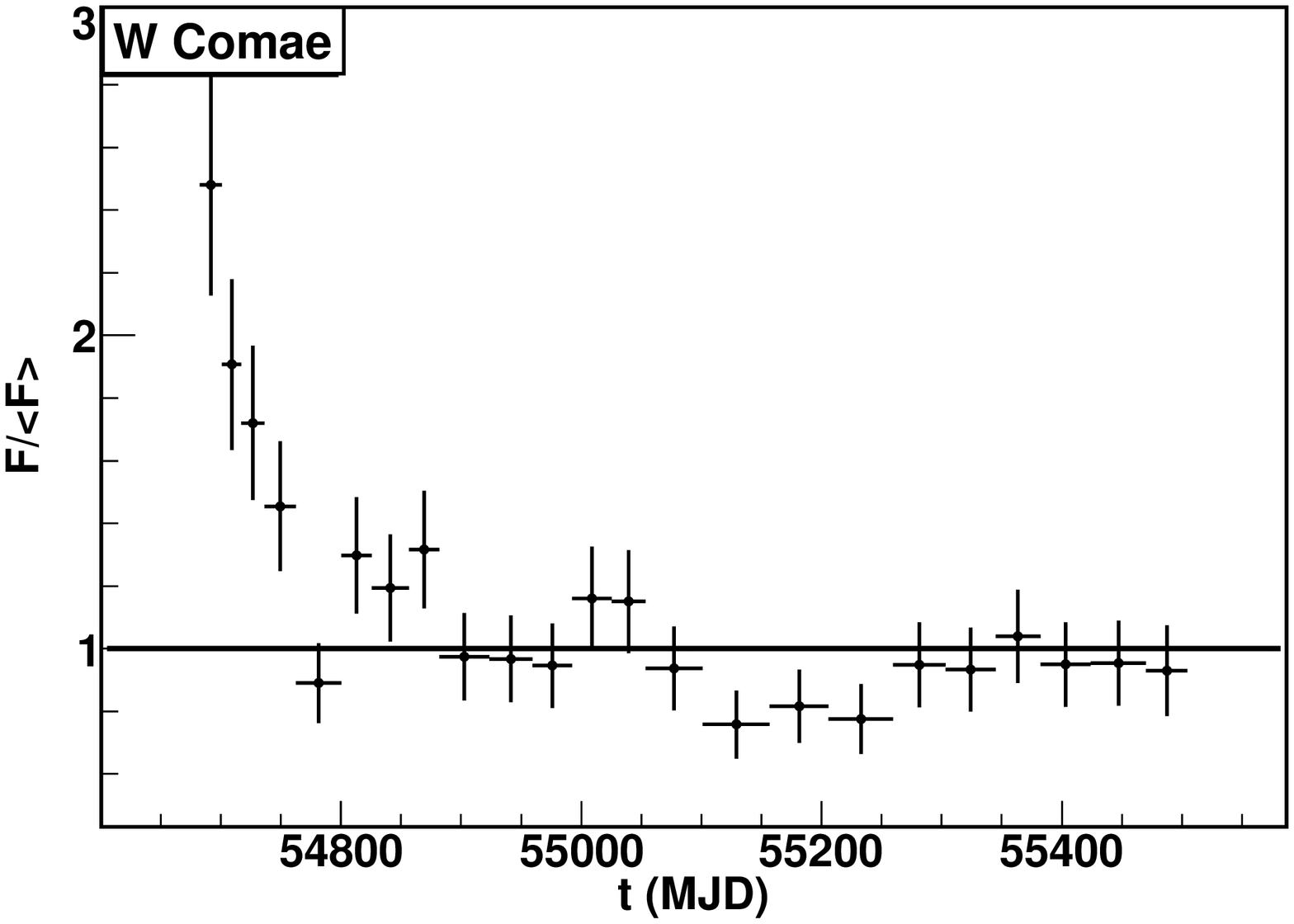}
\includegraphics[width=80mm]{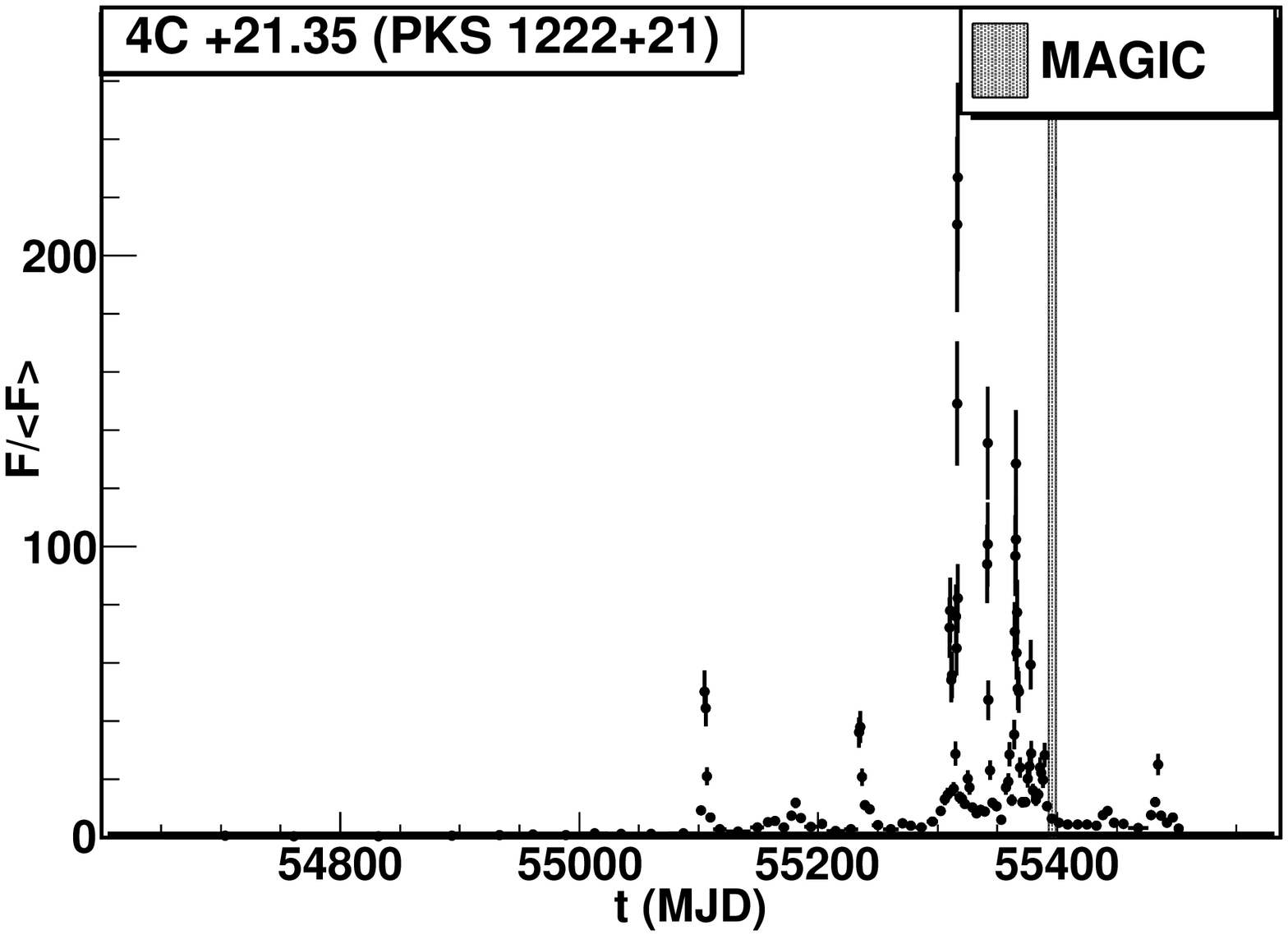}
\includegraphics[width=80mm]{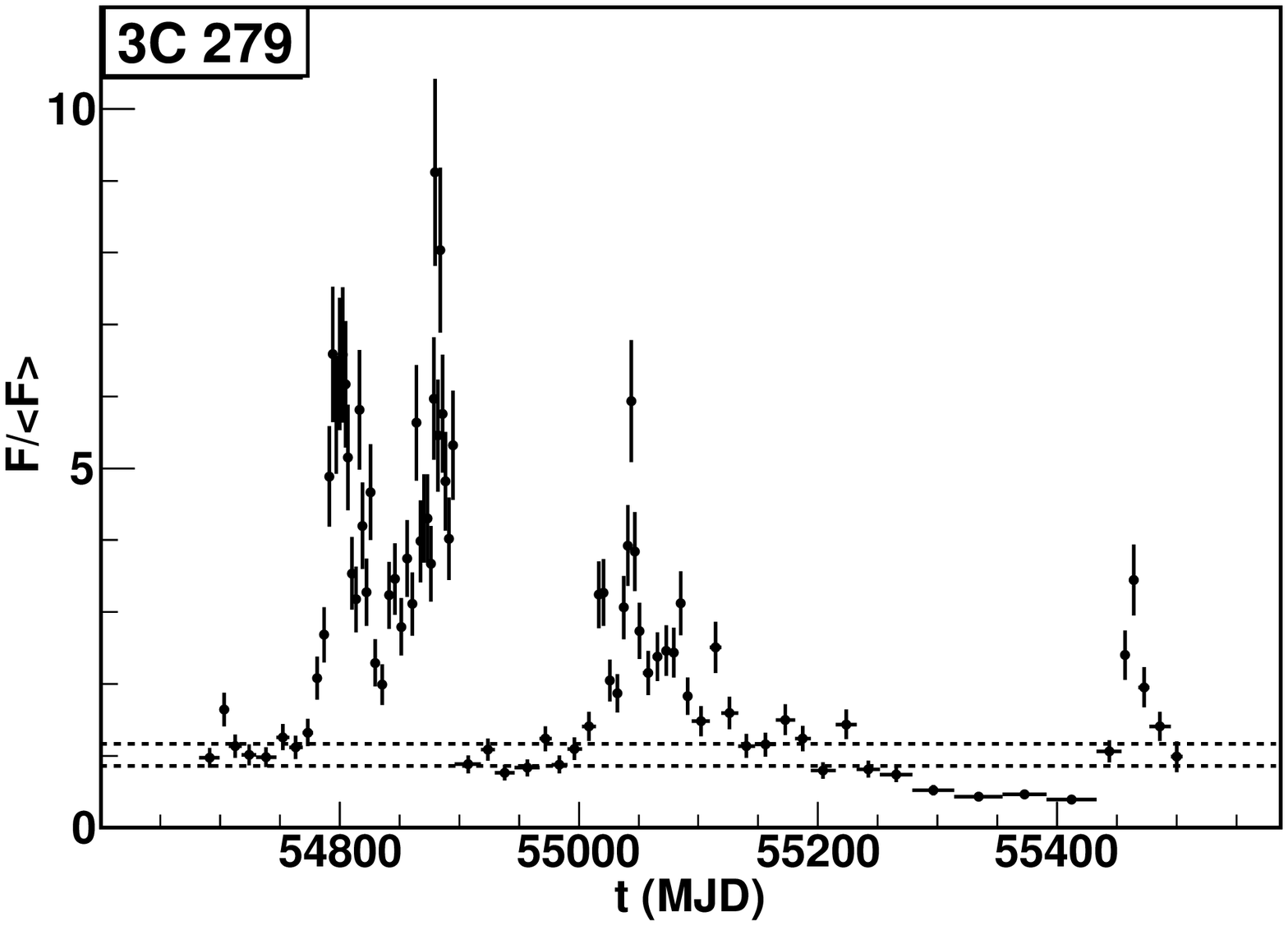}
\end{figure*}

\begin{figure*}
\centering
\includegraphics[width=80mm]{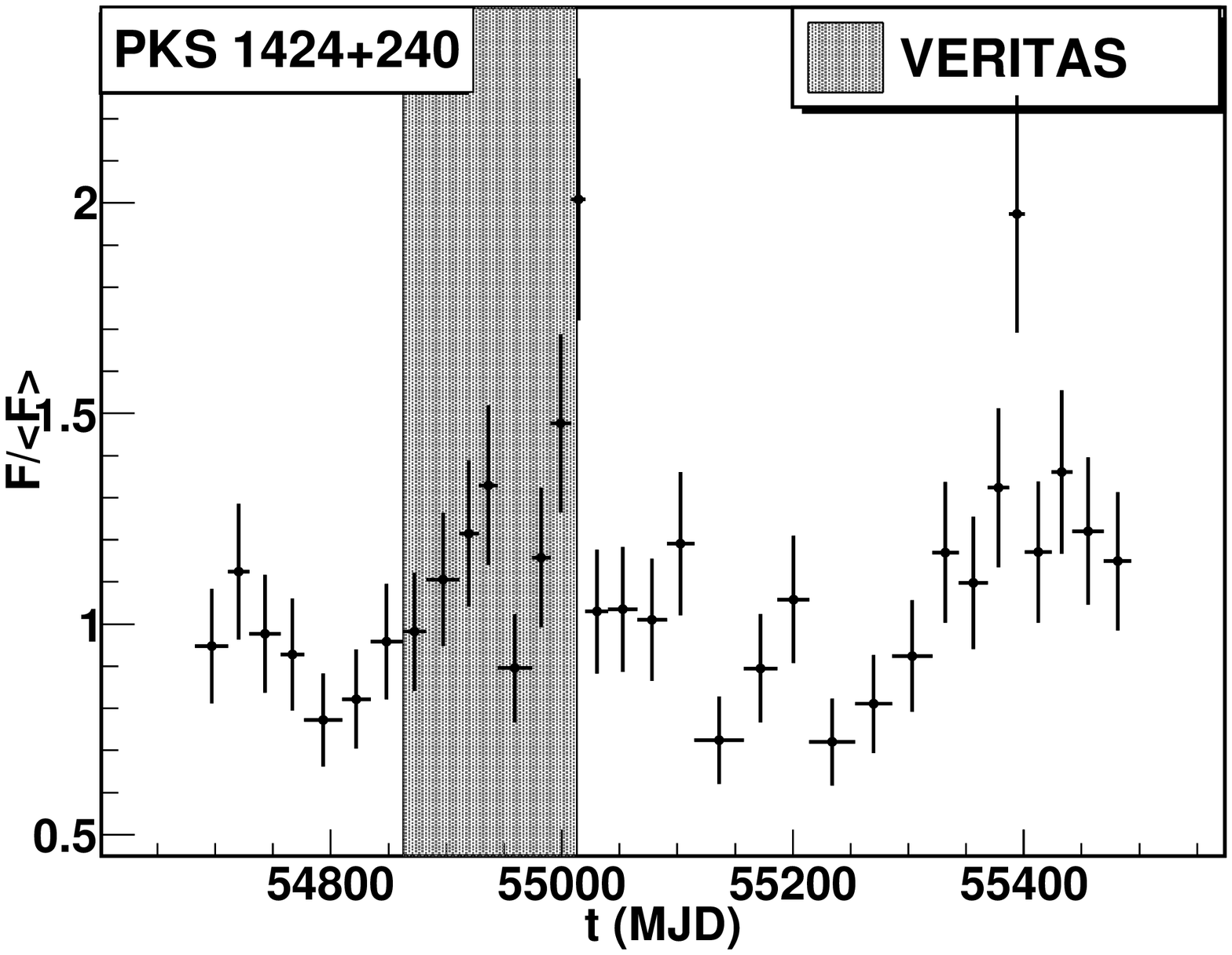}
\includegraphics[width=80mm]{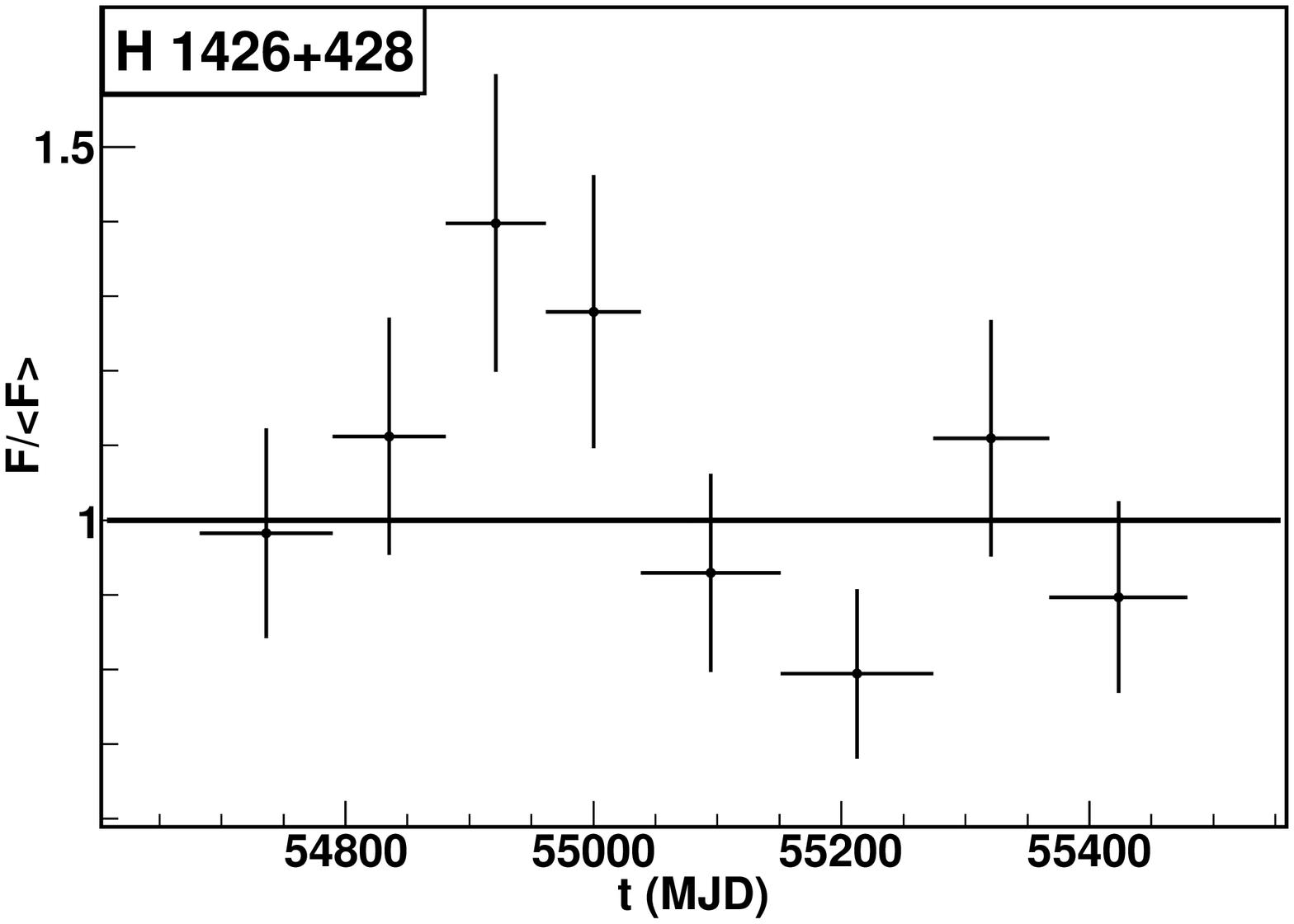}
\includegraphics[width=80mm]{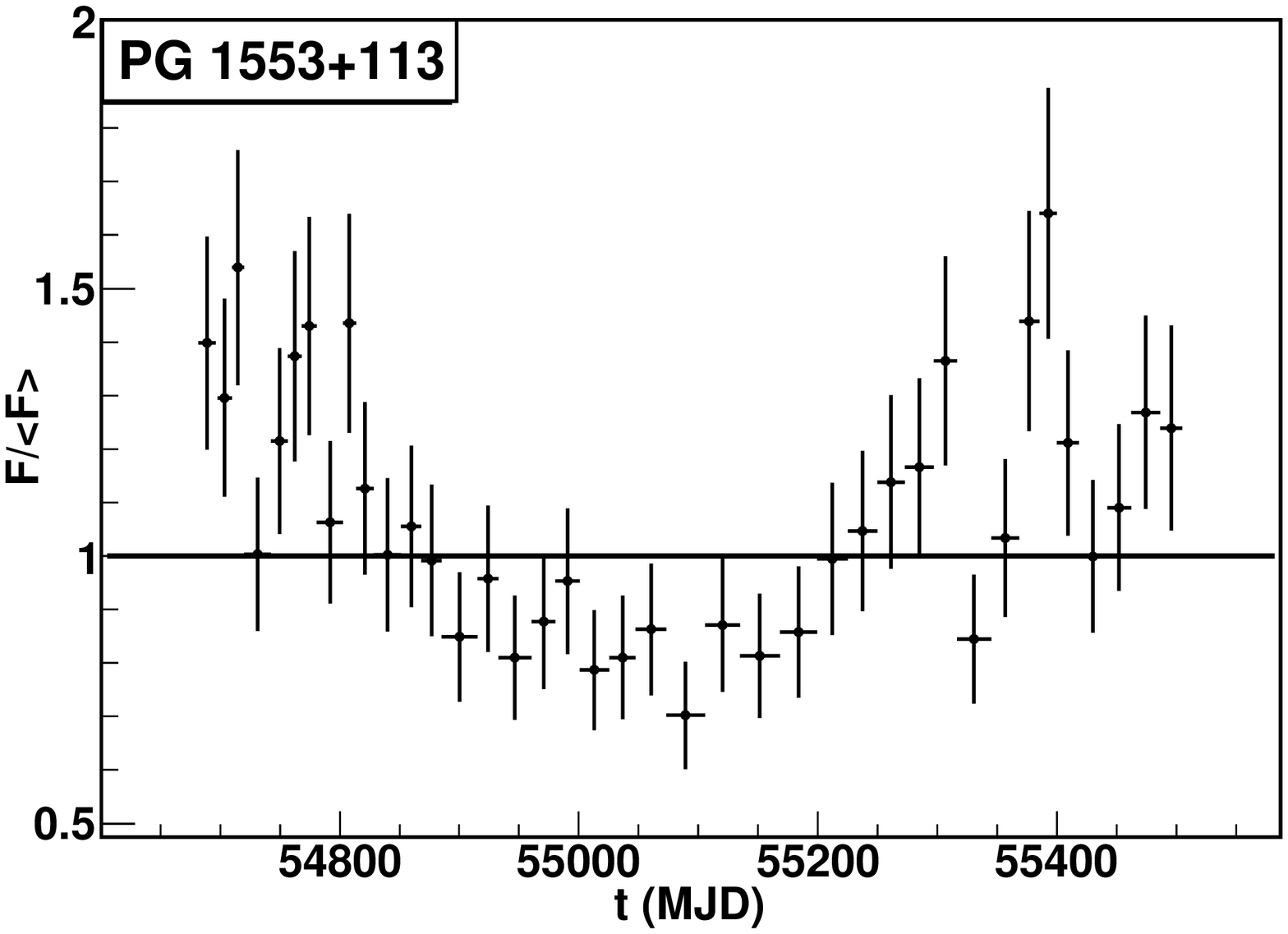}
\includegraphics[width=80mm]{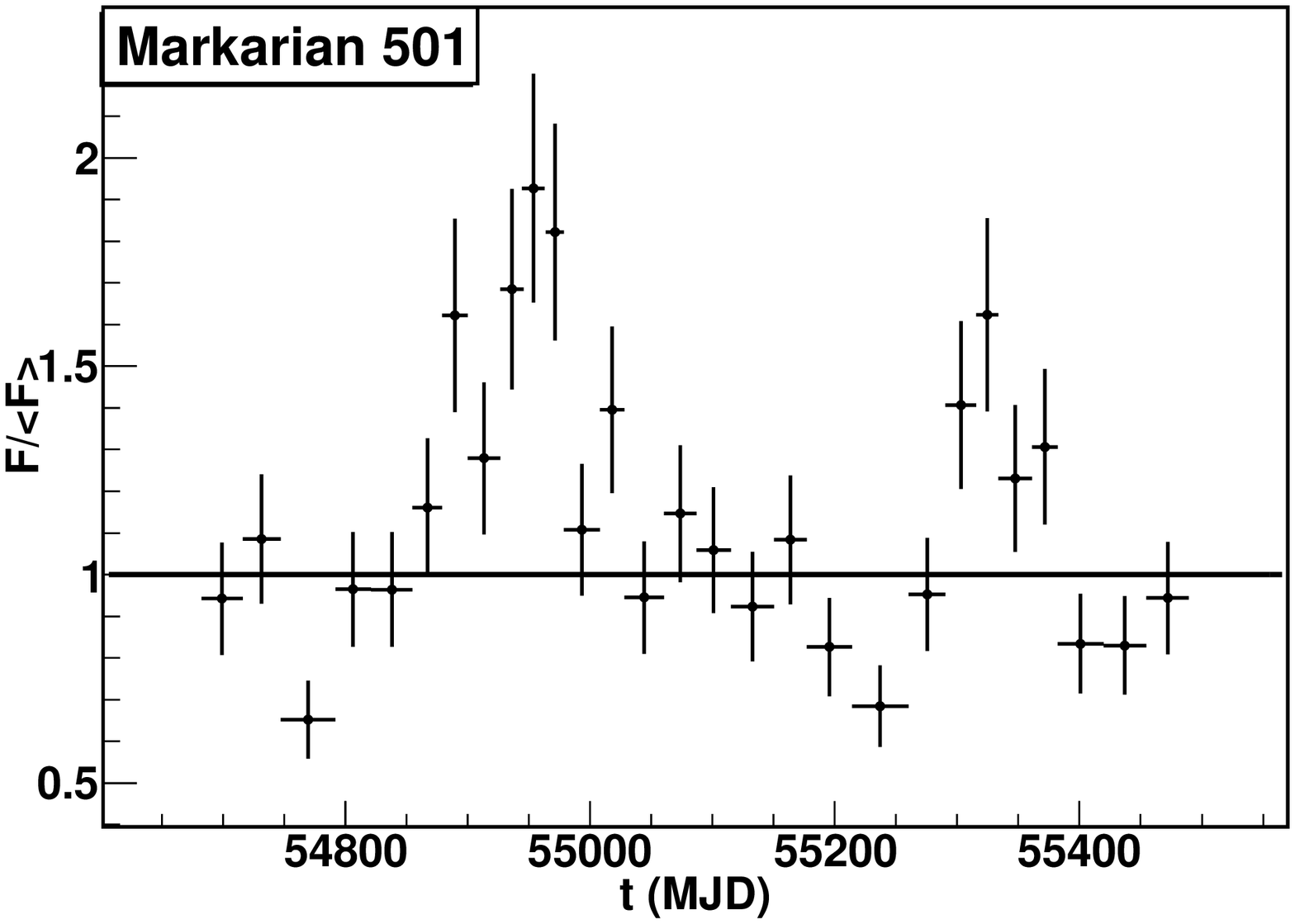}
\includegraphics[width=80mm]{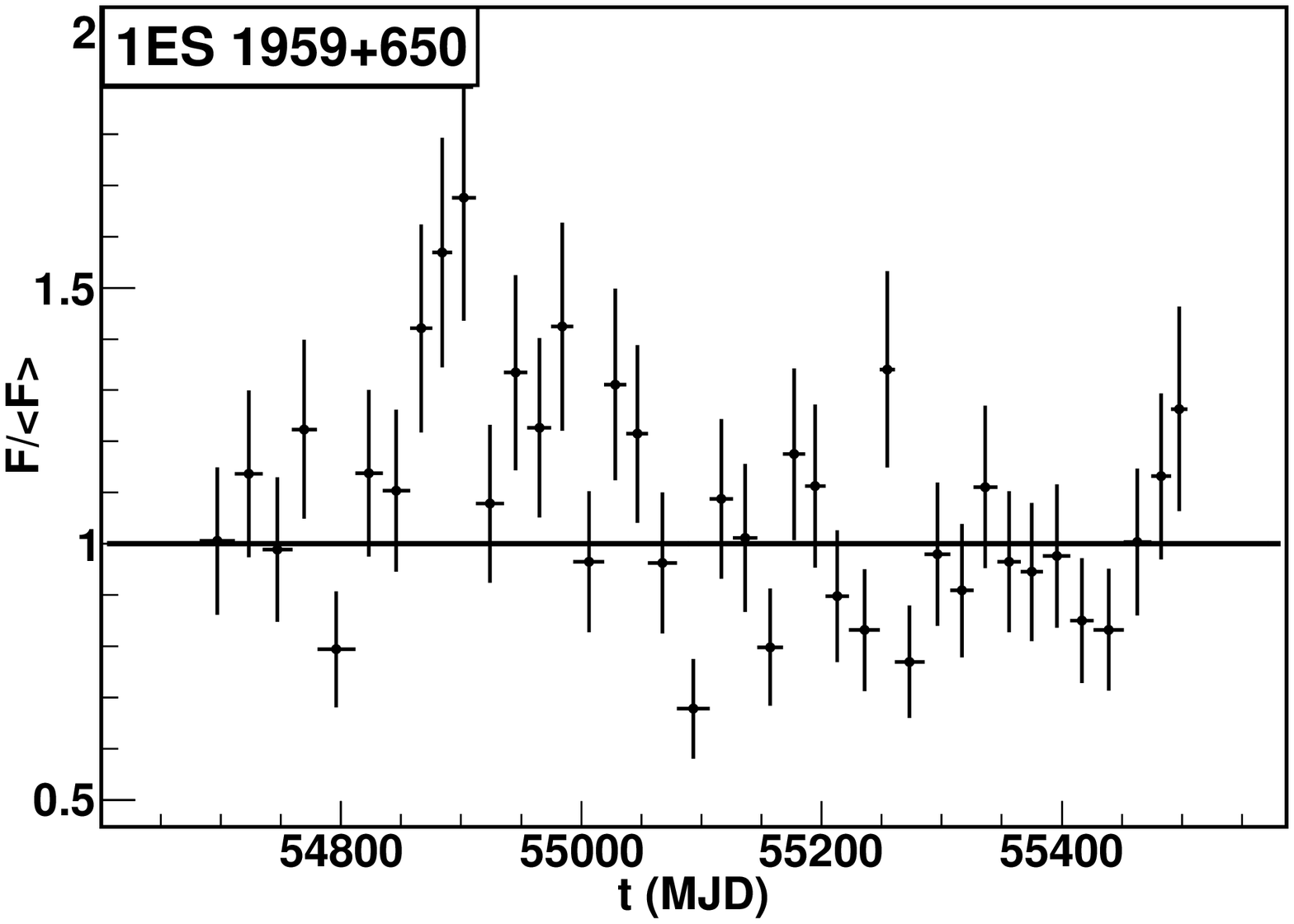}
\includegraphics[width=80mm]{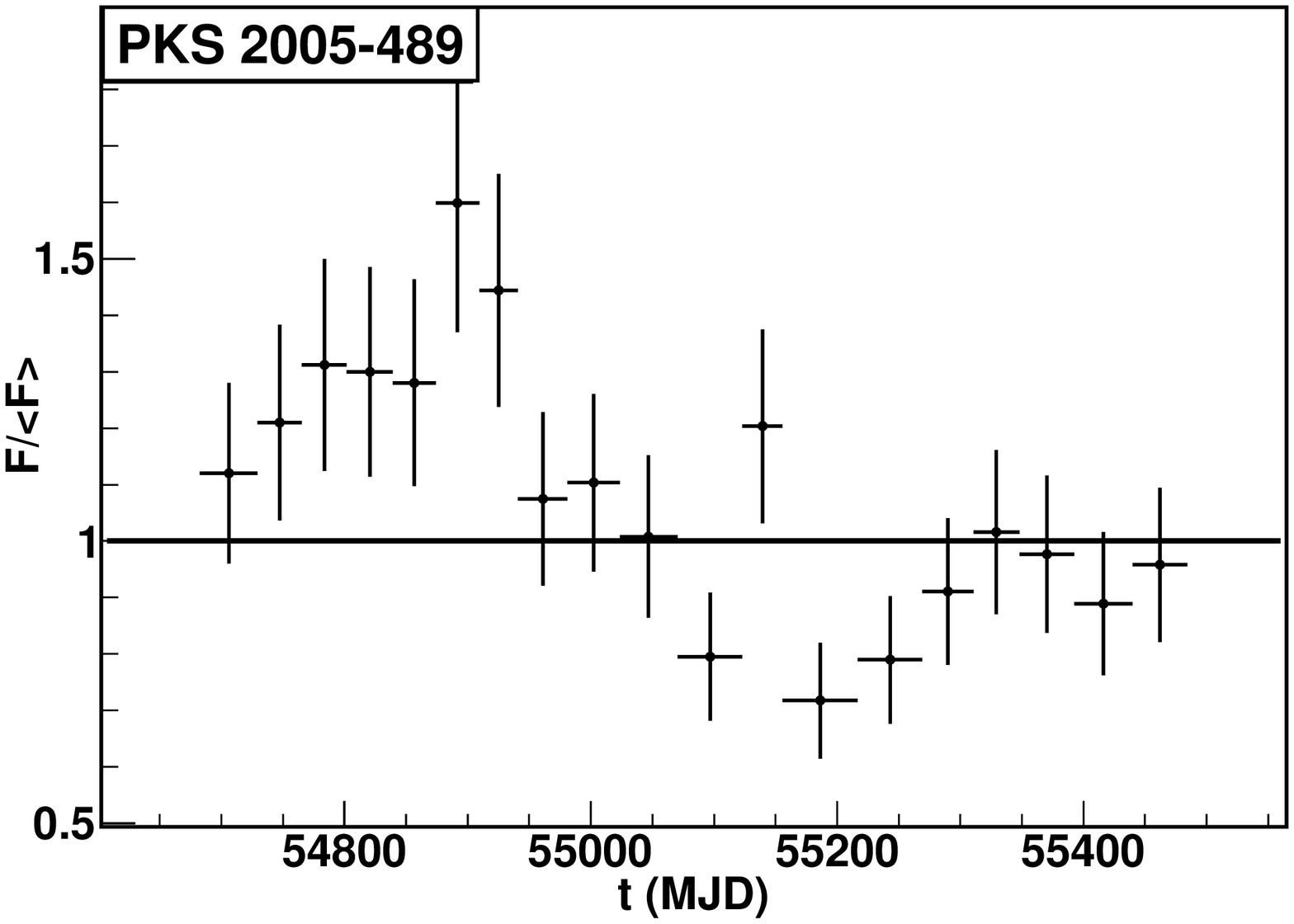}
\includegraphics[width=80mm]{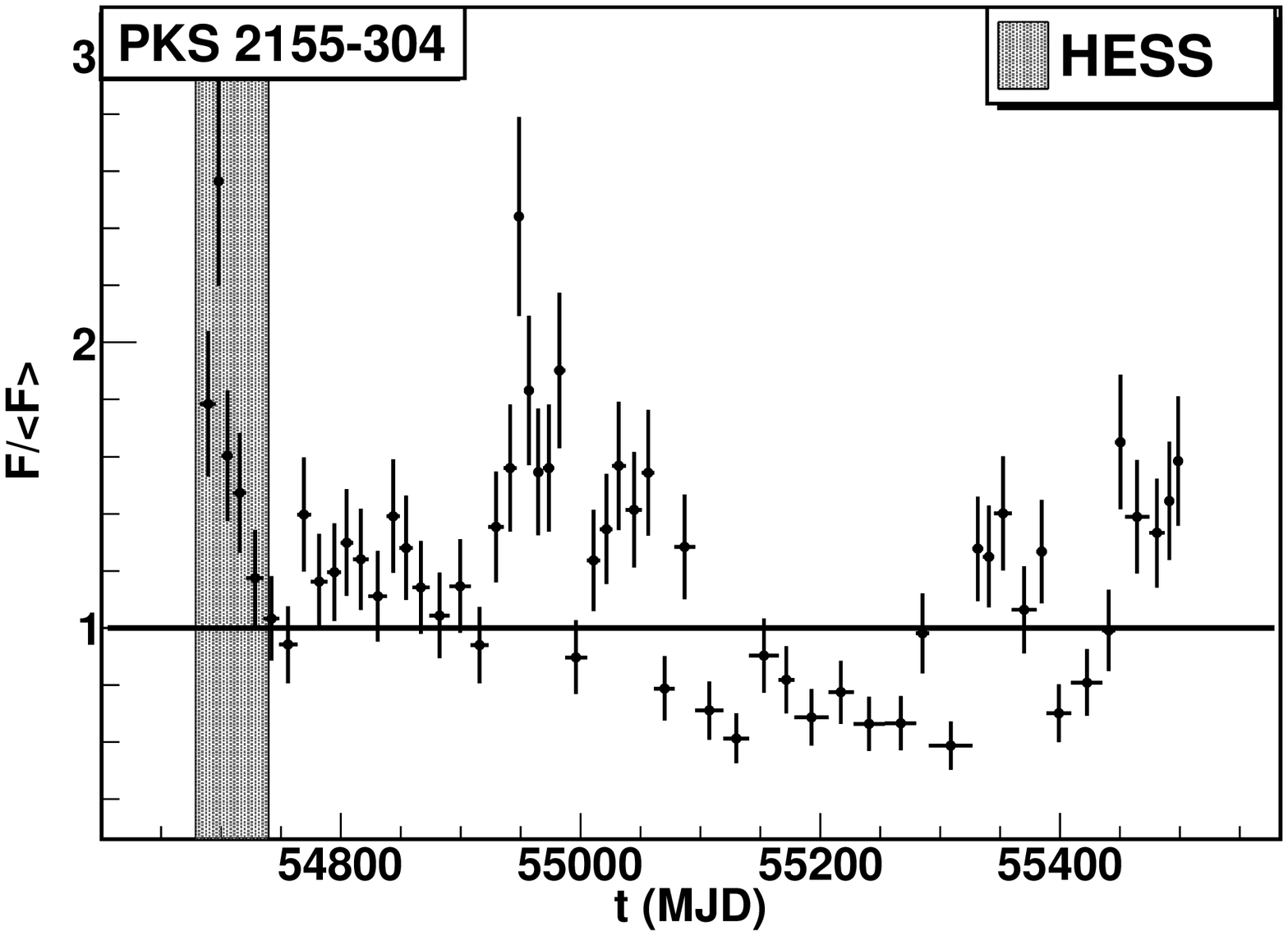}
\includegraphics[width=80mm]{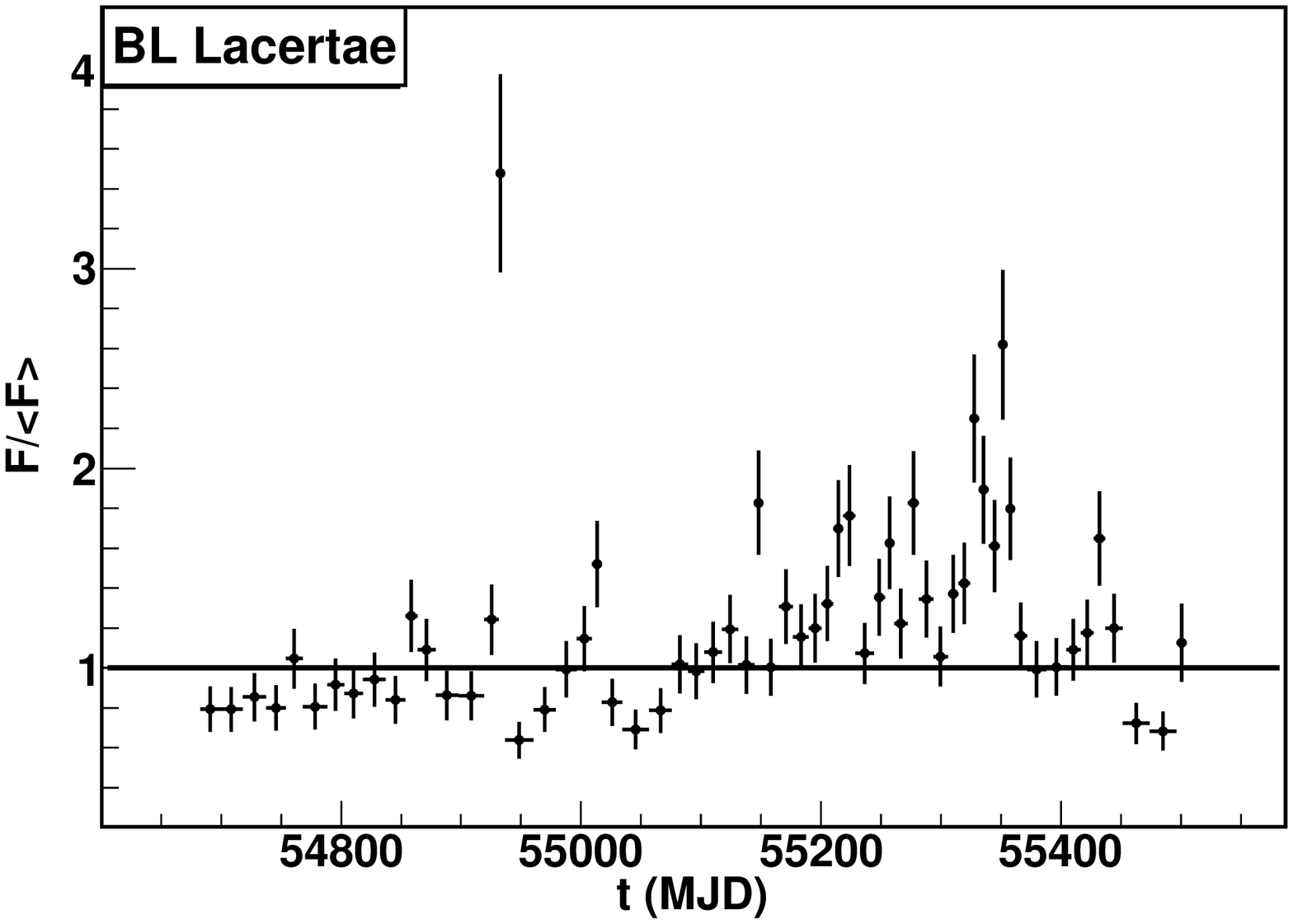}
\end{figure*}

\begin{figure*}
\centering
\includegraphics[width=80mm]{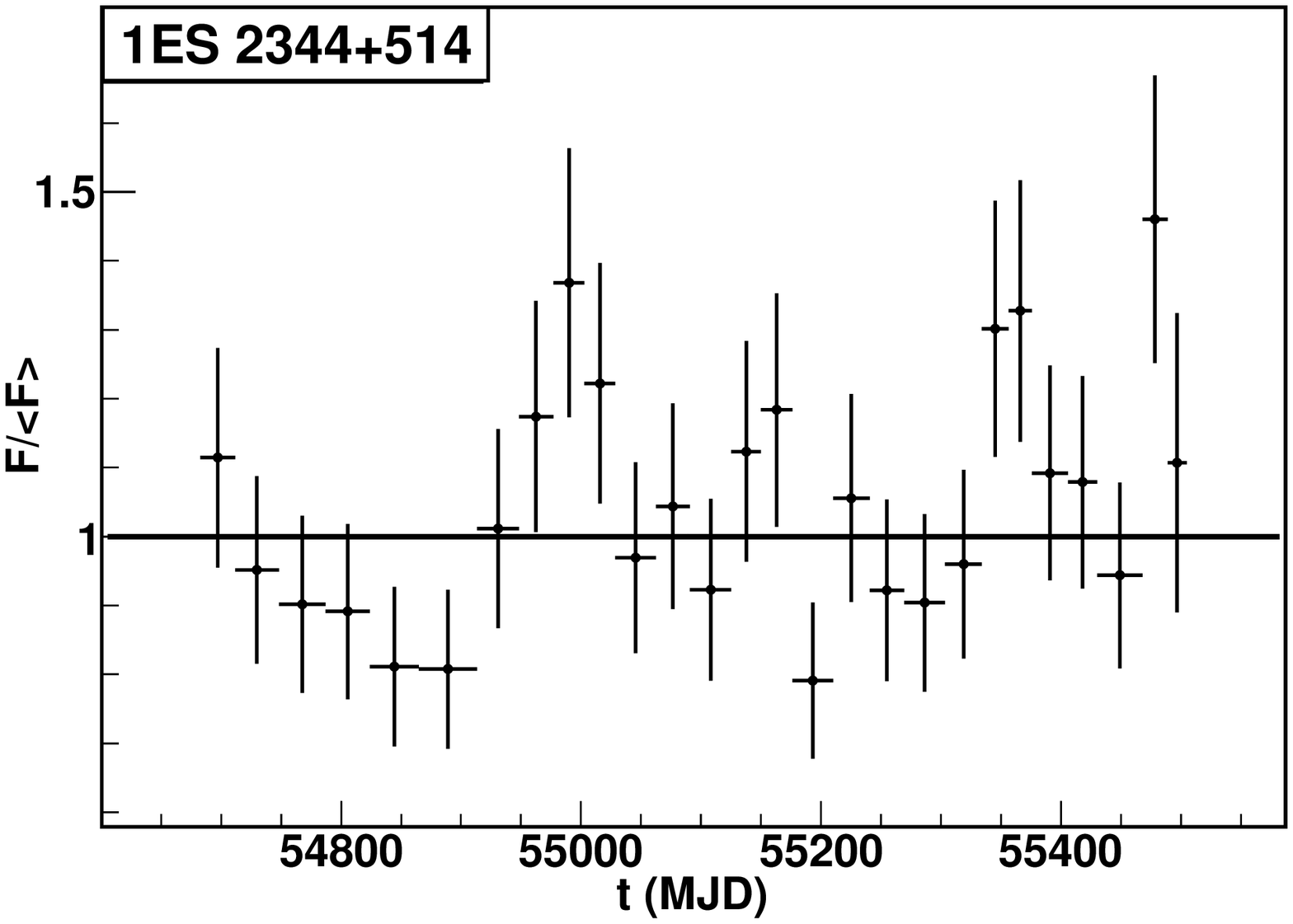}
\includegraphics[width=80mm]{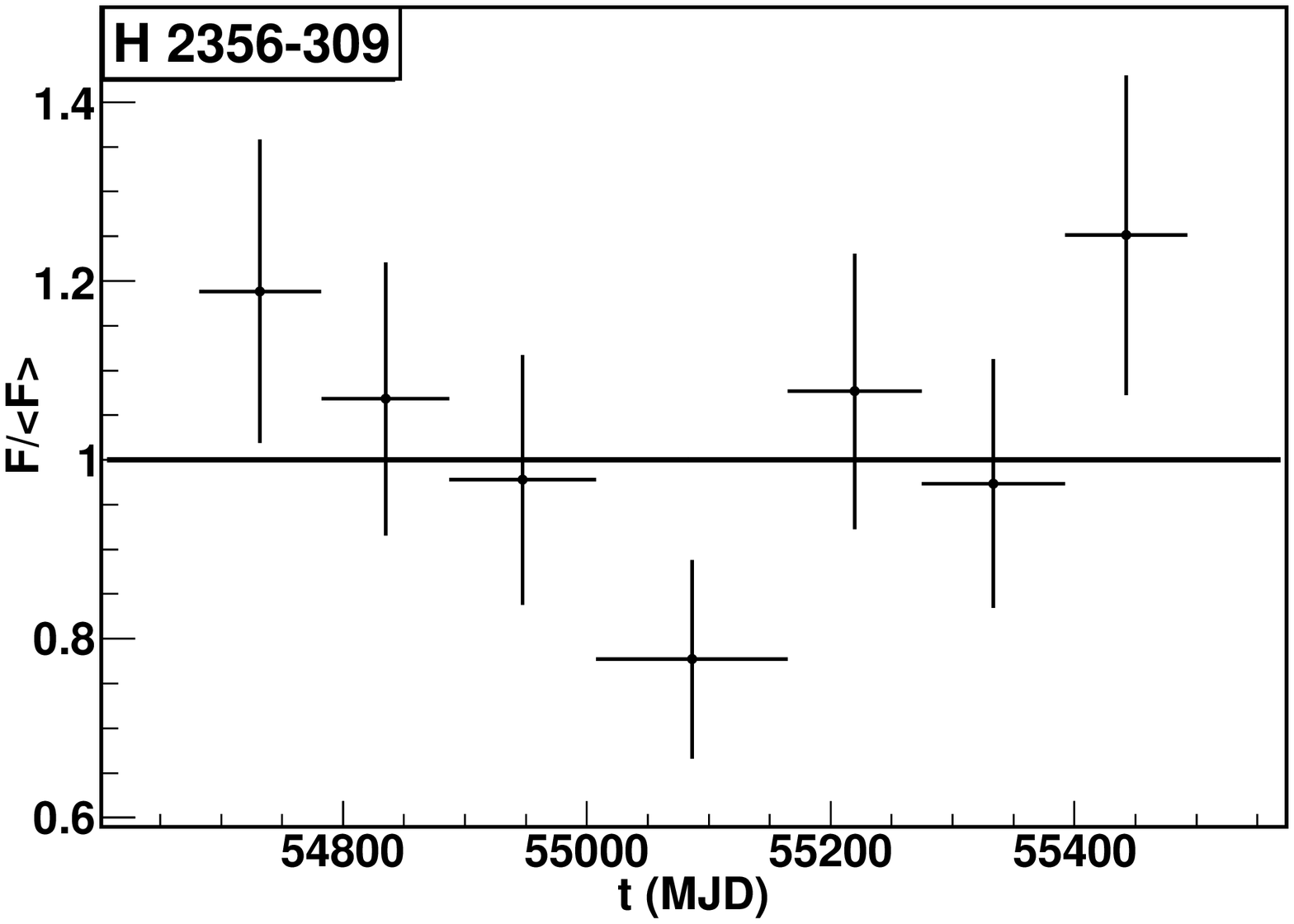}
\caption{{\footnotesize \emph{Fermi}-LAT aperture photometry light curves with no background subtraction, normalized to arbitrary units. The solid lines, representing the
average counts per area per time, separate ``low" and ``high" flux states, that are later on used to produce ``low" and ``high" state spectra. In the case of Markarian\,421 and
3C\,279 light curves, the dashed lines represent 1 $\sigma$ deviation from the average, dividing the data set into three separate flux states (``low", ``medium", and ``high").
The shaded areas show the contemporaneous time windows with the corresponding TeV instruments.}}
\label{fig:lc}
\end{figure*}
\par
For VHE data that were taken during the \fermi era, time periods of a few months that cover the corresponding VHE observations were selected for the \emph{Fermi} spectral analysis. 
For blazars that have VHE spectra measured before the \emph{Fermi} era, the first 27-month of \emph{Fermi} data were analyzed (from 2008 August 4 to 2010 November 4). 
In all the analysis steps, an energy selection from 300\,MeV to 100\,GeV was applied to the data.
\par
The \fermi data were analyzed in the following way. 
First, a 27-month light curve analysis was performed for each source using an aperture photometry technique.
\emph{Diffuse class} events from a region of $1^{\circ}$ radius from the target location were selected and counts were plotted as a function of time, each time bin containing 49 counts, corresponding to a signal to noise ratio of 7. 
For sources with high statistics, low- and high- flux states were identified and separated using the average count rate as a threshold. 
Figure~\ref{fig:lc} shows the resulting light curves for all sources, with fluxes normalized to arbitrary units.
It should be noted that in this analysis, no background subtraction was performed and therefore the resulting light curves merely give an estimate of high- and low-state time slices.
\par
Next, a spectral analysis was done for each data set. 
\emph{Diffuse class} events from a region of interest of $8^{\circ}$ radius were selected and analyzed with Fermi Science Tools v9r18p6\footnote{\scriptsize http://fermi.gsfc.nasa.gov/ssc/data/analysis/scitools/overview.html}, using instrument response functions P3\_V6\_DIFFUSE. 
Sources from the first \emph{Fermi}-LAT (1FGL) catalog~\citep{abdo2010}, bright spots with test statistics $>25$ and standard galactic and isotropic diffuse emission background components\footnote{\scriptsize http://fermi.gsfc.nasa.gov/ssc/data/access/lat/BackgroundModels.html} within the region of interest were included in the source model files. 
Unbinned maximum-likelihood analysis as described in~\citep{cash79,mattox96} was applied to each data set, assuming a power-law (PL) spectrum as given in Equation~(\ref{eq:PL}).
\begin{equation} 
dN/dE=N_{0}(E/E_{0})^{-\Gamma}
\label{eq:PL}
\end{equation}
Additionally, to look for possible spectral features in the data, spectral points were calculated and fitted with different power-law functions, and the results were compared. See Section~\ref{spectralFeatures} for more details.
\par
Finally, combined GeV-TeV SED data sets were constructed using archival TeV spectra and the corresponding flux state information from references shown in Table~\ref{tab:sample}. 
With each TeV spectrum, the most suitable \fermi data subset (average, low- or high- state) was used for further study.

\section{Results and Discussion}
Twelve out of 26 blazars did not have enough statistics for a temporal separation of the \emph{Fermi} data set into different flux states. 
Therefore for this subsample, an average spectrum was calculated using the entire data set. 
Data from another subsample with 12 blazars were split into high and low-flux states as described in Section~\ref{fermi}. 
Data from the two brightest blazars (Markarian 421 and 3C 279) were split into three subsets, with low, medium and high-flux states. 
See Table~\ref{tab:results} for a summary of our \fermi data analysis results.
\par
\begin{figure*}
\centering\includegraphics[width=80mm]{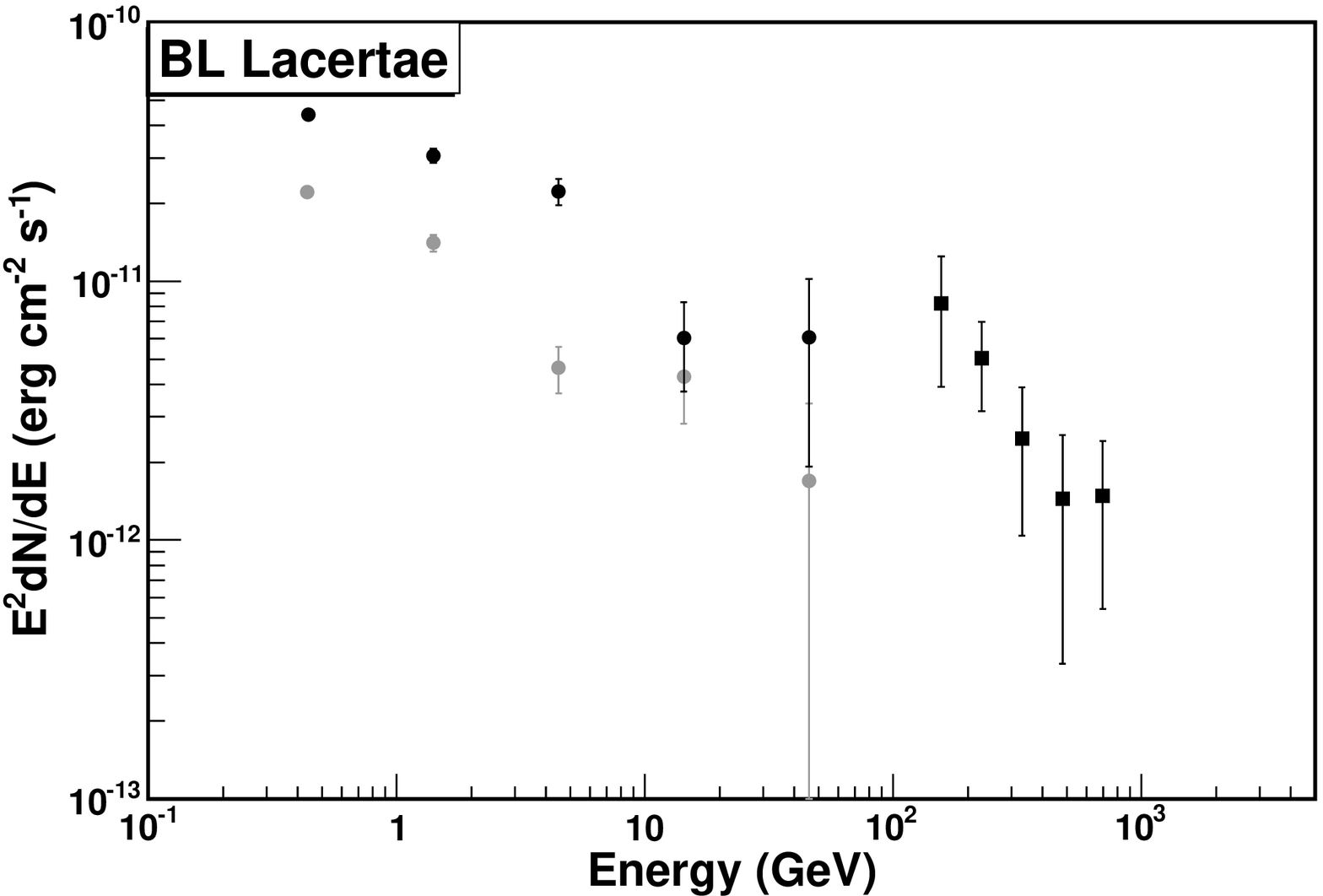}
\includegraphics[width=80mm]{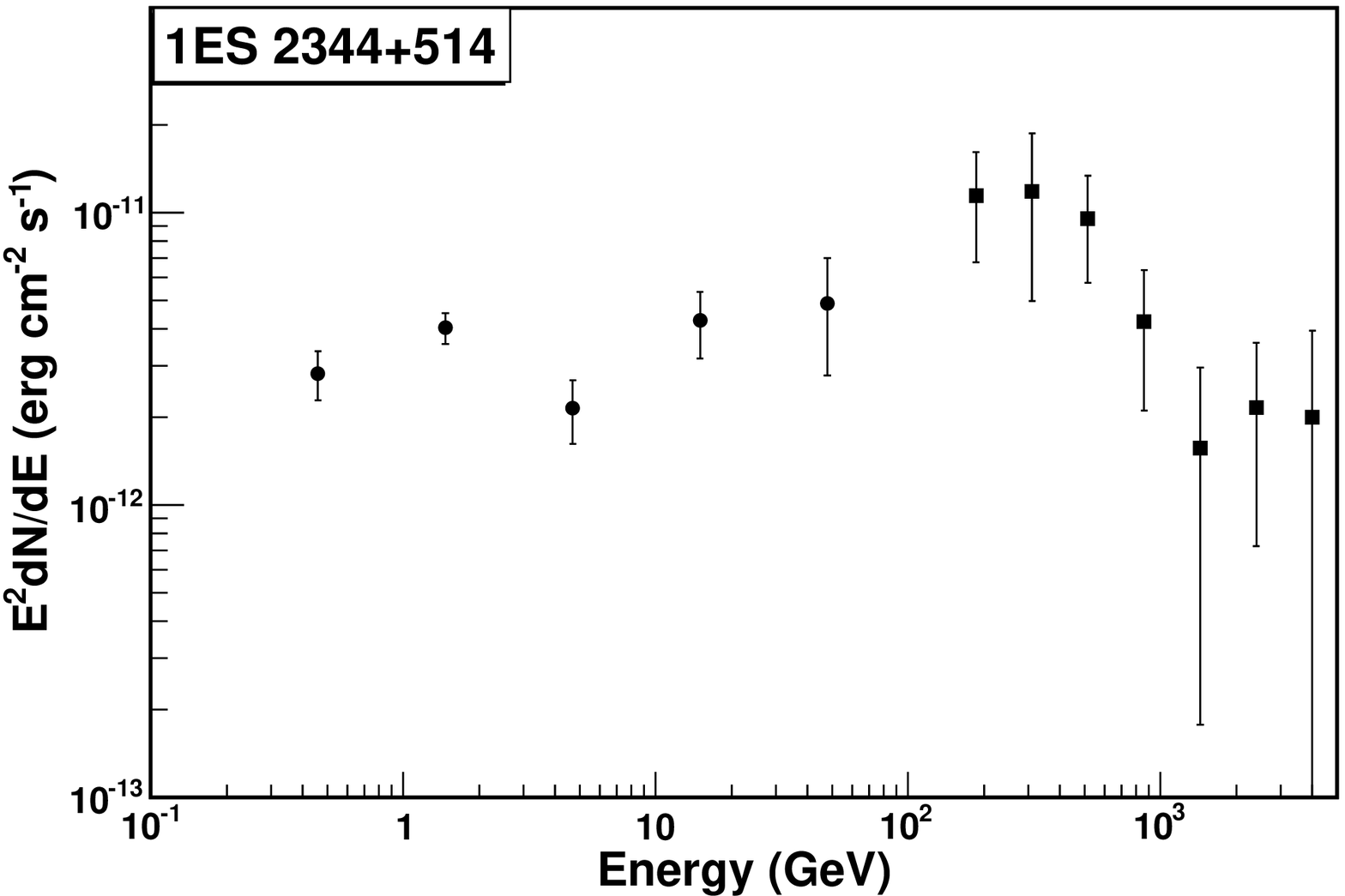}
\includegraphics[width=80mm]{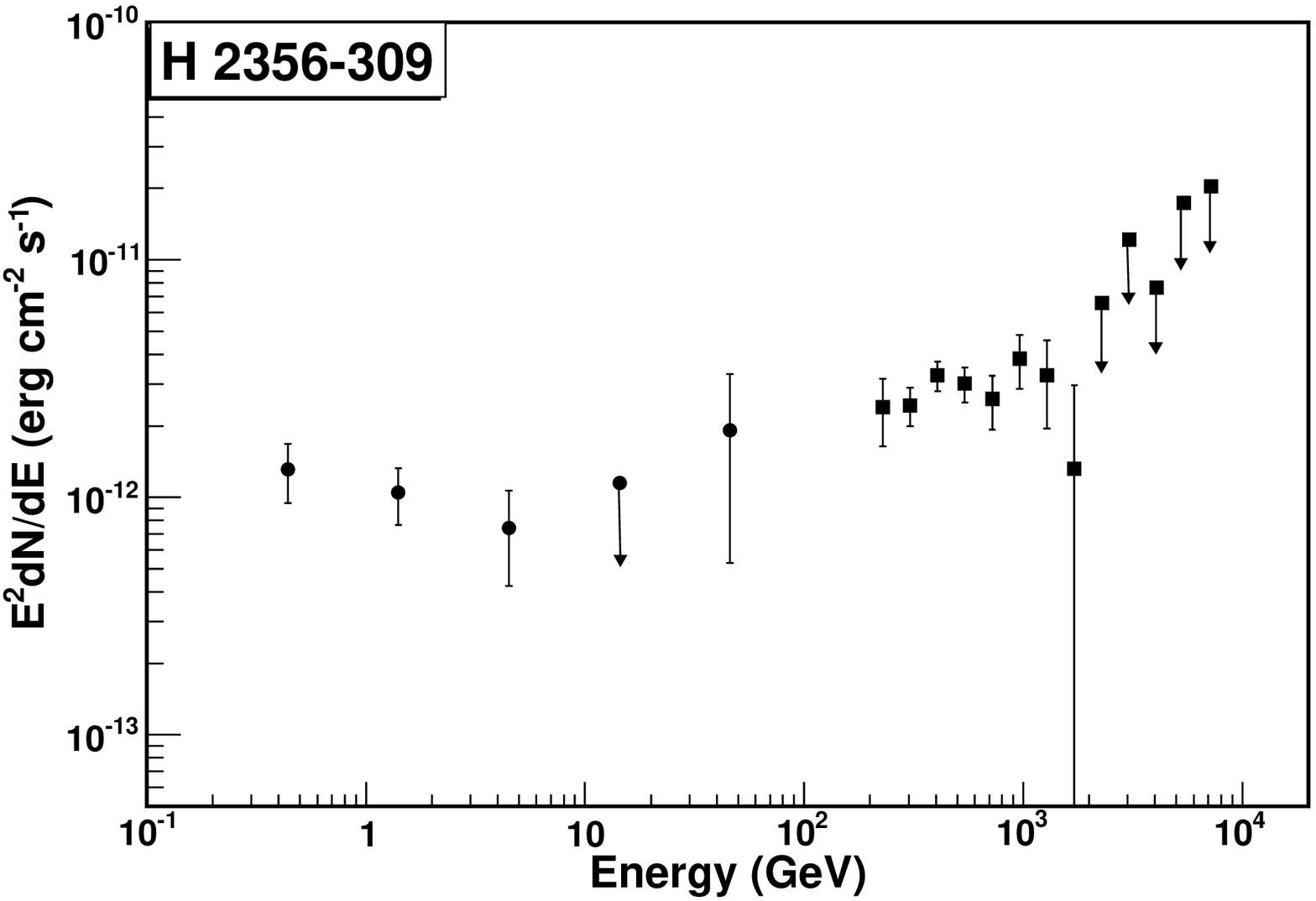}
\caption{{\footnotesize GeV-TeV spectra for the sample of blazars in this study. Filled circles represent the \fermi spectra and the filled squares the TeV spectra. Considering the TeV flux state
information given in the TeV papers, the best matching GeV and TeV spectral points are used for the combined analysis (shown in black). When available, spectral points belonging to other flux states (in both bands) are plotted in gray. 3C66A (VERITAS and MAGIC respectively), RGB J0710+591, 1ES 1218+304, PKS 1222+21 (4C +21.35), PKS 1424+240, and PKS 2155-304 spectra are quasi-simultaneous.}}
\label{fig:combined_spectra}
\end{figure*}

Our analysis results are consistent with the 2FGL catalog~\citep{2fgl}. 
We used the combined GeV-TeV SEDs (see Figure~\ref{fig:combined_spectra}) to estimate the IC peak frequency band of each blazar (see Section~\ref{IC_peak}).
Our sample contains a handful of candidate ``TeV-peaked" blazars that we discuss in Section~\ref{tev_peaked}.
In addition, considering the fact that \fermi spectral indices do not vary significantly between low- and high-states, we studied the change in spectral index from GeV to TeV as a function of the redshift, thus confirming the EBL effect on TeV spectra with a model-independent approach (see Section~\ref{flux_var}).
On the other hand, interesting spectral features in the GeV band are observed. 
To probe these features, the data were fitted with three different functions and the corresponding fit improvements were calculated (see Section~\ref{spectralFeatures}). 
Finally, in Section~\ref{sim}, we extended this study to contemporaneous combined SEDs. 

\begin{table*}

\centering
\begin{threeparttable}
\centering
\begin{scriptsize}
\begin{tabular}{|c|cccccccc|}
\hline
Name 		& SED type 	& $z$ 	& $F_\mathrm{var}$	& \fermi state 	&
$\Gamma_\mathrm{GeV}$	& $F_\mathrm{1-100}(\mathrm{cm^{-2}s^{-1}})$ & TS	& Live time (day)\\
		& (1)		& (2)	& (3)			& (4)		& (5)			& (6)		& (7)	& (8)\\
\hline
\hline
RGB J0152+017	& HBL 		& 0.080	& 0.19	& average	& $2.09\pm0.14$		& $(7.70\pm1.26)\times10^{-10}$	& 106 & 822\\
\hline
3C 66A*		& IBL		& 0.444\tnote{a}& 0.59 & MAGIC	&  $2.09\pm0.06$	& $(2.41\pm0.19)\times10^{-8}$ & 953 & 62  \\
		&		&	&	& VERITAS	& $1.91\pm0.05$ & $(2.50\pm0.16)\times10^{-8}$	& 1485 & 91 \\
\hline
1ES 0229+200 	& HBL 		& 0.140	& 0.07  & average 	& $2.23\pm0.34$ &	$(2.96\pm1.07)\times10^{-10}$ & 21 & 822\\
\hline
1ES 0347-121 	& HBL 		& 0.188 & $<0.12$ &average 	& $0.85\pm0.54$	&	$(5.33\pm3.90)\times10^{-11}$ & 16 & 822\\
\hline
PKS 0548-322	& HBL		& 0.069	& $<0.14$ &average	& $1.65\pm0.25$		& $(2.93\pm1.00)\times10^{-10}$ & 40 & 822\\
\hline
RGB J0710+591	& HBL		& 0.125	& $<0.11$ &VERITAS	& $1.44\pm0.33$ 	& $(9.94\pm4.09)\times10^{-10}$ & 33 & 121\\
\hline
S5 0716+714	& LBL		& 0.300	& 0.44 &high		& $2.13\pm0.04$		& $(2.01\pm0.09)\times10^{-8}$ & 3644 & 342\\
\hline
1ES 0806+524	& HBL		& 0.138	& 0.13 &average	& $1.77\pm0.07$ 	& $(1.45\pm0.14)\times10^{-9}$ & 400 & 822\\
\hline
1ES 1011+496	& HBL		& 0.212	& 0.15 &high		& $1.97\pm0.04$		& $(7.82\pm0.47)\times10^{-9}$ & 1705 & 332\\
\hline
1ES 1101-232	& HBL		& 0.186	& $<0.10$ &average	& $1.88\pm0.26$ 	& $(4.59\pm1.27)\times10^{-10}$ & 47 & 822\\
\hline
Markarian 421*	& HBL		& 0.031	& 0.22 &medium	& $1.78\pm0.02$		& $(2.64\pm0.08)\times10^{-8}$ & 7943 & 350\\
\hline
Markarian 180	& HBL		& 0.046	& 0.22 &average	& $1.87\pm0.08$		& $1.22\pm0.12\times10^{-9}$ & 356 & 822\\
\hline
1ES 1218+304	& HBL		& 0.182	& 0.44 &average	& $1.69\pm0.06$		& $(2.80\pm0.23)\times10^{-9}$ & 708 & 822\\
		&		&	& & VERITAS	& $1.84\pm0.11$ 	& $(3.30\pm0.49)\times10^{-9}$ & 187 & 182\\
\hline
W Comae*	& IBL		& 0.102 & 0.32 &high		& $2.07\pm0.06$		& $(8.34\pm0.59)\times10^{-9}$ & 1101 & 222\\
\hline
PKS 1222+21	& FSRQ		& 0.432	& 1.42 &MAGIC	& $2.17\pm0.04$		& $(7.24\pm0.47)\times10^{-6}$ & 4267& 6 \\
\hline
3C 279		& FSRQ		& 0.536 & 0.65	&high		& $2.37\pm0.02$		& $(5.25\pm0.15)\times10^{-8}$ & 13558 & 218\\	
\hline
PKS 1424+240	& IBL		& 0.260\tnote{b}& 0.22 &VERITAS		& $1.85\pm0.05$ & $(1.21\pm0.09)\times10^{-8}$ & 1116 & 150\\
\hline
H 1426+428	& HBL		& 0.129	& 0.07 & average	& $1.12\pm0.16$		& $(4.05\pm0.86)\times10^{-10}$ & 197 & 822\\
\hline
PG 1553+113	& HBL		& 0.4\tnote{c}	& 0.16 & high		& $1.74\pm0.03$	& $(1.66\pm0.07)\times10^{-8}$ & 3339 & 344\\
\hline
Markarian 501*	& HBL		& 0.034	& 0.25 &low		& $1.84\pm0.05$		& $(5.78\pm0.36)\times10^{-9}$ & 1280 & 458\\
\hline
1ES 1959+650*	& HBL		& 0.048	& 0.14 &low		& $2.04\pm0.06$		& $(4.45\pm0.30)\times10^{-9}$ & 834 & 443\\
\hline
PKS 2005-489	& HBL		& 0.071	& 0.15 &average	& $1.82\pm0.05$		& $(3.37\pm0.23)\times10^{-9}$ & 834 & 822\\
\hline
PKS 2155-304*	& HBL		& 0.117	& 0.29 &HESS	& $1.89\pm0.05$		& $(3.20\pm0.24)\times10^{-8}$ & 1308 & 61\\
		&		&	& & low		& $1.95\pm0.03$		& $(1.55\pm0.06)\times10^{-8}$ & 4118 & 420\\
\hline
BL Lacertae	& LBL		& 0.069	& 0.37 & high		& $2.34\pm0.04$ 	& $(1.52\pm0.07)\times10^{-8}$ & 2517 & 283\\
\hline
1ES 2344+514*	& HBL		& 0.044	& 0.09 & average	& $1.97\pm0.07$		& $(2.09\pm0.19)\times10^{-9}$ & 407 & 822\\
\hline
H 2356-309	& HBL		& 0.165	& 0.04 &average	& $2.40\pm0.18$		& $(5.63\pm0.81)\times10^{-10}$ & 108 & 822\\
\hline
\end{tabular}
\begin{tablenotes}
\item[a]\citep{miller78,lanzetta93}
\item[b]\citep{prandini2011}
\item[c]\citep{mazin07}
\item[*] Blazars that are reported as variable in the TeV band, according to TeVCat(http://tevcat.uchicago.edu/).
\end{tablenotes}
\end{scriptsize}
\caption{{\small \fermi analysis results for the sample. Columns (1)  and (2) are the same as in Table~\ref{tab:sample}. $F_\mathrm{var}$ (3) is the
calculated flux variability amplitude (see Section~\ref{flux_var}) for the 27 month period. \fermi states (4) are as described in Section~\ref{fermi}. In cases where \fermi data are contemporaneous with TeV observations, the corresponding TeV instruments are listed in Column (4). $\Gamma_\mathrm{GeV}$ (5) represents the photon index and $F_\mathrm{1-100}$ (6) the integral flux for 1--100~GeV. Test statistics (TS) and live time corresponding to the listed flux state are given in Columns (7) and (8), respectively.}}
\label{tab:results}
\end{threeparttable}
\end{table*}

\begin{figure*}
\centering\includegraphics[width=80mm]{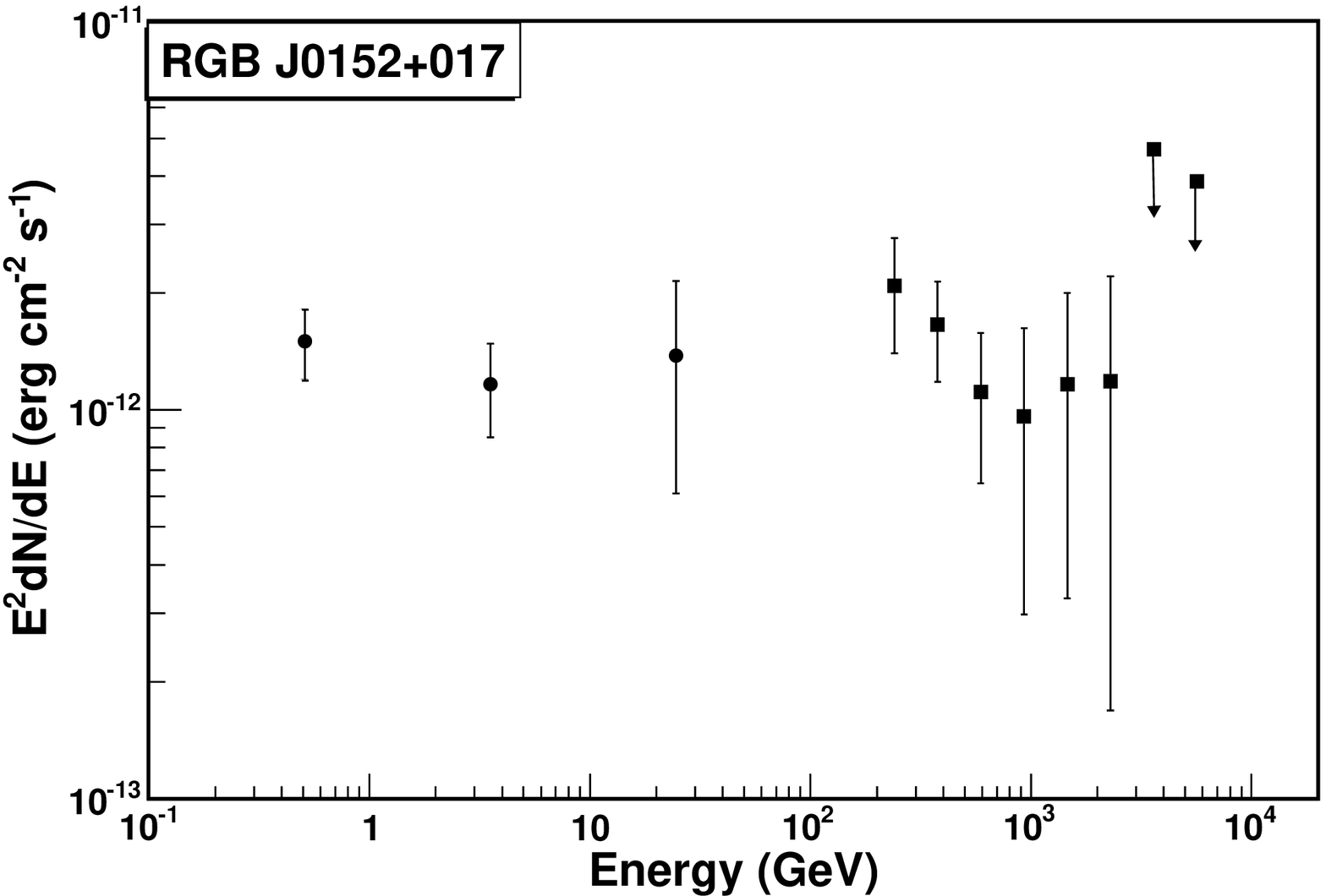}
\includegraphics[width=80mm]{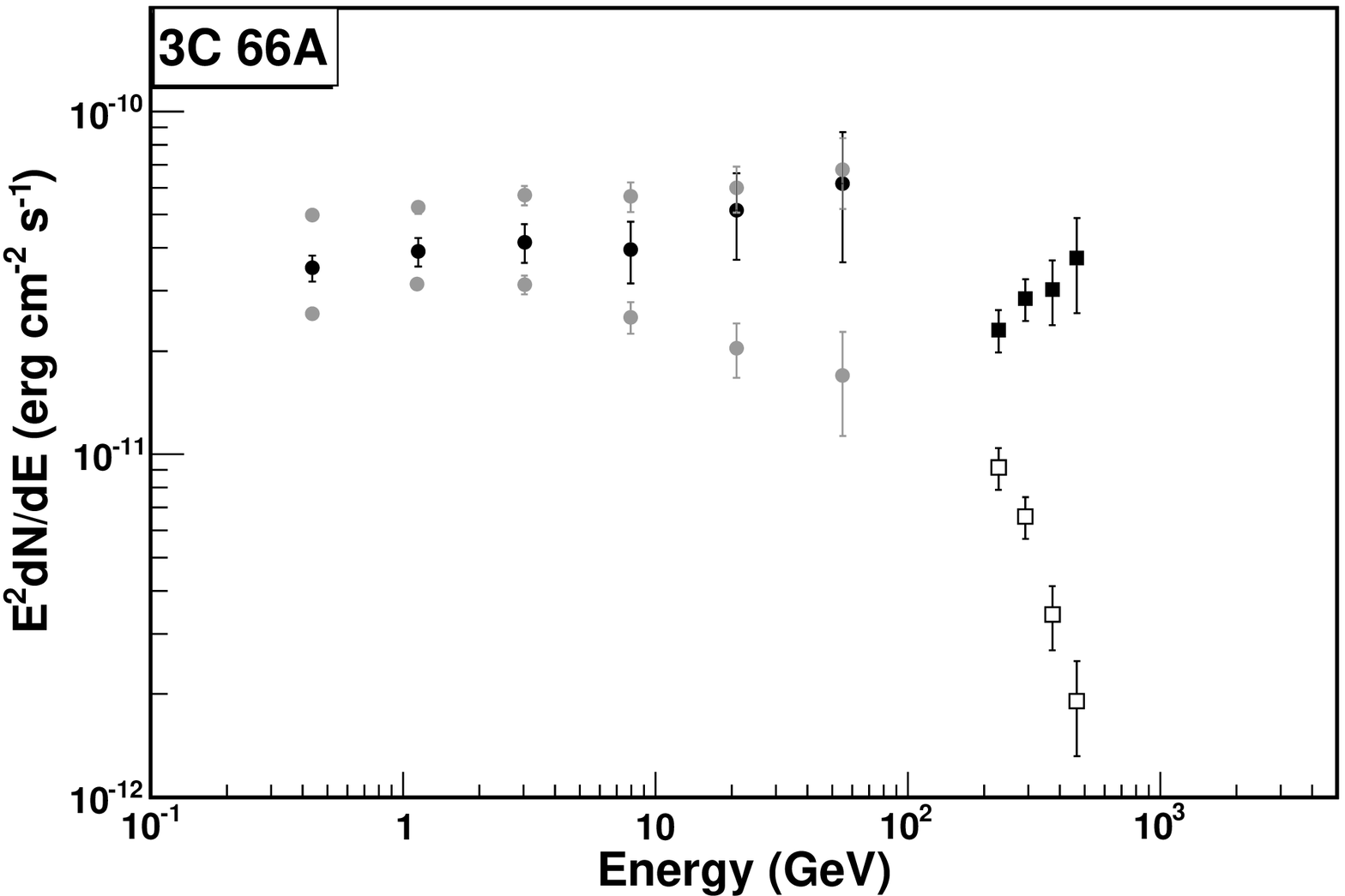}
\includegraphics[width=80mm]{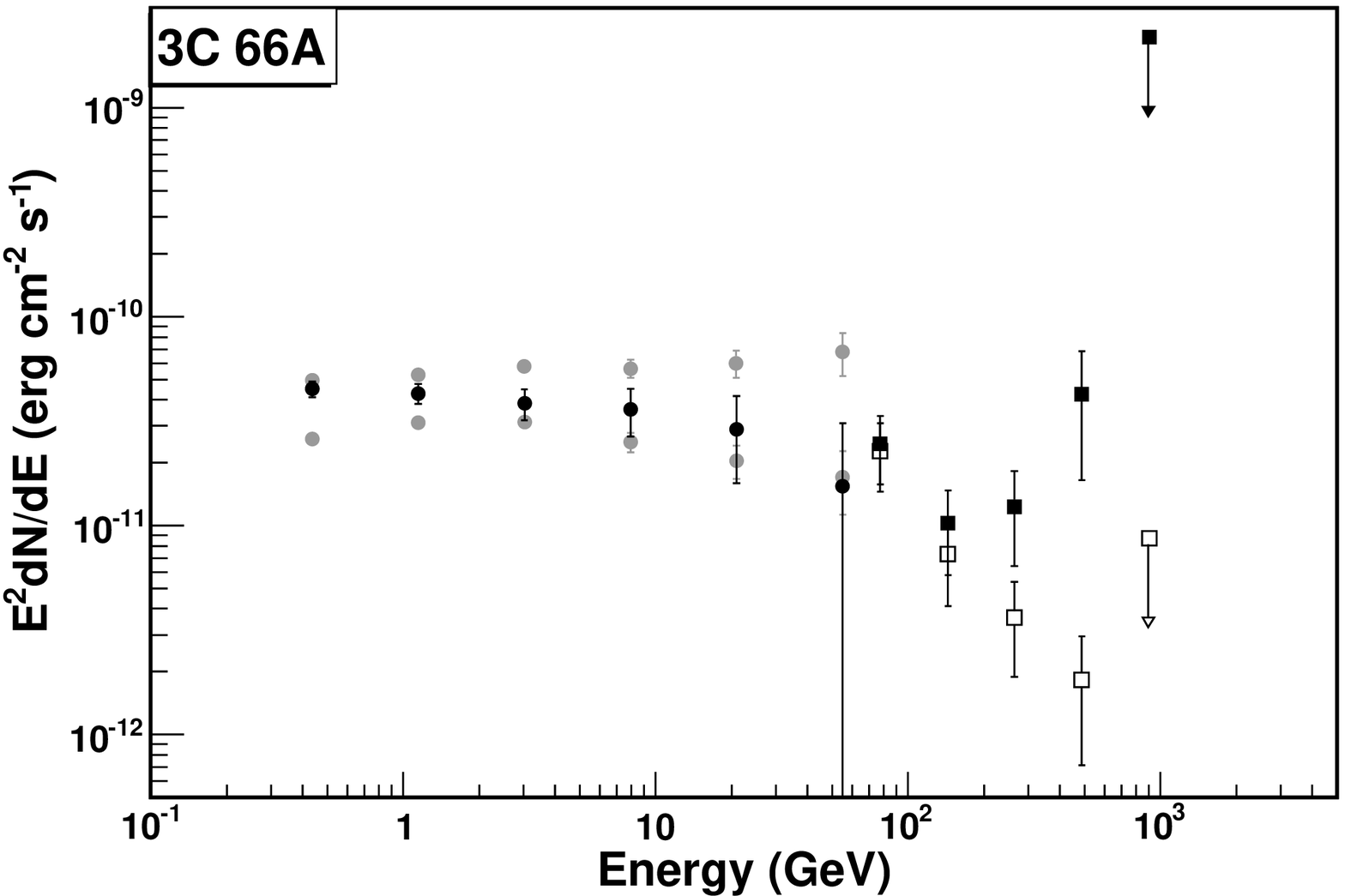}
\includegraphics[width=80mm]{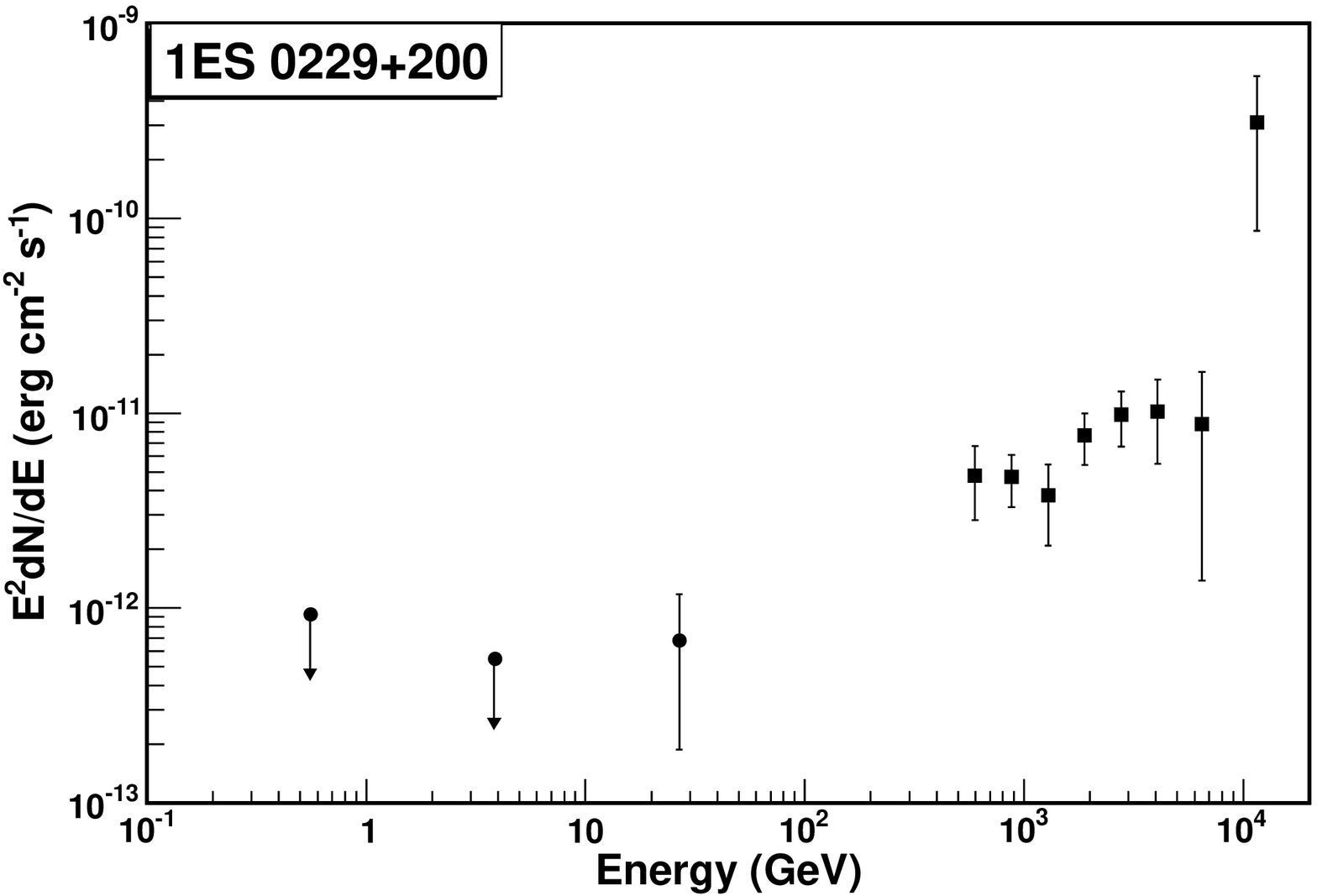}
\includegraphics[width=80mm]{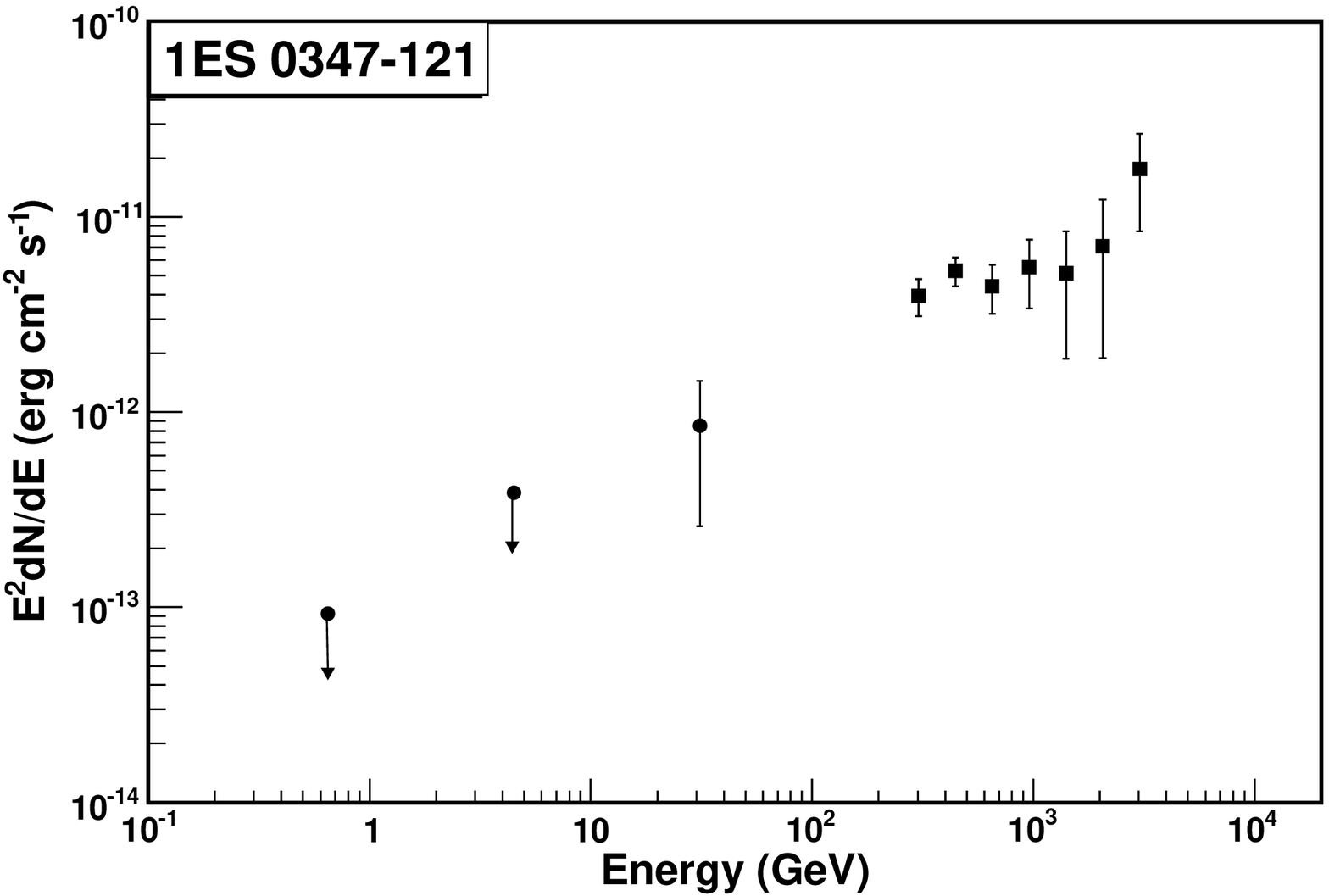}
\includegraphics[width=80mm]{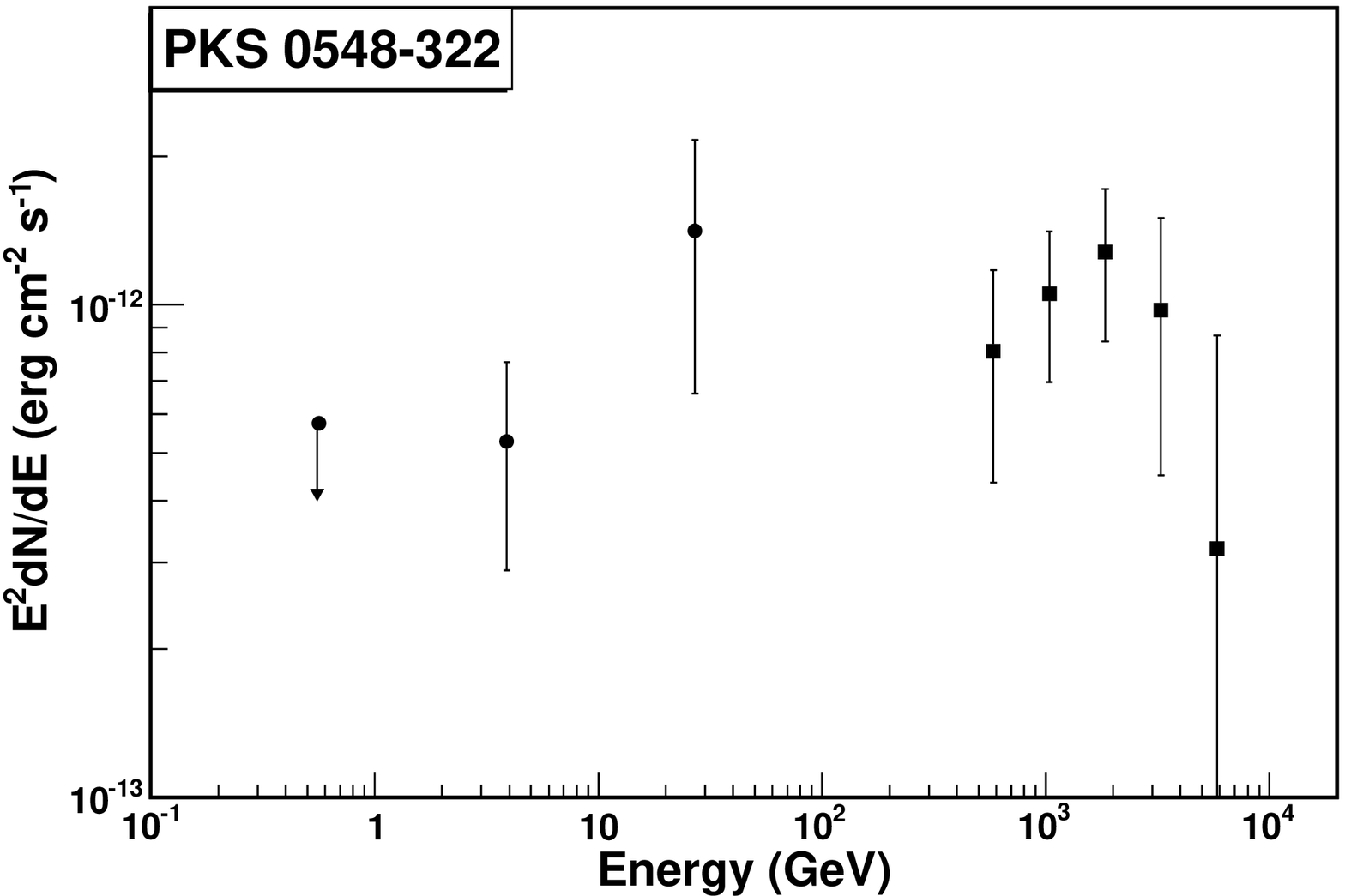}
\includegraphics[width=80mm]{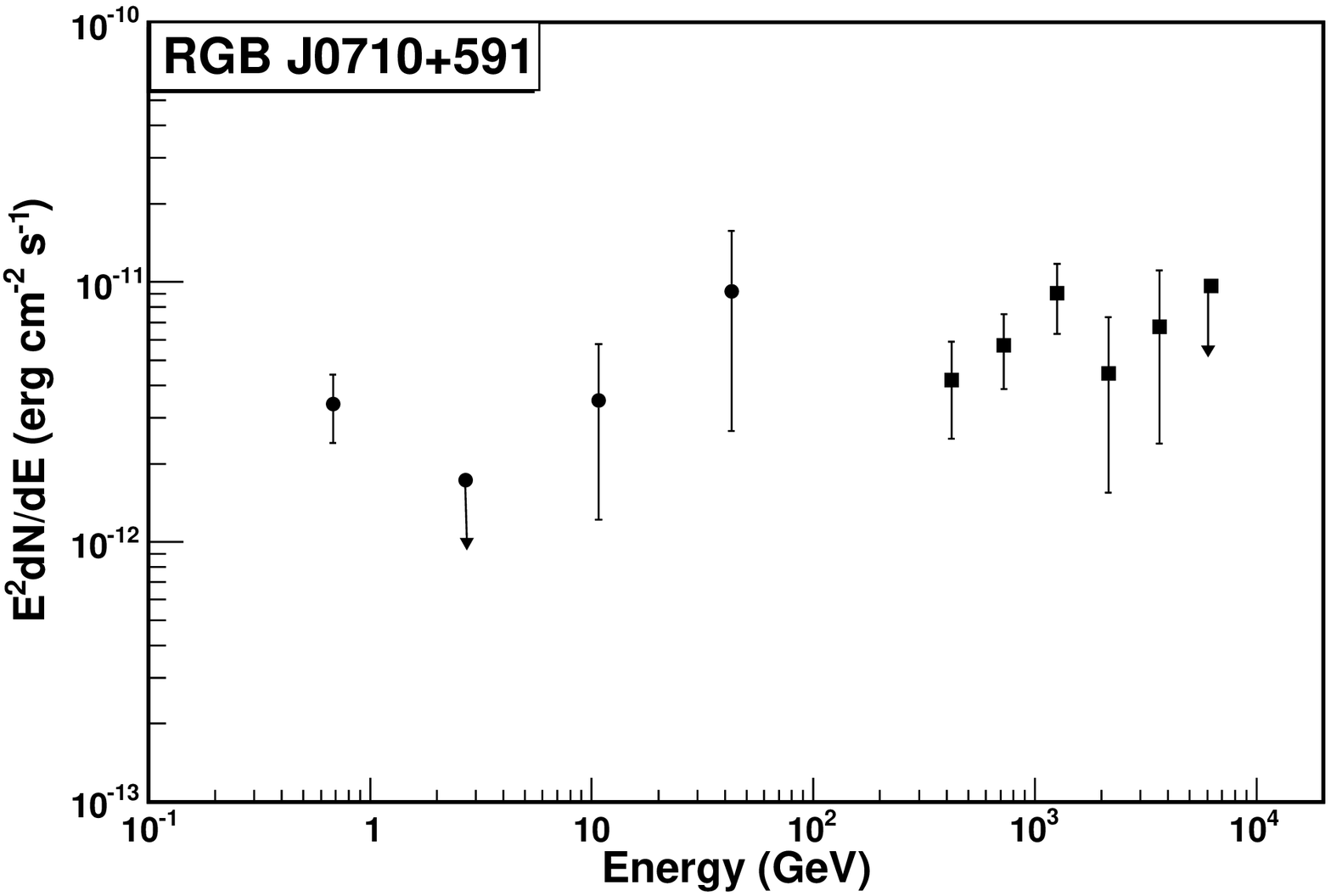}
\includegraphics[width=80mm]{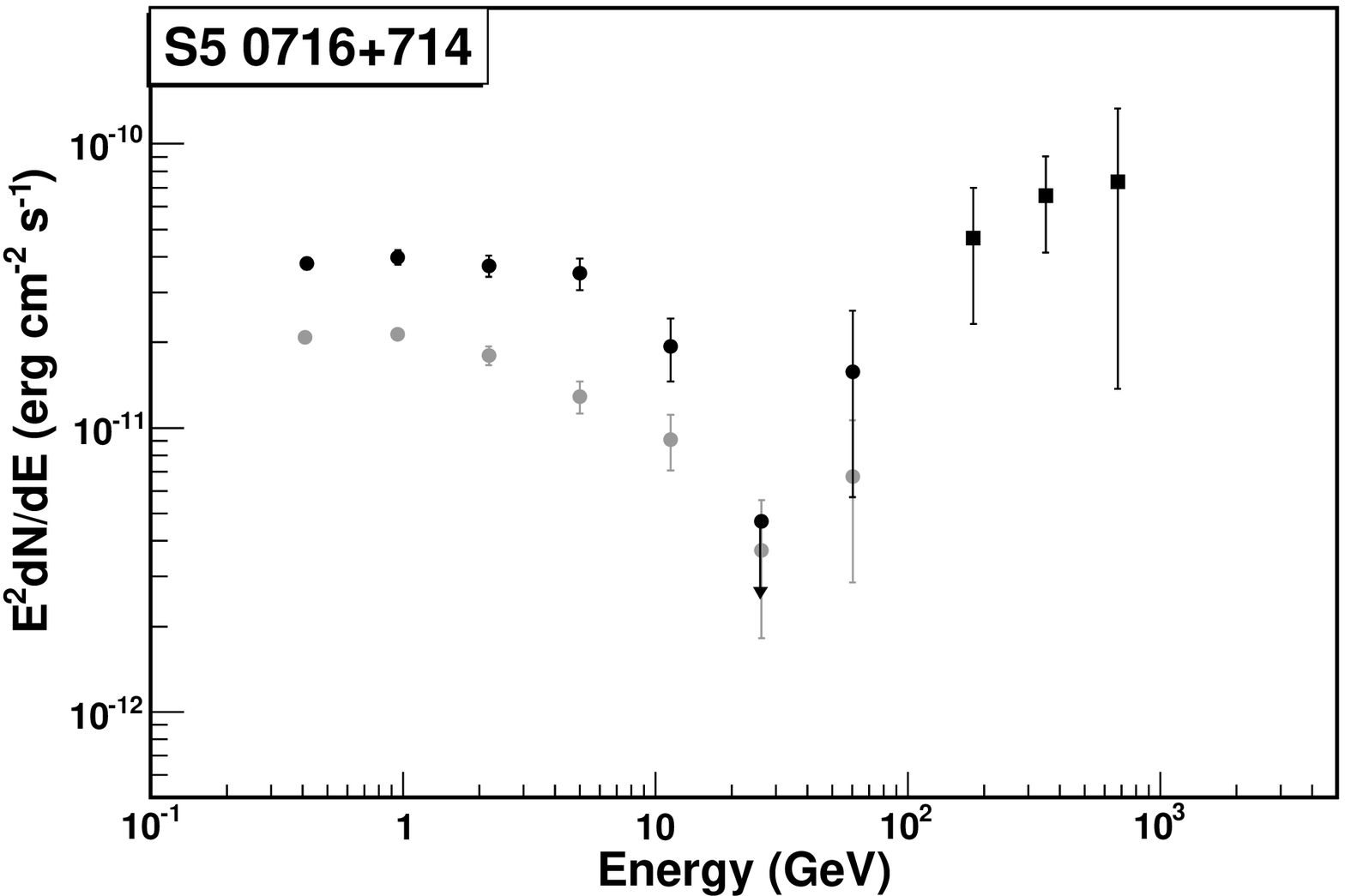}
\end{figure*}
\begin{figure*}
\centering
\includegraphics[width=80mm]{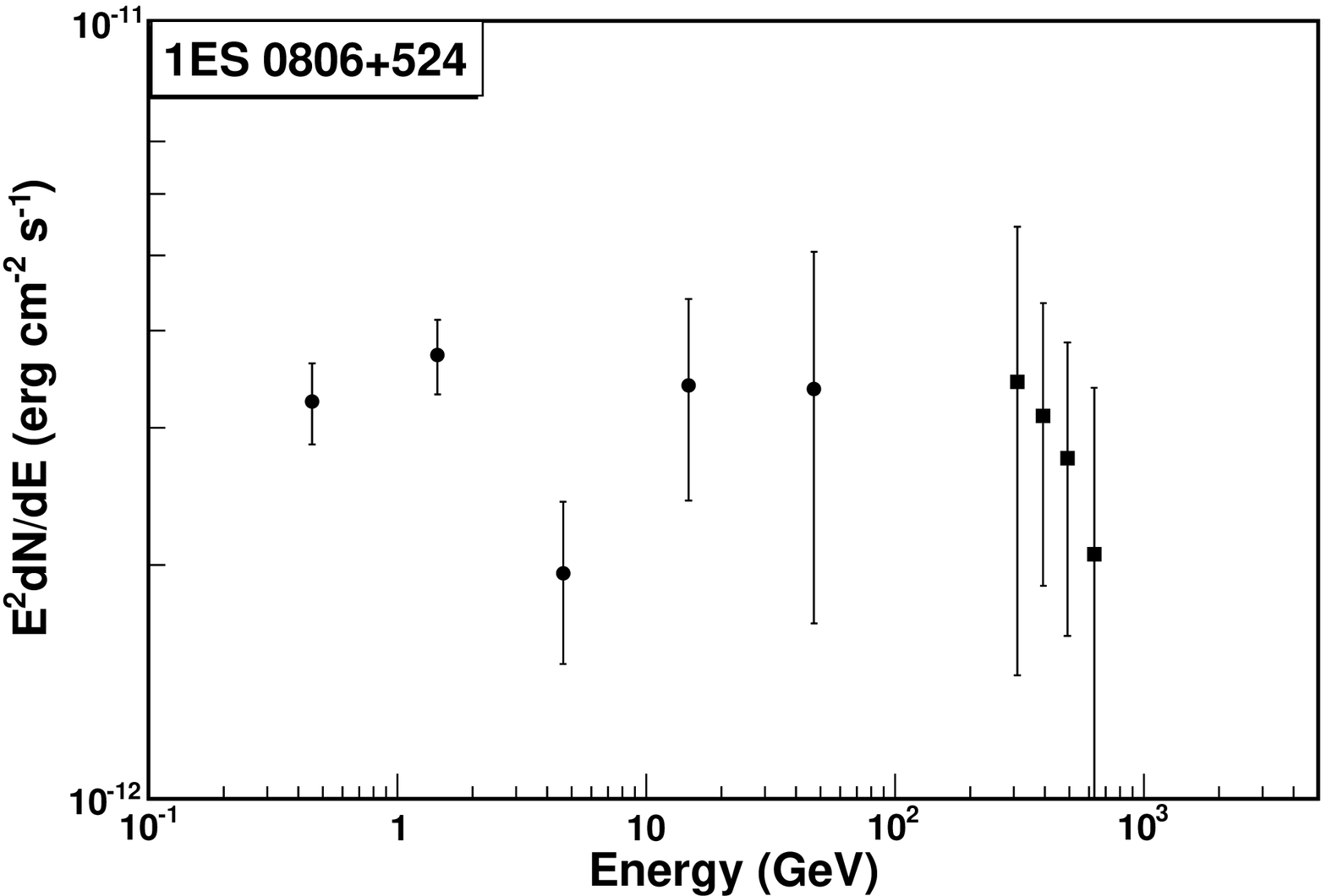}
\includegraphics[width=80mm]{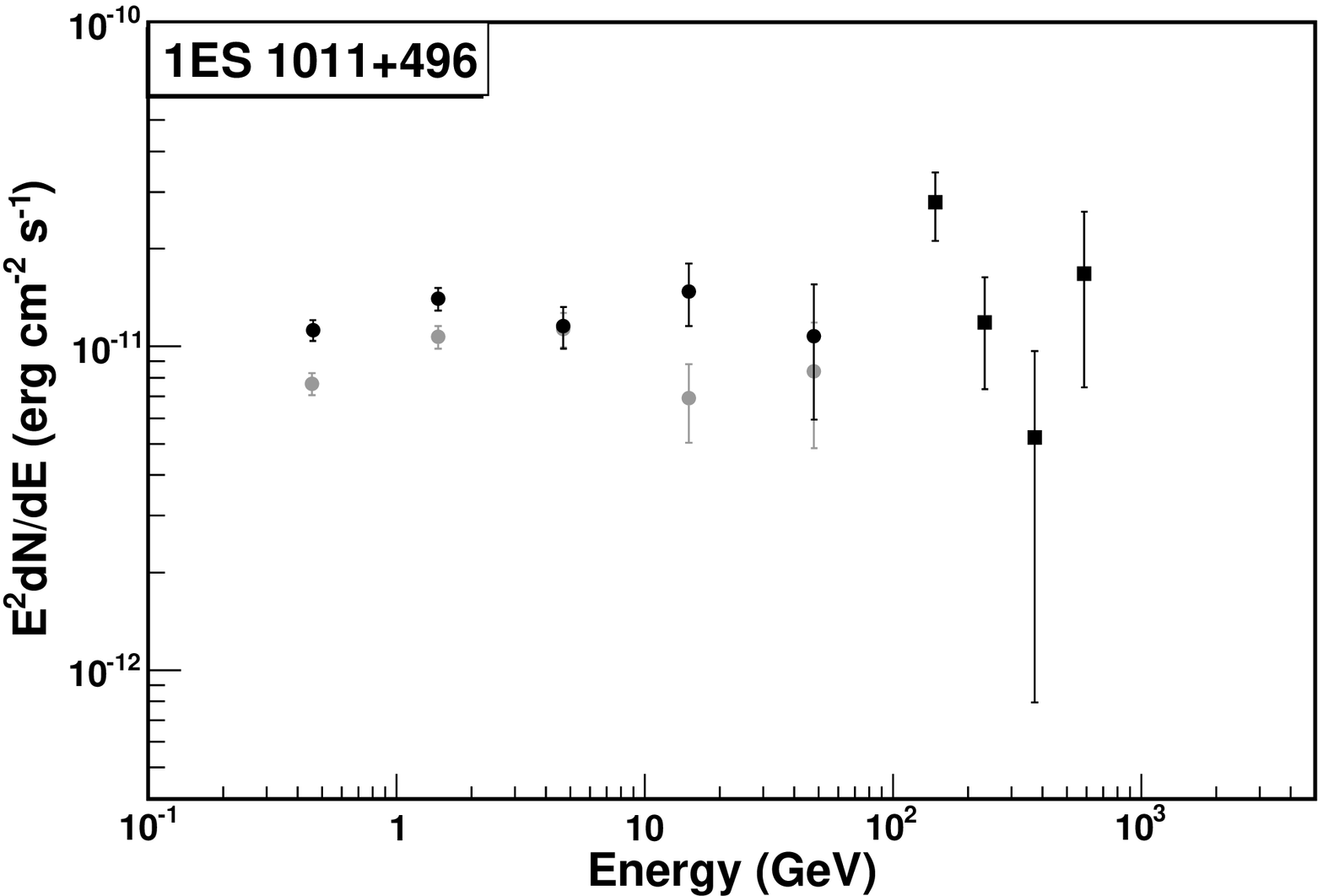}
\includegraphics[width=80mm]{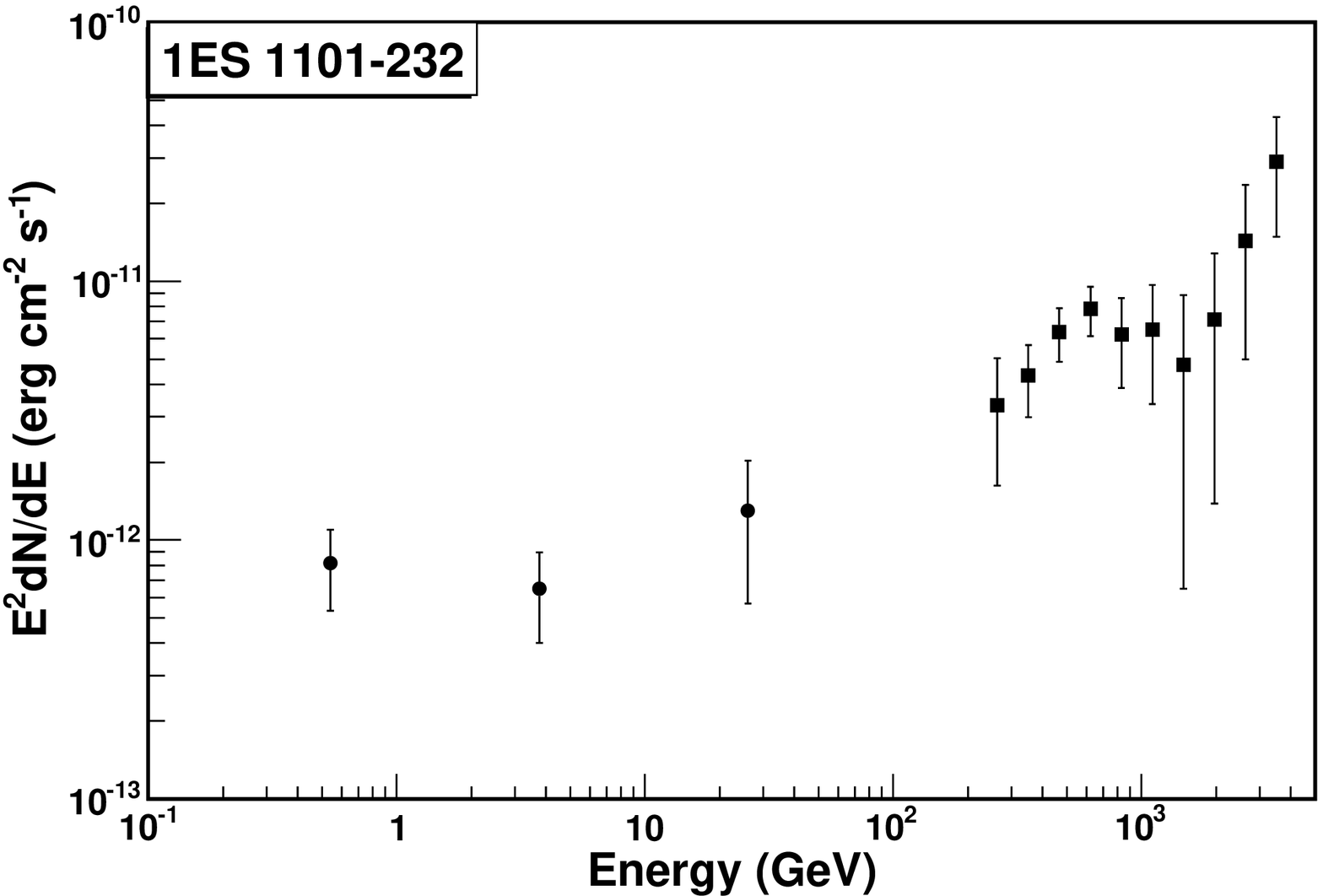}
\includegraphics[width=80mm]{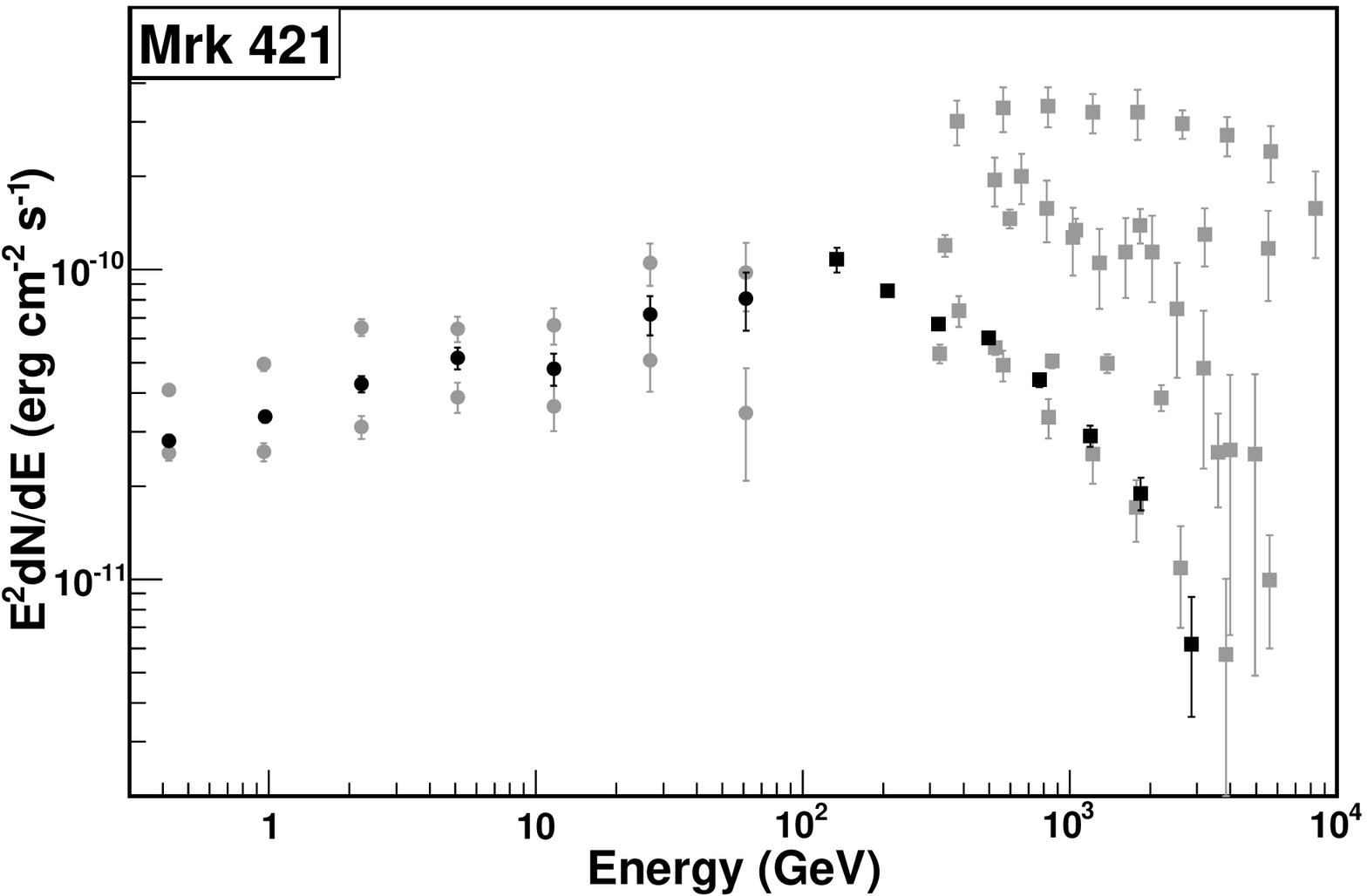}
\includegraphics[width=80mm]{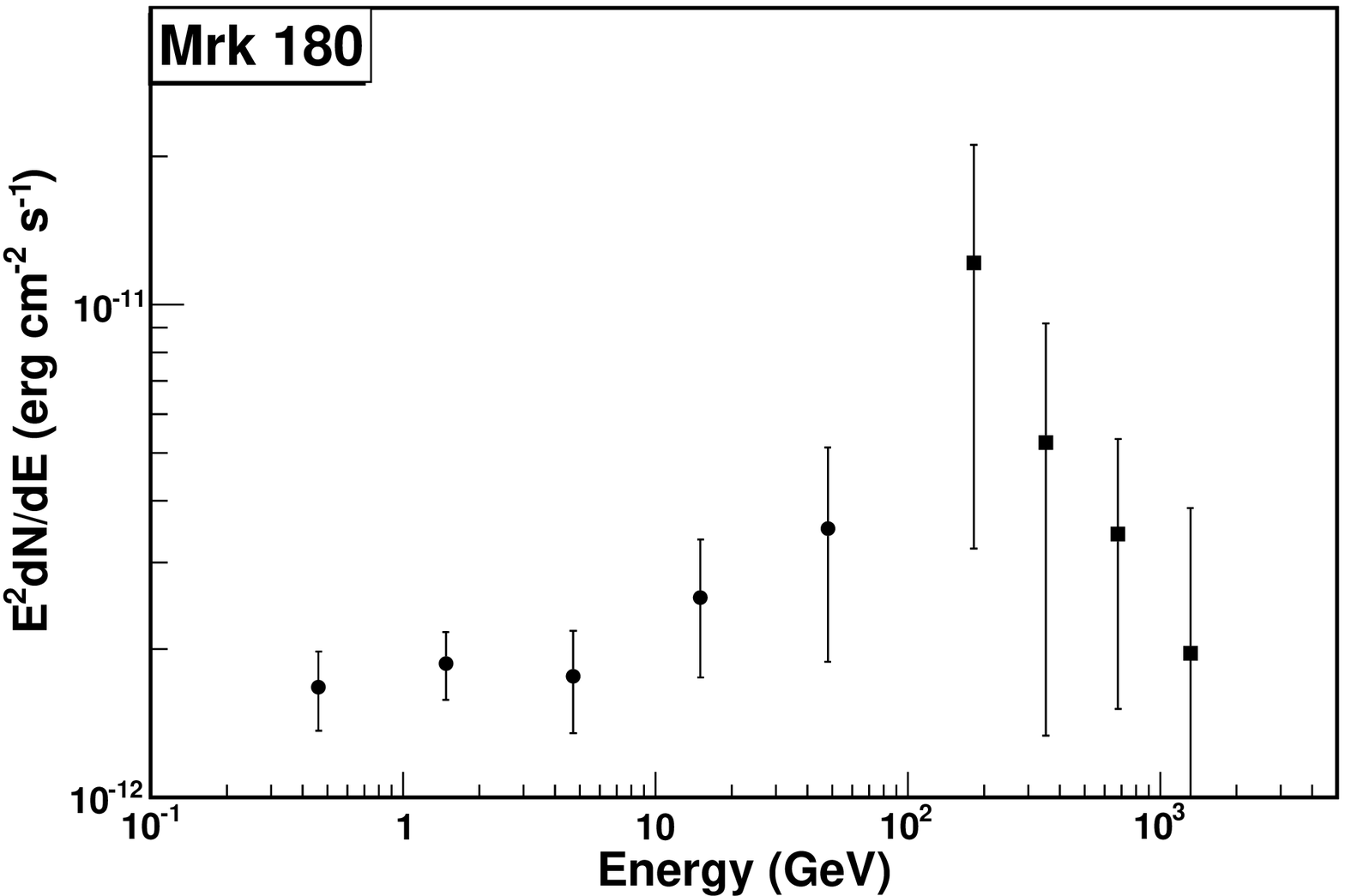}
\includegraphics[width=80mm]{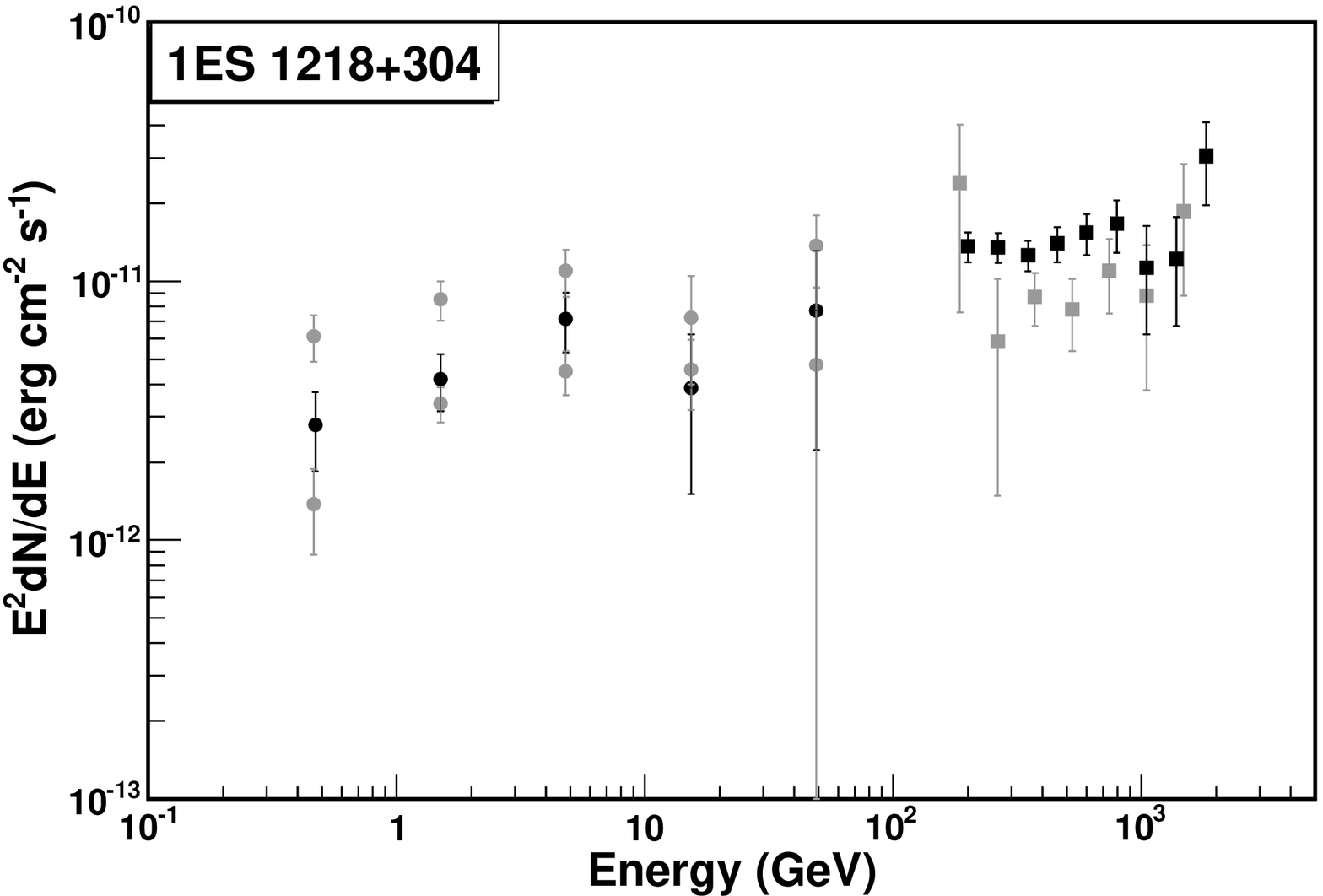}
\includegraphics[width=80mm]{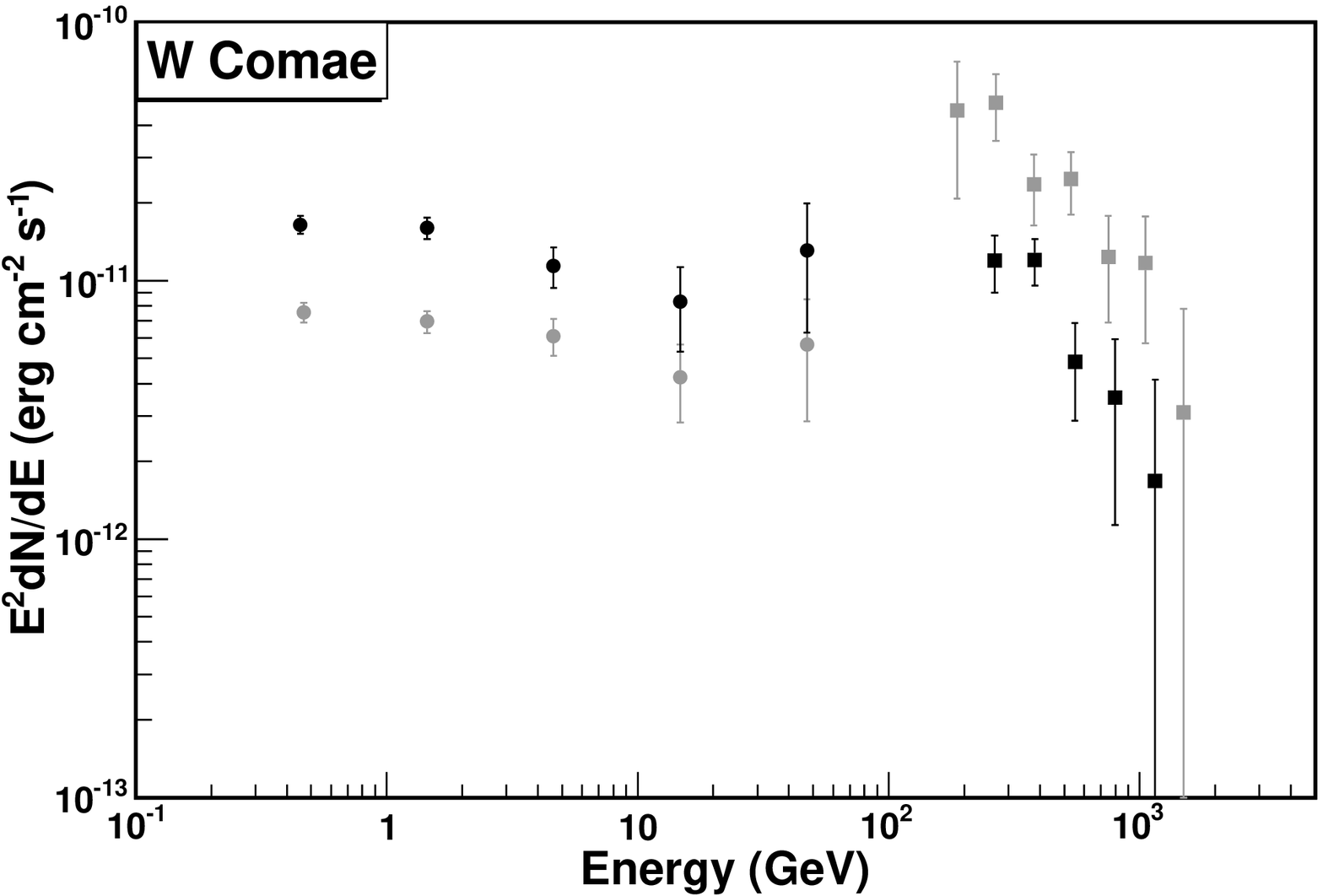}
\includegraphics[width=80mm]{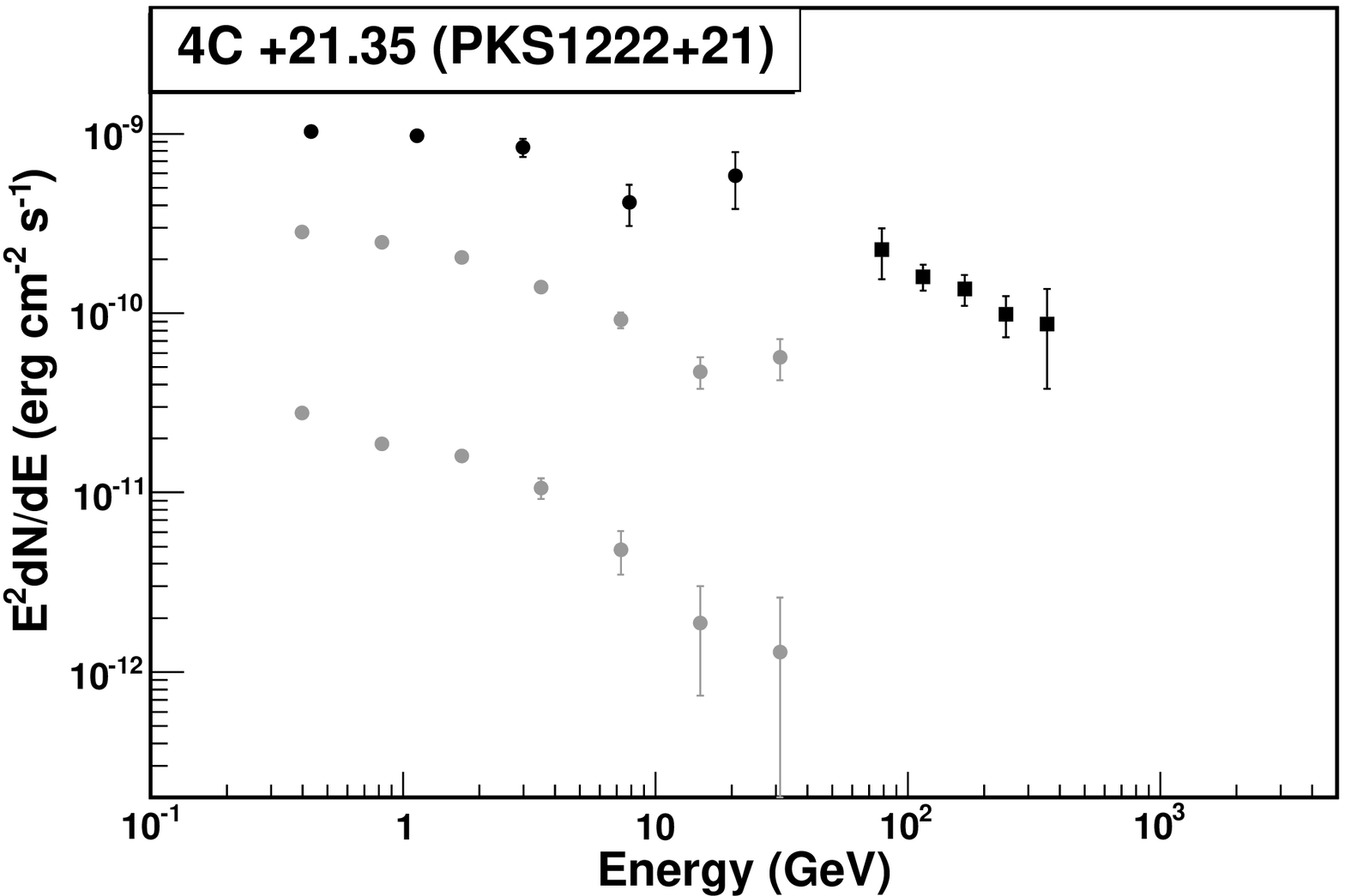}
\end{figure*}
\begin{figure*}
\centering
\includegraphics[width=80mm]{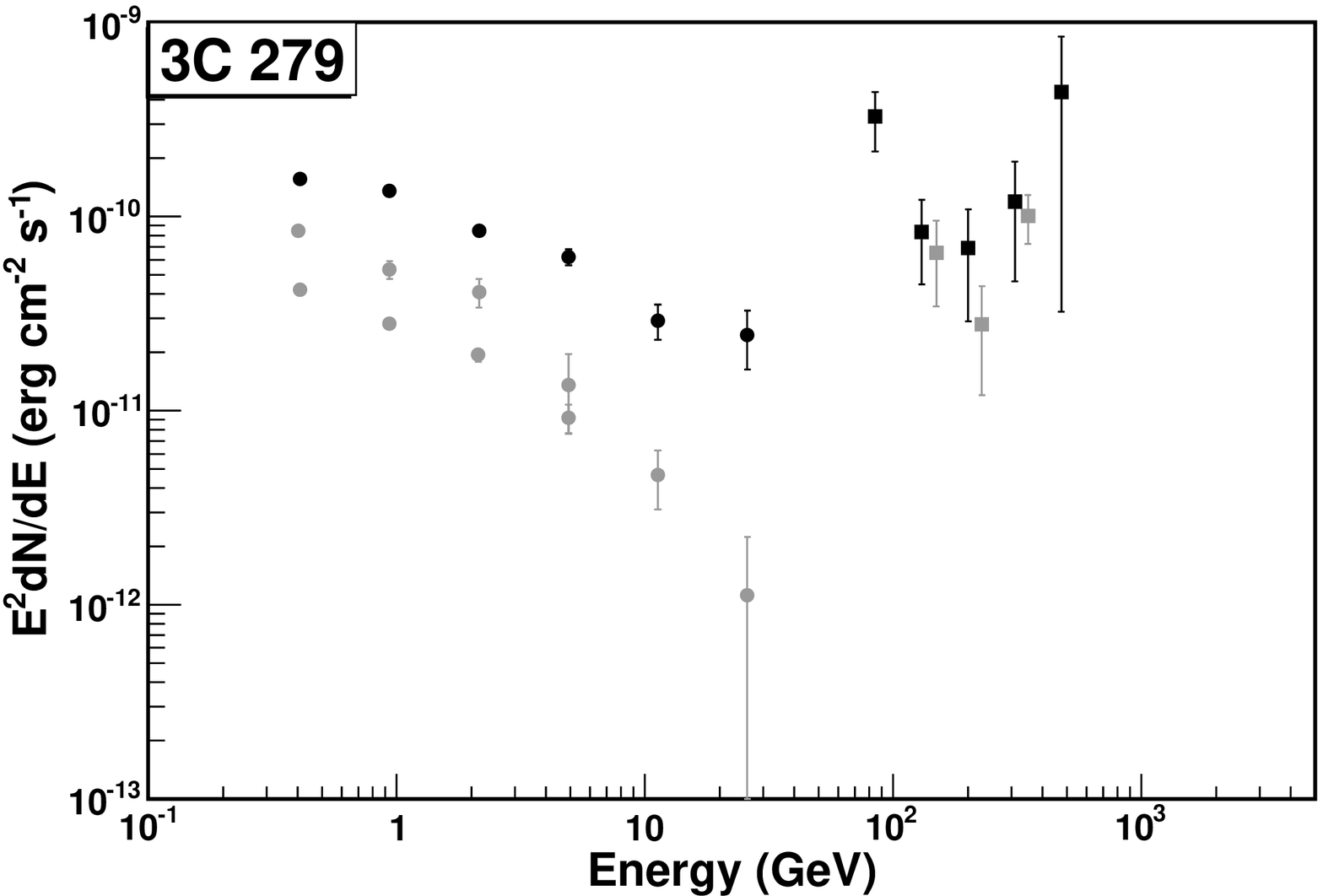}
\includegraphics[width=80mm]{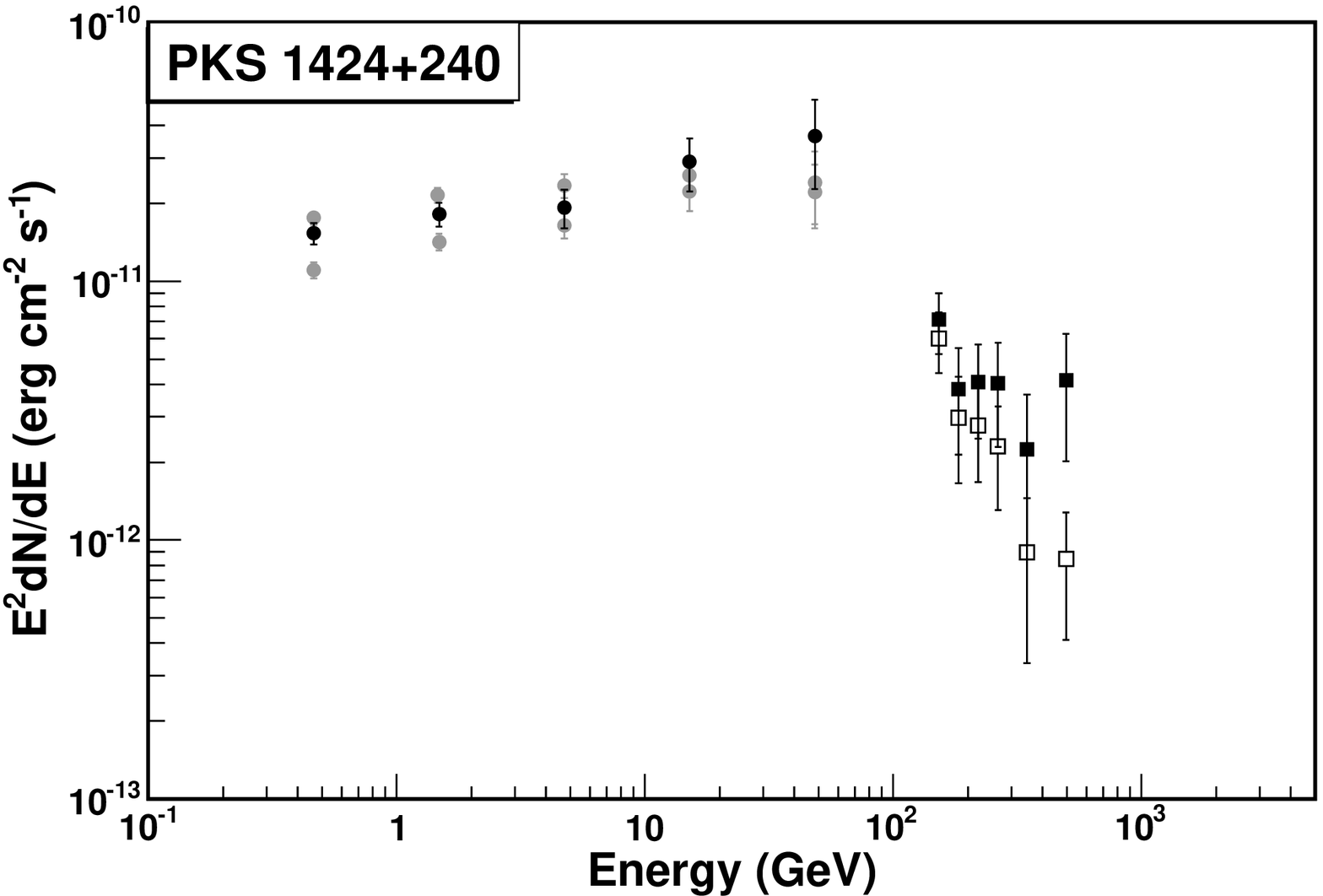}
\includegraphics[width=80mm]{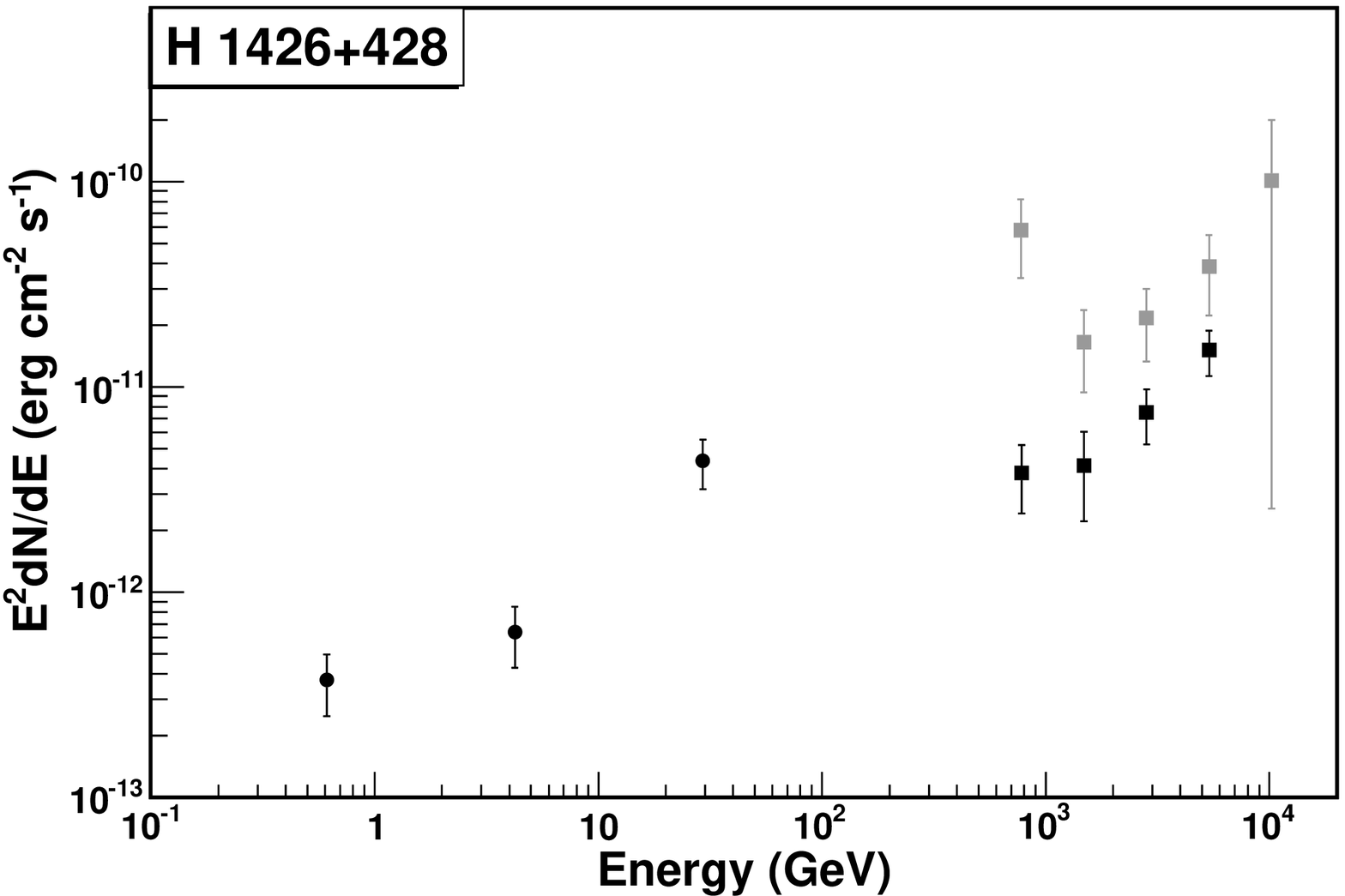}
\includegraphics[width=80mm]{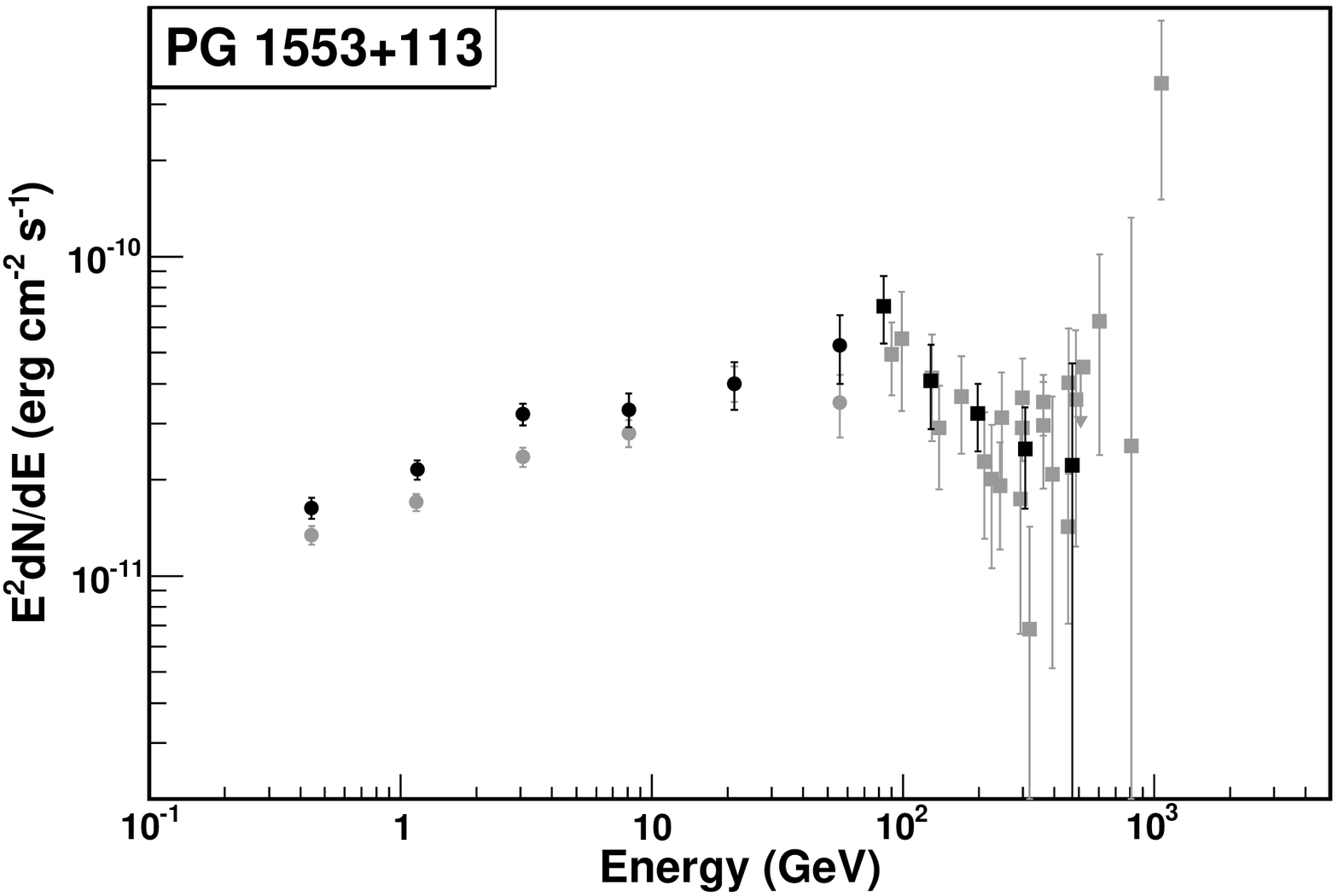}
\includegraphics[width=80mm]{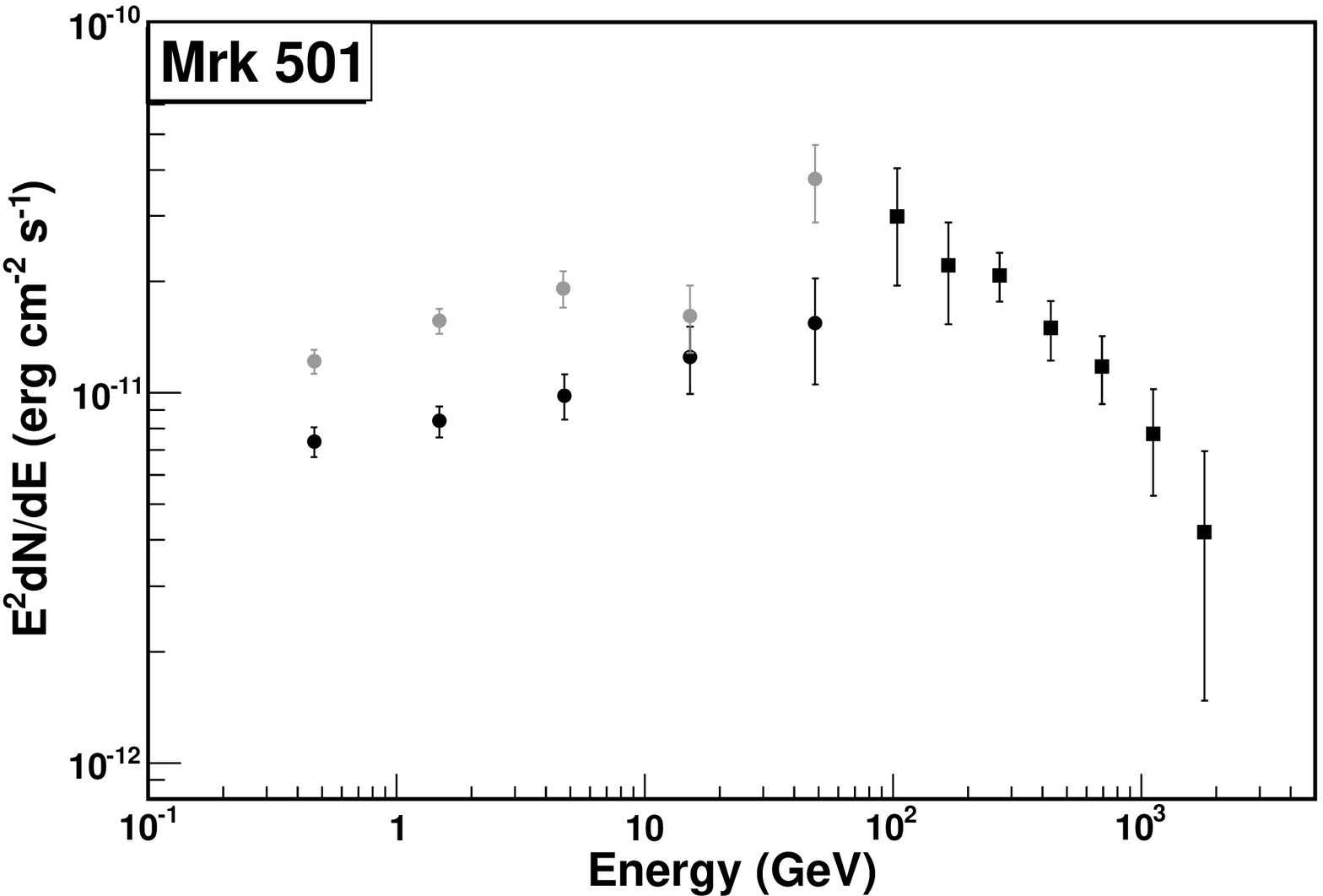}
\includegraphics[width=80mm]{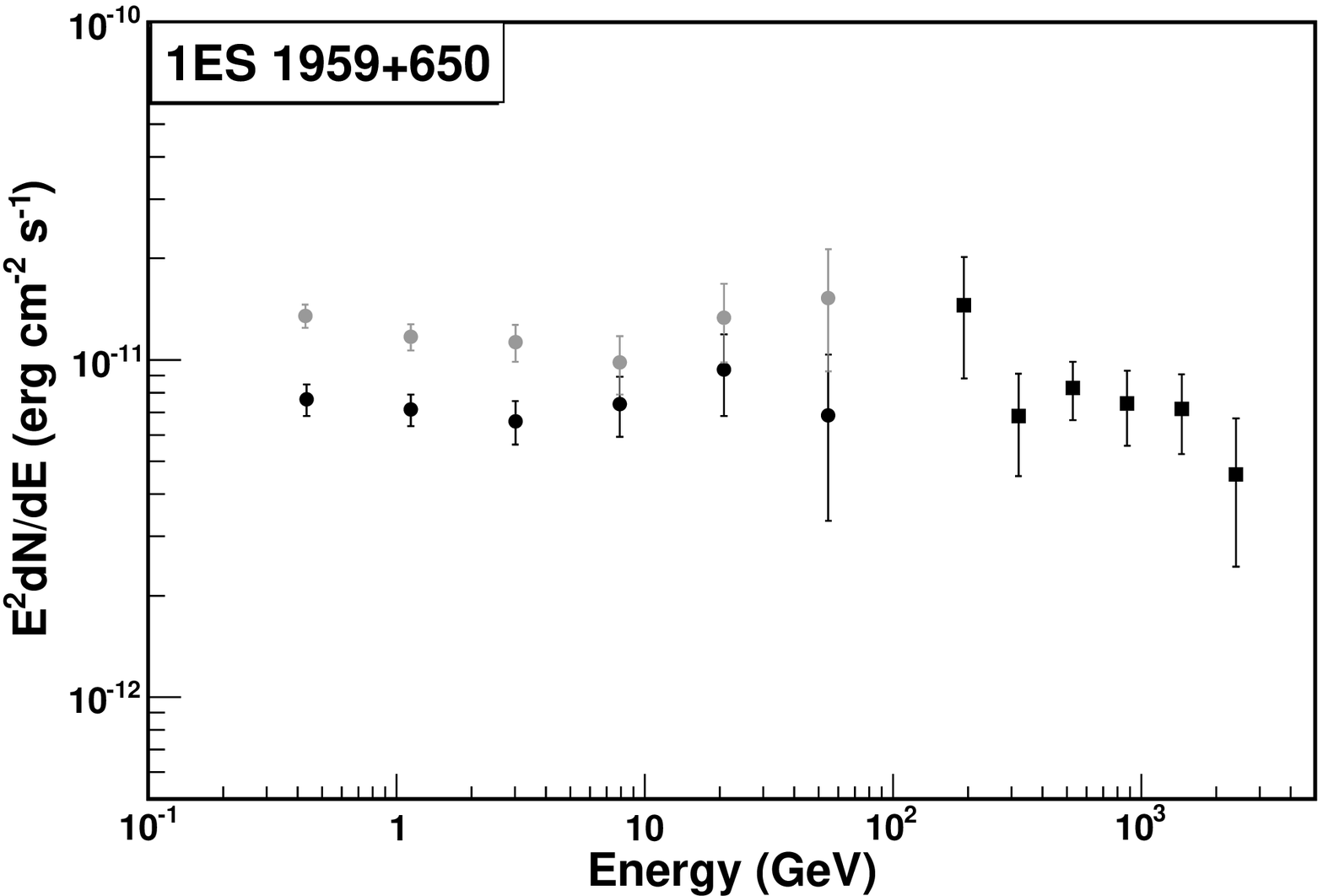}
\includegraphics[width=80mm]{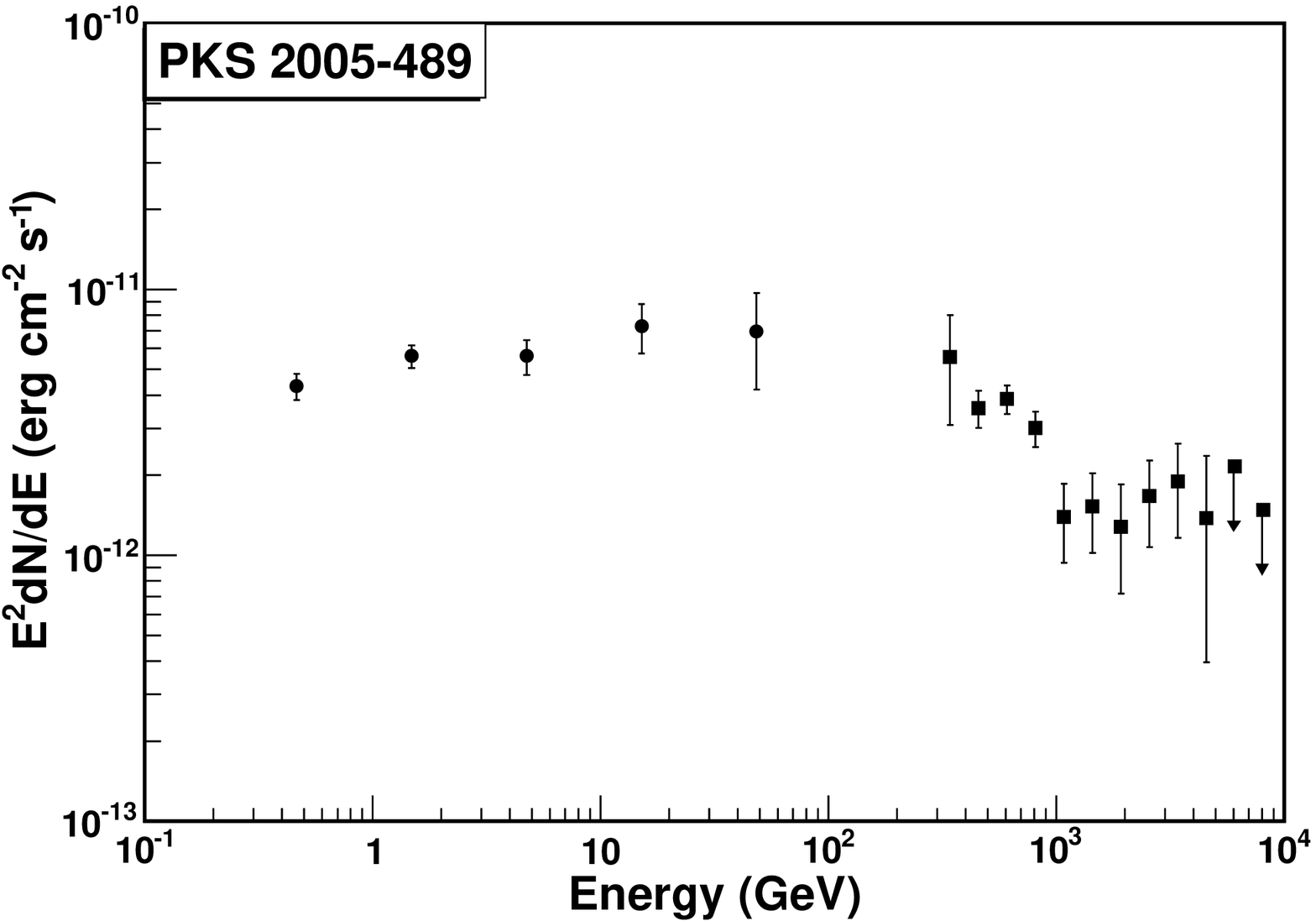}
\includegraphics[width=80mm]{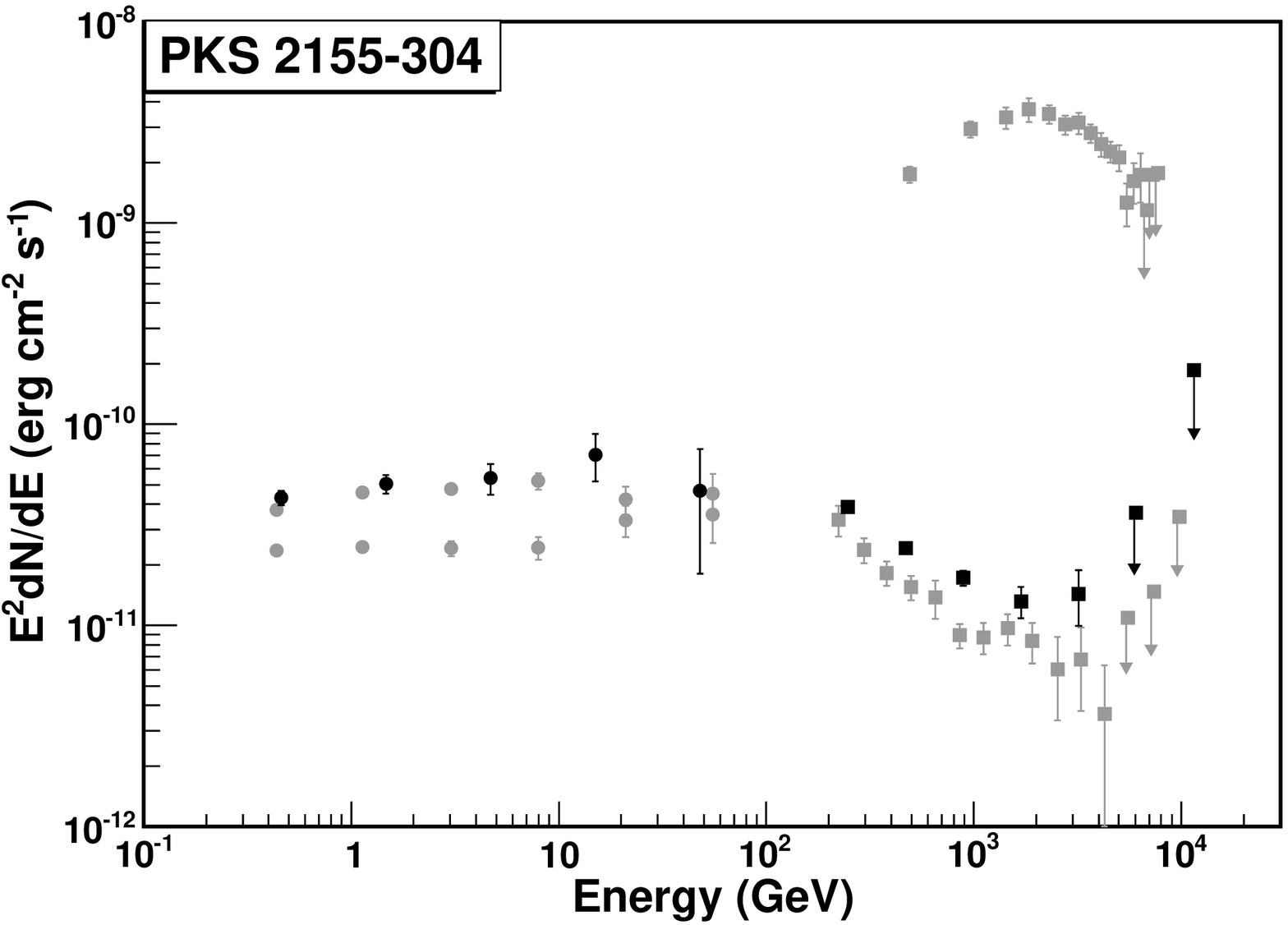}
\end{figure*}

\subsection{IC Peak Frequency}\label{IC_peak}
\begin{figure}
\centering
\includegraphics[width=80mm]{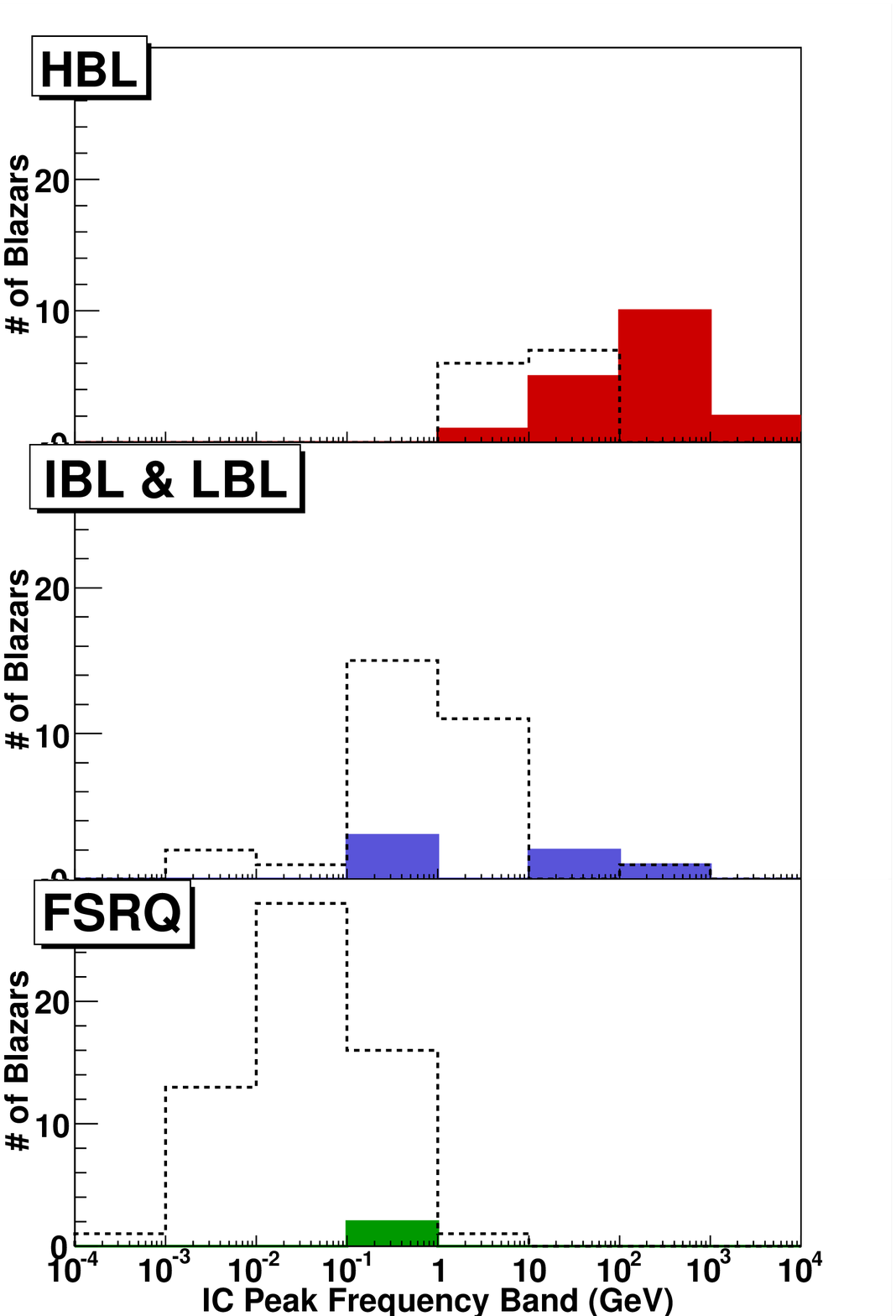}
\caption{{\footnotesize Distribution of the IC peak bands, defined as the energy decade in which the largest amount of power is emitted, in combined GeV-TeV spectra. Top, middle and bottom panels show HBLs, IBLs+LBLs, and FSRQs respectively. HBLs tend to peak at higher frequencies, in accordance with their respective synchrotron peak frequencies $\nu_\mathrm{syn}$ (see Table~\ref{tab:sample}), and a decreasing trend in IC peak bands from top to bottom panel is seen. The dashed lines represent the same distribution for the blazar sample from \citep{steve_LBAS}.}}
\label{fig:IC_peak_dist}
\end{figure}

The peak frequency of the IC component is a salient parameter for describing blazar non-thermal continua and studying population trends. Systematic studies for measuring the IC peak frequency mostly suffer from the lack of statistics and simultaneous data. A similar work was carried out in~\citep{zhang2011}, where archival multiwavelength data were used to model TeV blazar SEDs and determine the IC peak frequency ($\nu_\mathrm{IC}$). A positive correlation between $\nu_\mathrm{syn}$ and $\nu_\mathrm{IC}$ was reported. In this work, we focus on finding the IC ``peak frequency band" rather than the ``peak frequency", using a model independent approach. 
For each blazar SED shown in Figure~\ref{fig:combined_spectra}, we identify the energy decade in which the largest amount of power is emitted. 
Note that the spectral points used in the VHE spectra are EBL-corrected. 
Figure~\ref{fig:IC_peak_dist} shows the distribution of the IC peak bands for different blazar types.
We observe that the FSRQs, LBLs and IBLs have the maximum of their emission mostly below 1 GeV. 
On the other hand, HBLs tend to peak in the TeV range. 
This positive correlation between the synchrotron ($\nu_\mathrm{syn}$) and the IC
peak frequencies ($\nu_\mathrm{IC}$), is in accordance with simple SSC models that
predict a positive correlation  
between $\nu_\mathrm{syn}$ and $\nu_\mathrm{IC}$~\citep{steve_LBAS}. 
The dashed lines represent the same distributions with the bright AGN sample from the first three months of \fermi data~\citep{steve_LBAS}. 
Our results tend to span the high frequency sides of all distributions and one clearly sees a shift to higher frequencies in the case of HBLs.
This is expected since our sample consists of TeV-selected objects, that mostly correspond to relatively weak sources in the GeV data, and are therefore less likely to appear in a bright AGN sample.
It should also be noted that we use a model independent method using only \fermi and VHE data, whereas \citep{steve_LBAS} uses multiwavelength data and some modeling in cases where the soft X-ray band is dominated by the synchrotron component, a typical feature for our blazar sample.
\begin{figure}
\centering
\includegraphics[width=80mm]{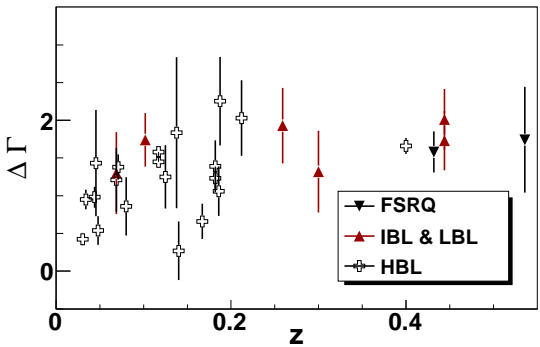}
\caption{{\footnotesize $\Delta\Gamma$ ($\Gamma_\mathrm{TeV}-\Gamma_\mathrm{GeV}$)
vs. redshift. Empty crosses, triangles and inverse triangles represent HBLs,
IBLs+LBLs, and FSRQs, respectively. The fact that a constant function does not provide a good description for the data could be interpreted as a model independent indication for the EBL absorption.}}
\label{fig:delta_gamma_vs_z}
\end{figure}
\begin{figure}
\centering
\includegraphics[width=80mm]{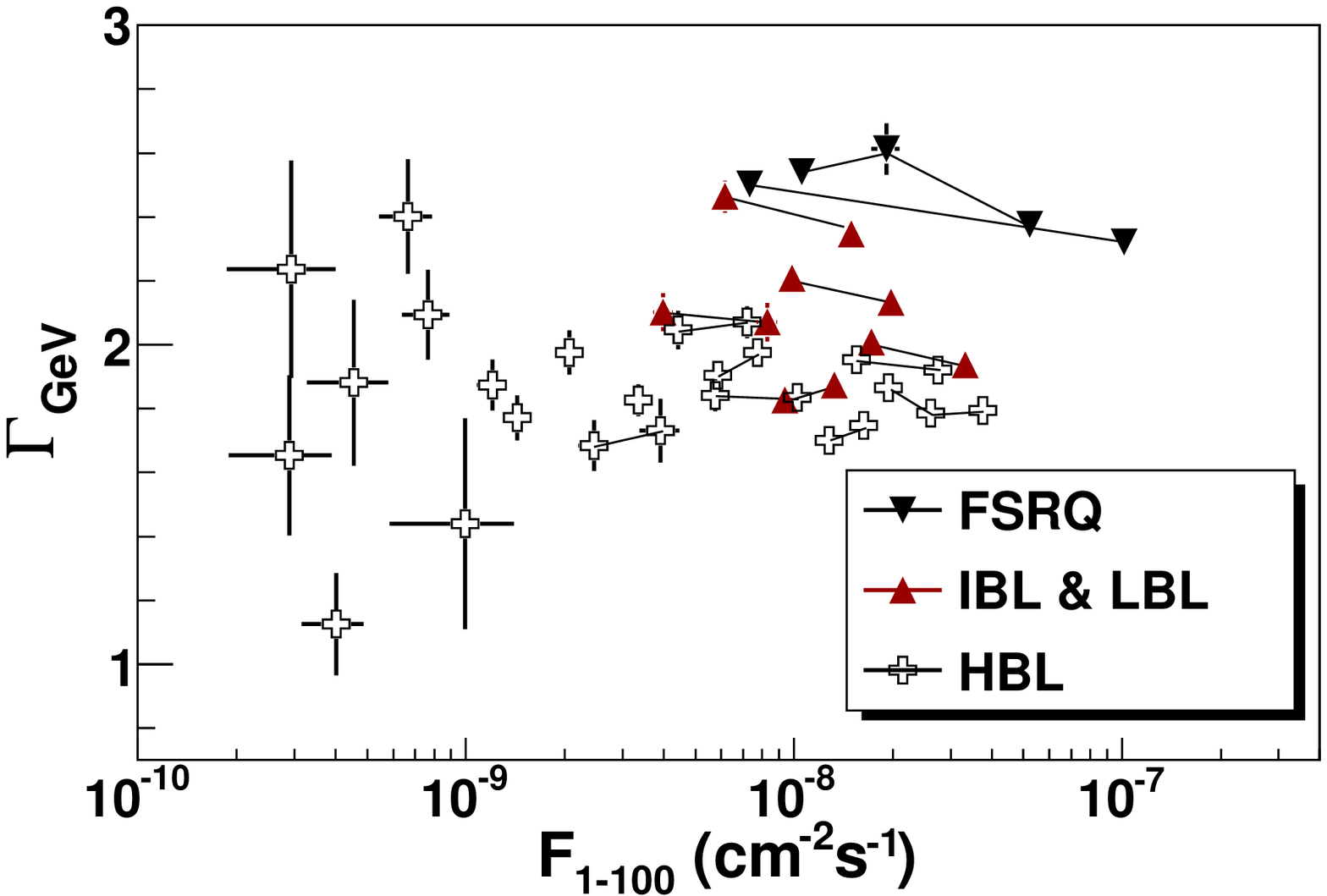}
\caption{{\footnotesize $\Gamma_\mathrm{GeV}$ vs. flux in the energy band 1-100\,GeV from the analysis of 27 month \fermi data with a power law fit (see Table~\ref{tab:results}). Empty crosses, triangles and inverse triangles represent HBLs, IBLs+LBLs, and FSRQs, respectively. Solid lines connect different states of the same blazar.}}
\label{fig:gamma_vs_N}
\end{figure}
\subsection{Hard TeV BL Lac Objects}\label{tev_peaked}
The combined GeV-TeV spectra of some blazars in our sample (1ES 0229+200, 1ES 0347-121, 1ES 1101-232, 1ES 1218+304, H 1426+428) suggest a $\nu_\mathrm{IC}$ beyond $\sim1$ TeV. 
These blazars are mostly weak or non-detected in the \fermi range, with a hard spectral index in both GeV and TeV bands. 
It follows that they may belong to the so-called ultra-high-frequency-peaked BL Lac sub class (UHBLs; see, e.g., \citep{costamante_rome}) that would constitute the extreme end of the population, and is expected to dominate the TeV luminosity of the universe. 
Several mechanisms have been set forth to explain the formation of these hard $\gamma$-ray spectra~\citep{ultraHE}.
Extensive spectral analysis of these objects would be valuable for EBL and intergalactic magnetic field measurements. 
It should be noted that at energies of a few TeV and beyond, our spectra become EBL-model-dependent. 
For this reason, we have compared our adopted EBL model with two other models from recent studies~\citep{finke_bg,ebl_absorption}. 
We have found that for the data samples mentioned above, if we used any of the other two EBL models, the dispersion in highest energy flux points would be less than 20\%, and consequently the observed spectral upturns would not be affected significantly.
\par
With additional data, a deeper variability study carried on these blazars would relate to arguments that support the cosmic ray production as the origin of TeV blazar emission, since in that scenario no short timescale variability would be expected to be observed~\citep{uhe_cr}.
Among the UHBL candidates, the ones that are present in the 1FGL catalog (1ES~1101-232, 1ES~1218+304, H~1426+428) have relatively small \fermi variability indices (see Table~\ref{tab:sample}). 
In addition to that, our calculations of $F_\mathrm{var}$ for all five blazars using 27 months of \fermi data do not indicate a significant hint of variability either (see Table~\ref{tab:results}). 
\begin{table*}
\centering
\begin{small}
\begin{tabular}{|c|cccc|}
\hline
Name 		&SED type 	& Increase in $F_\mathrm{1-100}$ (\%)	&$\Gamma_\mathrm{low}$	&$\Gamma_\mathrm{high}$ 	\\
		&(1)		&(2)					&(3)			&(4)				\\
\hline
\hline
3C66A		&IBL		&95						&$2.00\pm0.02$	&$1.93\pm0.02$		\\
\hline
S5 0716+714	&LBL		&100						&$2.20\pm0.03$	&$2.13\pm0.04$		\\
\hline
1ES 1011+496	&HBL		&30						&$1.90\pm0.04$	&$1.97\pm0.04$		\\
\hline
Mrk 421		&HBL		&90						&$1.86\pm0.03$	&$1.79\pm0.02$		\\
\hline
1ES 1218+304	&HBL		&60						&$1.68\pm0.08$	&$1.73\pm0.10$		\\
\hline
W Comae		&IBL		&110						&$2.10\pm0.06$	&$2.07\pm0.06$		\\
\hline
PKS 1222+21	&FSRQ		&1290						&$2.50\pm0.04$	&$2.32\pm0.02$		\\
\hline
3C 279		&FSRQ		&395						&$2.54\pm0.03$	&$2.37\pm0.02$		\\
\hline
PKS 1424+240	&IBL		&45						&$1.82\pm0.04$	&$1.87\pm0.03$		\\
\hline
PG 1553+113	&HBL		&30						&$1.70\pm0.03$	&$1.74\pm0.03$		\\
\hline
Mrk 501		&HBL		&80						&$1.84\pm0.05$	&$1.83\pm0.04$		\\
\hline
1ES 1959+650	&HBL		&60						&$2.04\pm0.06$	&$2.07\pm0.05$		\\
\hline
PKS 2155-304	&HBL		&80						&$1.95\pm0.03$	&$1.92\pm0.02$		\\
\hline
BL Lacertae	&LBL		&140						&$2.46\pm0.05$	&$2.34\pm0.04$		\\
\hline
\end{tabular}
\end{small}
\caption{{\small Spectral variations in the \emph{Fermi} data for blazars for which at least two different \fermi flux states are available. Column (1) shows the spectral energy distribution (SED) type. $F_\mathrm{1-100}$ is the integral flux for the 1--100~GeV band. Column (2) shows the percent increase in $F_\mathrm{1-100}$ from low to high \fermi state. FSRQs and LBLs seem to show the most significant flux variability in this energy range. Columns (3) and (4) list the GeV photon indices in low and high \fermi states, respectively. No significant changes in photon index are seen, except for the two FSRQs for which the index shows a slight hardening from low to high flux states.}}
\label{tab:spectral_variations}
\end{table*}
\subsection{Spectral Variability}\label{flux_var}
VHE emission from blazars is highly variable. 
This variability, manifested in irregular flares, is one of the most typical and promising blazar behaviors for studying the nature of underlying emission mechanisms. 
The observed flux change during a VHE flare can be as rapid as minute scales~\citep{magic_501_flare} and as large as 40 times the baseline emission~\citep{bl_lac_flare}. 
Blazars that have been reported to have a variable flux are marked with an asterisk in Table~\ref{tab:sample}.
On the other hand, \emph{Fermi} data do not exhibit flux variability as extreme as in the VHE band. 
In fact, having a smaller effective area than the ground-based VHE telescopes and operating mostly in survey mode rather
than pointing, \emph{Fermi}-LAT does not have the sensitivity to probe sub-hour timescale variability in blazars. 
Still, a possible correlation between GeV and TeV emission remains viable~\citep{3c_66a_MW,aleksic11b} and an enhanced activity in the high-energy tail of the \fermi band could therefore indicate a TeV flare. 
In this frame, monitoring GeV flares to trigger TeV observations is important~\citep{manel_ICRC}, and potentially could help in probing fast variability.
To examine variability within the \fermi data, we compared high- and low-state \emph{Fermi} spectra from 14 blazars (see Table~\ref{tab:spectral_variations}). 
Half of these blazars have their integral flux in 1-100\,GeV ($F_{\mathrm{1-100}}$) increased by at least 90\% in the high state. 
The largest flux increase is seen in the case of the two FSRQs 3C 279 and PKS 1222+21. 
As depicted by their respective light curves in Figure~\ref{fig:lc}, these two objects have undergone dramatic GeV flares. 
Such a large scale flux increase does not hold for the remainder of the blazars. 
However, one should keep in mind that for most of the TeV blazars, the \emph{Fermi} band is a relatively stable
region of the SED, since it samples the low energy part of the parent electrons, that have a longer cooling time.
Table~\ref{tab:spectral_variations} gives a summary of the results of the spectral variations seen in the \fermi data. 
We also calculated the variability amplitude ($F_\mathrm{var}$) within 27 months of \fermi data for each blazar, using the method described in~\citep{f_var}. 
$F_\mathrm{var}$ is a measure of the intrinsic source variance, calculated based on excess variance. 
For blazars with negative excess variance, 95\% confidence level upper limits are given.
The blazars 3C~66A, PKS 1222+21 and 3C~279 are the most variable ones according to this calculation ($F_\mathrm{var}>0.5$). 
Our results are in agreement with the 1FGL catalog (see Table~\ref{tab:sample}).
Comparing these results with the TeV variability flags, we do not find any obvious relation between GeV and TeV variabilities (see Table~\ref{tab:results}).
\par
Within the \fermi energy range, blazars in our sample do not exhibit dramatic changes in their spectral index between different flux states (see Table~\ref{tab:spectral_variations}). 
Consequently, this makes the photon index a reasonable parameter to use for studying the non-contemporaneous combined SEDs. 
Figure~\ref{fig:delta_gamma_vs_z} shows a scatter plot of observed $\Gamma_\mathrm{TeV}-\Gamma_\mathrm{GeV}$ versus redshift. 
A constant function does not provide a good description for the data, with $\chi^{2}/\mathrm{dof}=204/27$, which could be interpreted as a model-independent indication for the EBL absorption. 
The difference between TeV and GeV photon indices increases with redshift. 
This is expected since the VHE $\gamma$-ray photons pair produce with the EBL photons~\citep{ebl_absorption} and this effect becomes more enhanced at larger redshifts, making the universe opaque to TeV $\gamma$ rays at distances larger than $z\sim0.5$. 
HE spectra are not affected by the EBL, whereas VHE spectra become softer with increasing redshift.
A similar observation was reported by~\citep{steve_tev}, in a study carried out on a sample of TeV-selected AGNs detected with \emph{Fermi}. 

\par
Figure~\ref{fig:gamma_vs_N} shows the relation between the spectral index $\Gamma_{\mathrm{GeV}}$ and the flux normalization $F_{\mathrm{1-100}}$ obtained from power-law fits. 
FSRQs and two subgroups of BL Lacs are clearly separated in the parameter space. 
This is in accordance with the aforementioned positive correlation trend between $\nu_\mathrm{syn}$ and $\nu_\mathrm{IC}$, since 1\,GeV typically corresponds to the rising edge of the IC component in an HBL SED, sampling a relatively low flux with hard spectral index. 
On the other hand for an FSRQ, 1\,GeV will correspond to the peak or the falling edge of the IC component. 
The fact that FSRQs have relatively more luminous IC emission explains the softening trend with a larger normalization factor.
However, the pattern that we observe between different flux states of a given blazar is the opposite. 
In most cases, a slight spectral hardening accompanies high flux states, indicating a change in the spectral shape and enhanced flux increase at high-energy tail of the spectrum.

\subsection{Spectral Features}\label{spectralFeatures}
In most of the blazars in our sample, we observe interesting spectral features in the \emph{Fermi} band, that appear as dips in the 1--100 GeV energy range. 
In an attempt to find a quantitative description for these features, we fit the \emph{Fermi} spectral points with a simple power law (PL; Equation~(\ref{eq:PL})) and a broken power law (BPL; Equation~(\ref{eq:BPL})), and then compare the results.

\begin{equation} 
dN/dE=N_{0}\times\begin{cases}
(E/E_{b})^{-\Gamma_{1}}, & E<E_{b}\\
(E/E_{b})^{-\Gamma_{2}}, & \mathrm{otherwise}\\
\end{cases}
\label{eq:BPL}
\end{equation}

In the PL fit, the normalization $N_\mathrm{0}$ and the spectral index $\Gamma$ are free parameters, and the energy $E_{0}$ is fixed at 1\,GeV. 
In the BPL fit, the break energy $E_\mathrm{b}$ and the indices $\Gamma_{1}$ and $\Gamma_{2}$, along with the normalization $N_{0}$ are free. 
In Table~\ref{tab:dips_1}, we list the best-fit parameters from both functions and the likelihood ratio test results of BPL over PL. 
In 9 out of 33 cases, BPL yields a better fit over PL with more than $2\sigma$ significance. 
\par
There are several possible mechanisms that may cause the observed features in the SEDs. 
One possibility is a break in the electron spectrum caused by the synchrotron cooling effects, generally yielding a change in spectral index by 0.5 ~\citep{cooling}, which is in agreement with our results (see Table~\ref{tab:dips_1}). 
Another mechanism that could explain the observed breaks is the absorption by an external photon field~\citep{poutanen}. 
For those nine data sets where the BPL gives a better fit than the PL, the break energy ranges from $\sim2$~GeV to $\sim8$~GeV. 
In addition, 7 of these data sets belong to non-HBL blazars, that are usually characterized by broad emission
lines, thought to be originating from a region of molecular gas (broad line region; BLR) that is highly ionized by the optically thin accretion disk. 
This seems in accordance with the idea of relating the \fermi spectral features to absorption of GeV photons on
radiation from $\mathrm{H}_\mathrm{I}$ (13.6 eV) and $\mathrm{He}_\mathrm{II}$ (54.4 eV) recombination continua in the BLR, that are expected to cause jumps in $\gamma$-ray opacity around $\sim19.2$ and $\sim4.8$~GeV, respectively~\citep{poutanen}. 
We tested a general absorbed power-law (APL) function of the following form on the \fermi data:

\begin{equation} 
dN/dE=N_{0}(E/E_{0})^{-\Gamma}e^{-\tau_{\gamma\gamma}(E, z, E_\mathrm{abs})}
\label{eq:APL_free}
\end{equation}

where the free parameters are the normalization $N_{0}$ at $E_{0}=1\mathrm{GeV}$, photon index $\Gamma$, and absorption line energy $E_\mathrm{abs}$. 
$\tau_{\gamma\gamma}$ is the optical depth for the $\gamma$-$\gamma$ pair annihilation of photons with energies $E$ and $E_\mathrm{abs}$ at a redshift of $z$.
Within the \fermi energy band, BPL and APL functions fit the data equally well. 
Upturns at high-energy tails of \fermi spectra are observed (see, e.g., W Comae in Figure~\ref{fig:combined_spectra}), but they are not statistically significant enough to favor an absorption scenario over a BPL fit. 
Therefore, it appears that one cannot statistically distinguish between the BPL and APL fits, but possible absorption scenarios are worth investigating further. 
To address this issue, we make use of contemporaneous GeV-TeV spectra to test and compare BPL and four different APL scenarios (see Section~\ref{sim}). 
This permits us to test the APL over a larger energy range and investigate the apparent \fermi spectral absorption-like features with higher statistics. 
\par
\begin{figure*}
\centering
\includegraphics[width=80mm]{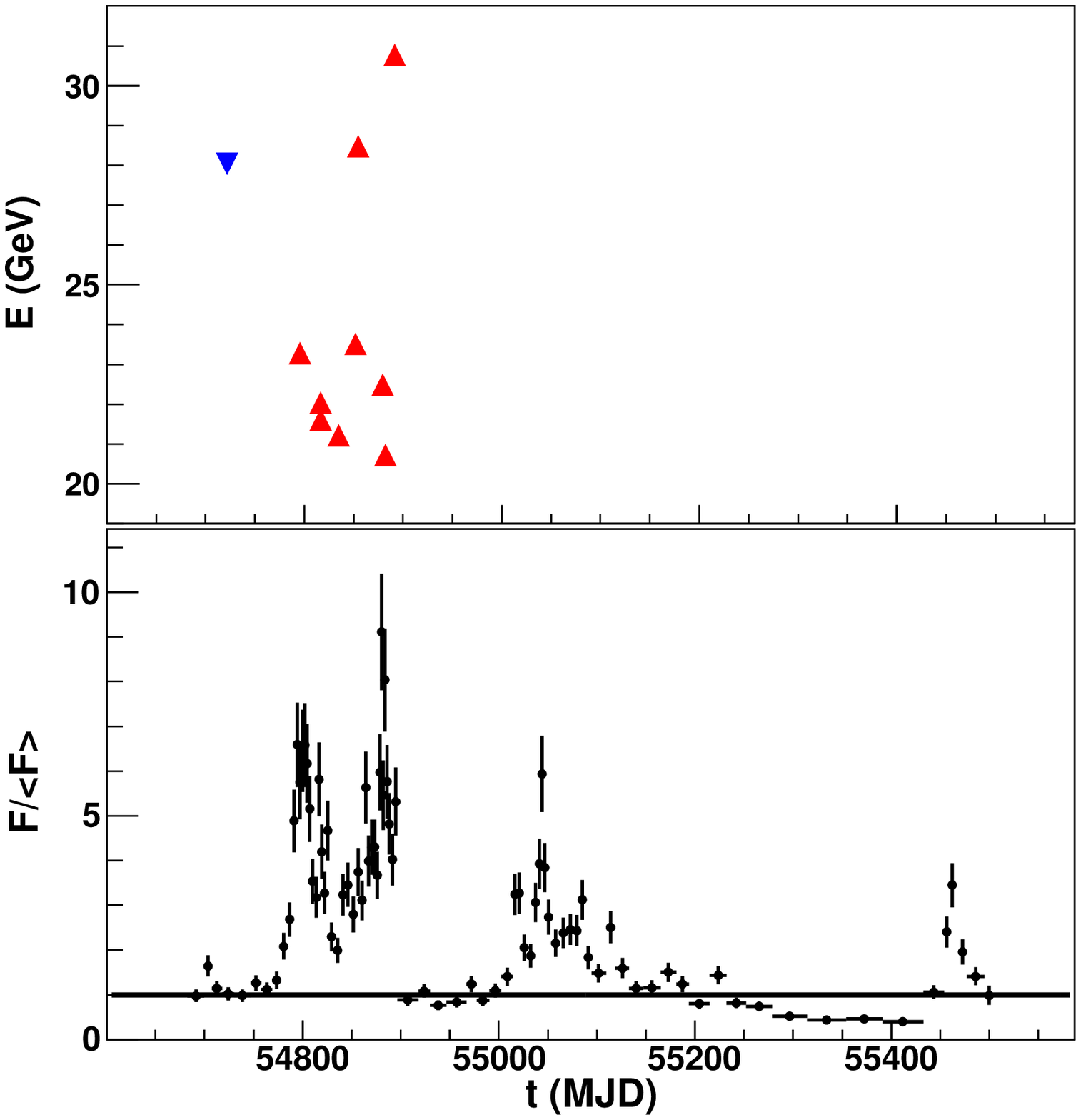}
\includegraphics[width=80mm]{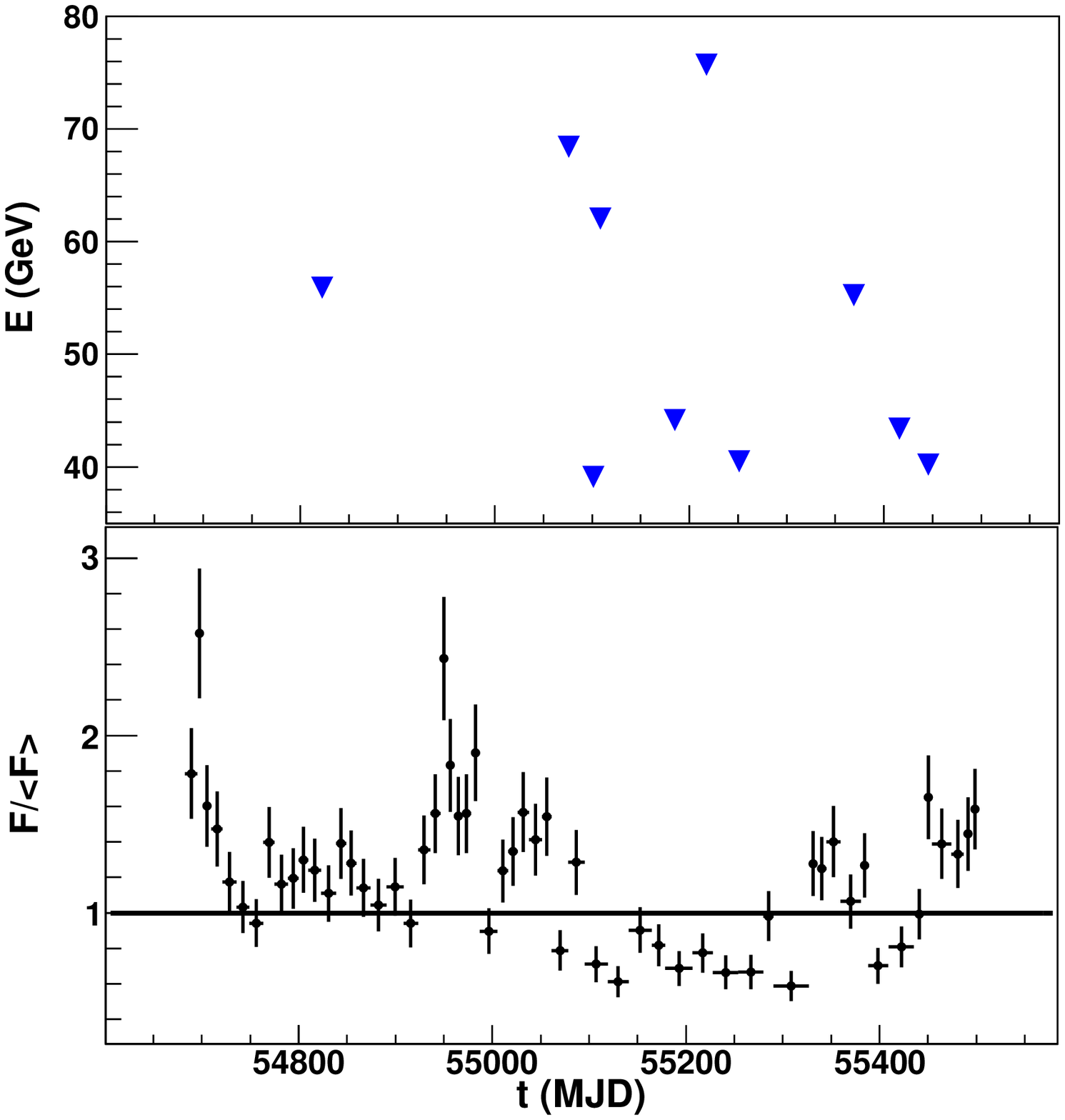}
\caption{{\footnotesize Arrival times of the photons in the highest energy bin (top panels) and aperture light curves with arbitrary flux units (bottom panels) for the blazars 3C~279 (left) and PKS~2155-304 (right). Blue upside-down triangles represent the low-state photons and red triangles the high-state ones. In both cases, the highest energy photons do not show any obvious clustering within their respective data sets. As described in Section~\ref{fermi}, low- and high-states are distinguished based on the flux averages (solid lines) in light curves.}}
\label{fig:photon_arrival_times}
\end{figure*}

Another caveat related to these spectral features is that the upturn seen at the highest \fermi energy bin might be coming from a group of photons clustered in time. 
In that case the dip would be an artifact of a flaring event, thus not representative of the time-averaged spectrum. 
To make sure this is not the case, we checked the arrival times of the highest-energy photons and did not find any obvious clustering (see Figure~\ref{fig:photon_arrival_times}). 
Note that the arrival time distributions should be considered within a given flux state. For instance, in the left panel of the figure, the red triangles represent the high
energy photons from the high flux state and are evenly distributed in a time window that belongs to the high state. Therefore, one concludes that no clustering is found.

\begin{sidewaystable*}
\centering
\begin{small}
\begin{tabular}{|c|cc|ccccc|}
\hline
	&  & Power Law &  & & Broken Power Law	 & & \\

Name	& $N(\times10^{-11})$	& $\Gamma$ 	& $F_{\mathrm{1GeV}}(\times10^{-12})$ 	& $\Gamma_{1}$ 	& $\Gamma_{2}$ 	& $E_{\mathrm{break}}$ (GeV) 	& $\sigma_{\mathrm{BPL}}$	\\
	&(1)			&(2)		&(3)					&(4)		&(5)		&(6)				&(7)			\\
\hline\hline
3C66A (low)	&$2.78\pm0.07$	&$2.01\pm0.02$	&$9.30\pm4.94$	&$1.81\pm0.06$	&$2.22\pm0.07$	&$1.92\pm0.51$	&4.12 \\
S5 0716+714 (low)	&$1.85\pm0.06$	&$2.23\pm0.03$	&$5.18\pm2.07$	&$2.14\pm0.04$	&$2.74\pm0.23$	&$5.43\pm1.01$	&3.07\\
S5 0716+714 (high)	&$3.68\pm0.13$	&$2.10\pm0.03$	&$2.29\pm1.78$	&$2.00\pm0.06$	&$2.57\pm0.30$	&$4.10\pm1.47$	&2.29\\
1ES 1011+496 (low)	&$0.88\pm0.04$	&$1.94\pm0.04$	&$1.70\pm1.34$	&$1.70\pm0.10$	&$2.28\pm0.20$	&$2.76\pm1.15$	&2.76\\
PKS 1222+21 (low)	&$1.72\pm0.06$	&$2.53\pm0.04$	&$0.99\pm0.70$	&$2.41\pm0.06$	&$3.11\pm0.28$	&$3.36\pm0.87$	&2.84\\
3C 279 (low)	&$2.55\pm0.08$	&$2.59\pm0.03$	&$0.15\pm0.28$	&$2.53\pm0.04$	&$3.72\pm1.54$	&$7.70\pm5.57$	&2.51\\
3C 279 (high)	&$1.16\pm0.02$	&$2.39\pm0.02$	&$8.14\pm10.72$	&$2.31\pm0.03$	&$2.74\pm0.26$	&$3.21\pm1.79$	&3.11\\
PKS 2155-304 (low)	&$4.19\pm0.10$	&$1.92\pm0.02$	&$1.75\pm1.98$	&$1.86\pm0.03$	&$2.13\pm0.14$	&$5.54\pm3.23$	&2.08\\
BL Lacertae (high)	&$3.26\pm0.11$	&$2.39\pm0.04$	&$1.08\pm0.09$	&$2.30\pm0.04$	&$3.00\pm0.30$	&$4.49\pm0.03$	&2.50\\
\hline
\end{tabular}
\end{small}
\caption{{\small Fit results for power law (PL) and broken power law (BPL), where BPL yields a better fit over PL with more than $2\sigma$ significance (9 out of 33 cases). Columns (1) and (2) show the PL parameters, flux normalization at 1 GeV and the photon index respectively. $F_{\mathrm{1GeV}}$ (3) is the flux normalization at 1~GeV for BPL. $N$ and $F_{\mathrm{1GeV}}$ are in erg cm$^{-2}$ s$^{-1}$}. Columns (4) and (5) show the photon indices for BPL, as given in Equation~(\ref{eq:BPL}). The break energy for BPL is listed in Column (6). $\sigma_\mathrm{BPL}$ (7) is the likelihood ratio test results of BPL over PL. For these 9 cases, the break energy $E_{\mathrm{break}}$ ranges from $\sim2$~GeV to $\sim8$~GeV. In addition, 7 of these data sets belong to non-HBL blazars. }
\label{tab:dips_1}
\end{sidewaystable*}

\begin{figure*}
\centering
\includegraphics[width=80mm]{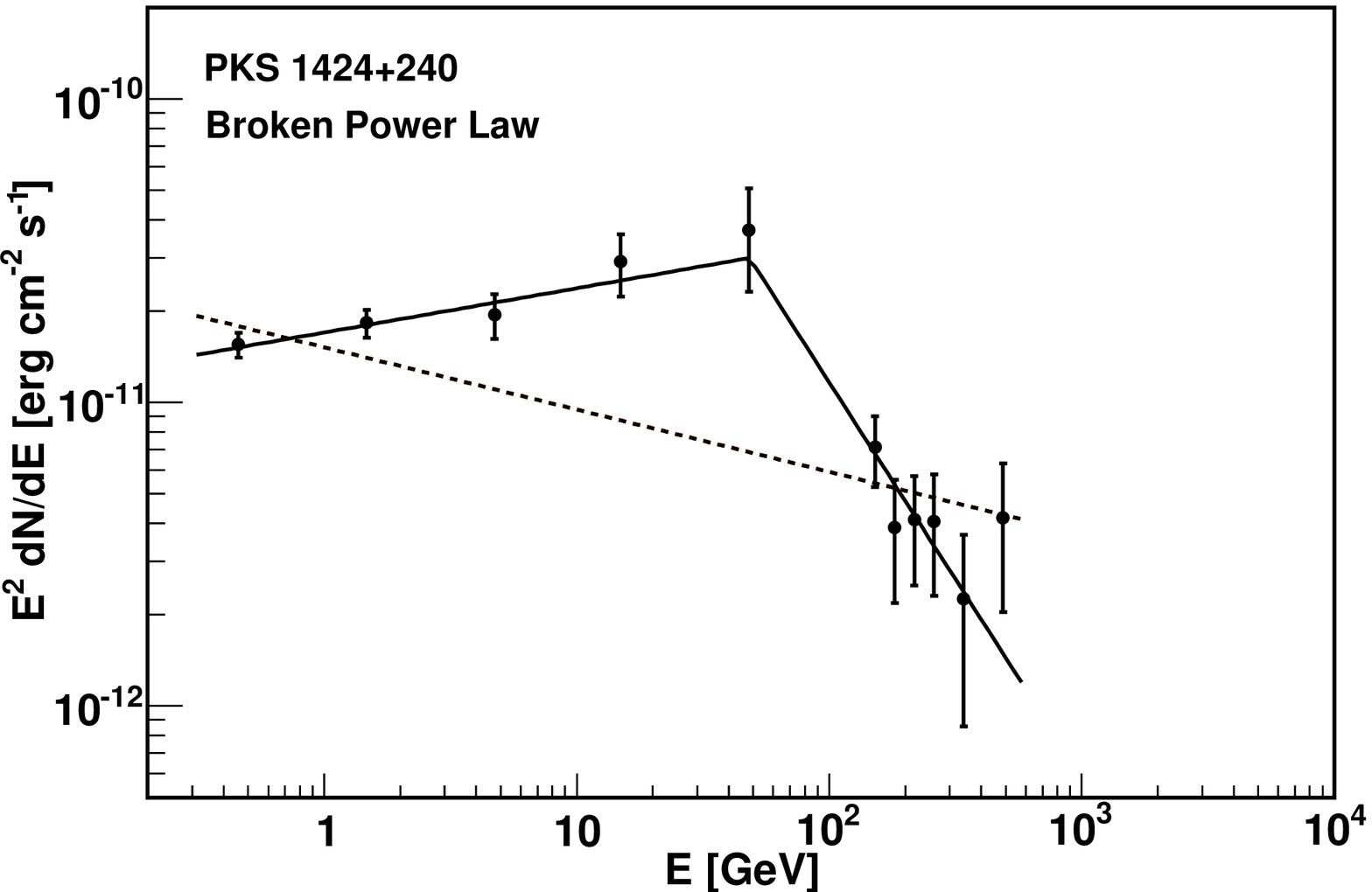}
\includegraphics[width=80mm]{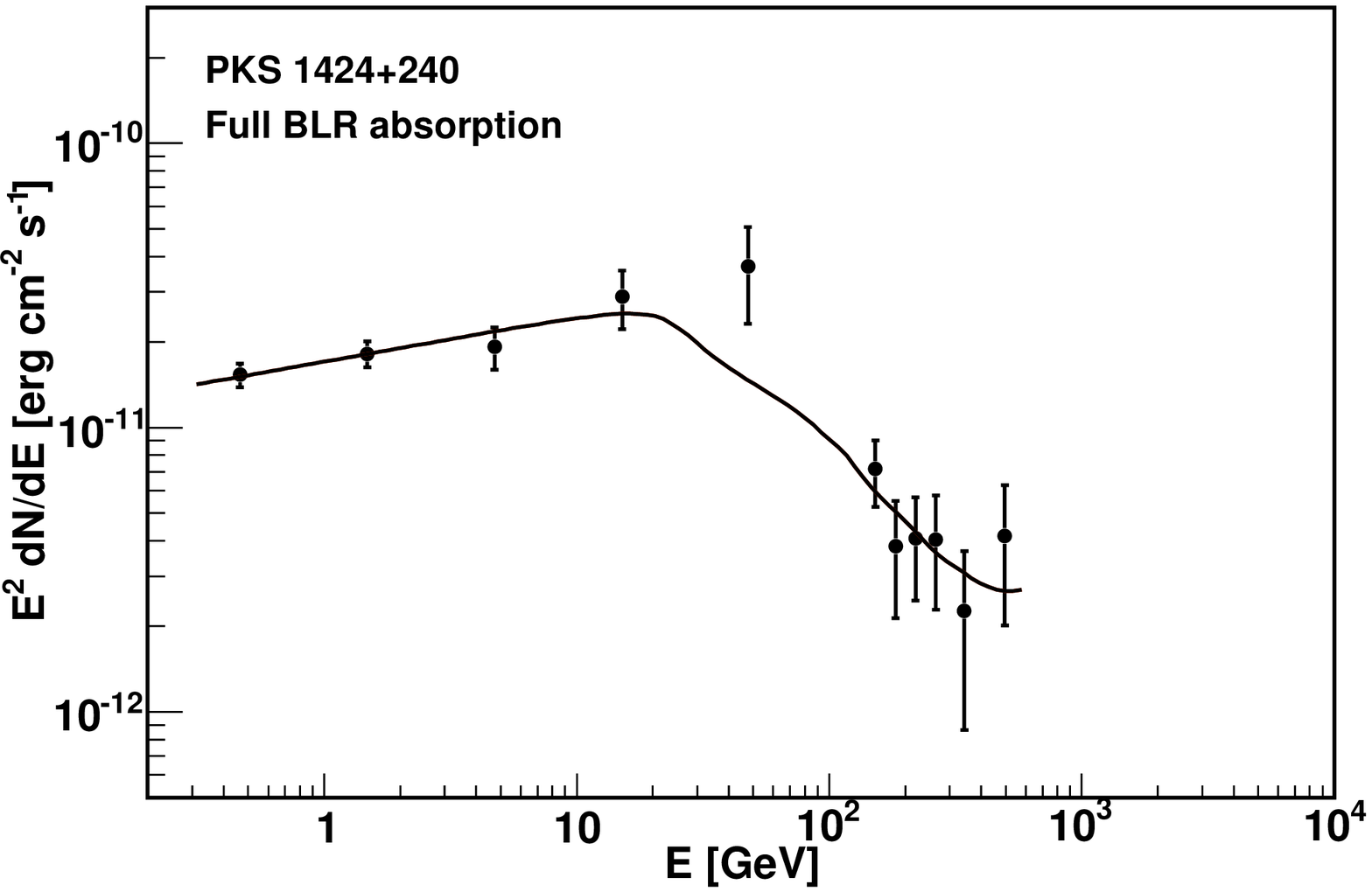}
\includegraphics[width=80mm]{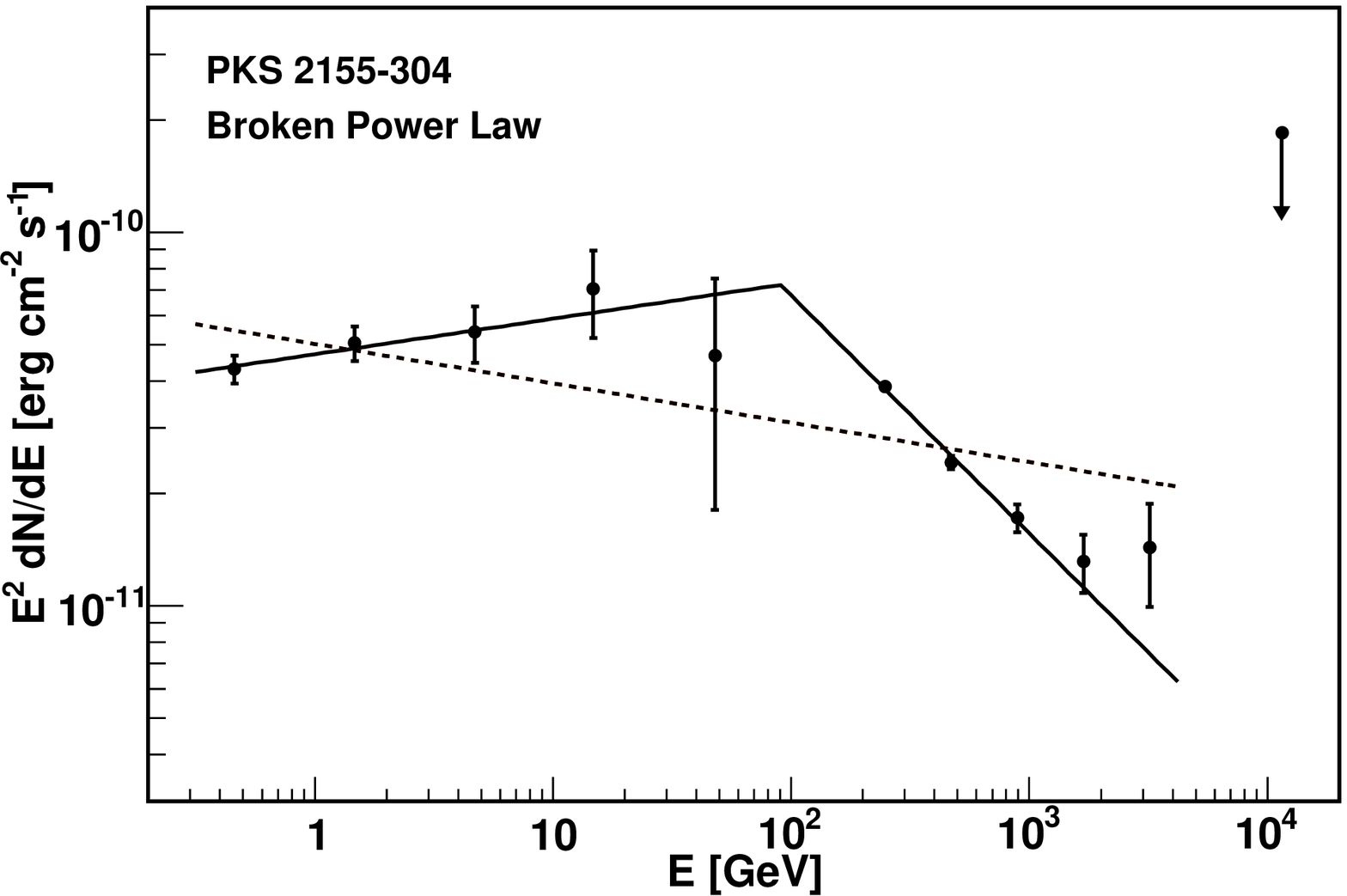}
\includegraphics[width=80mm]{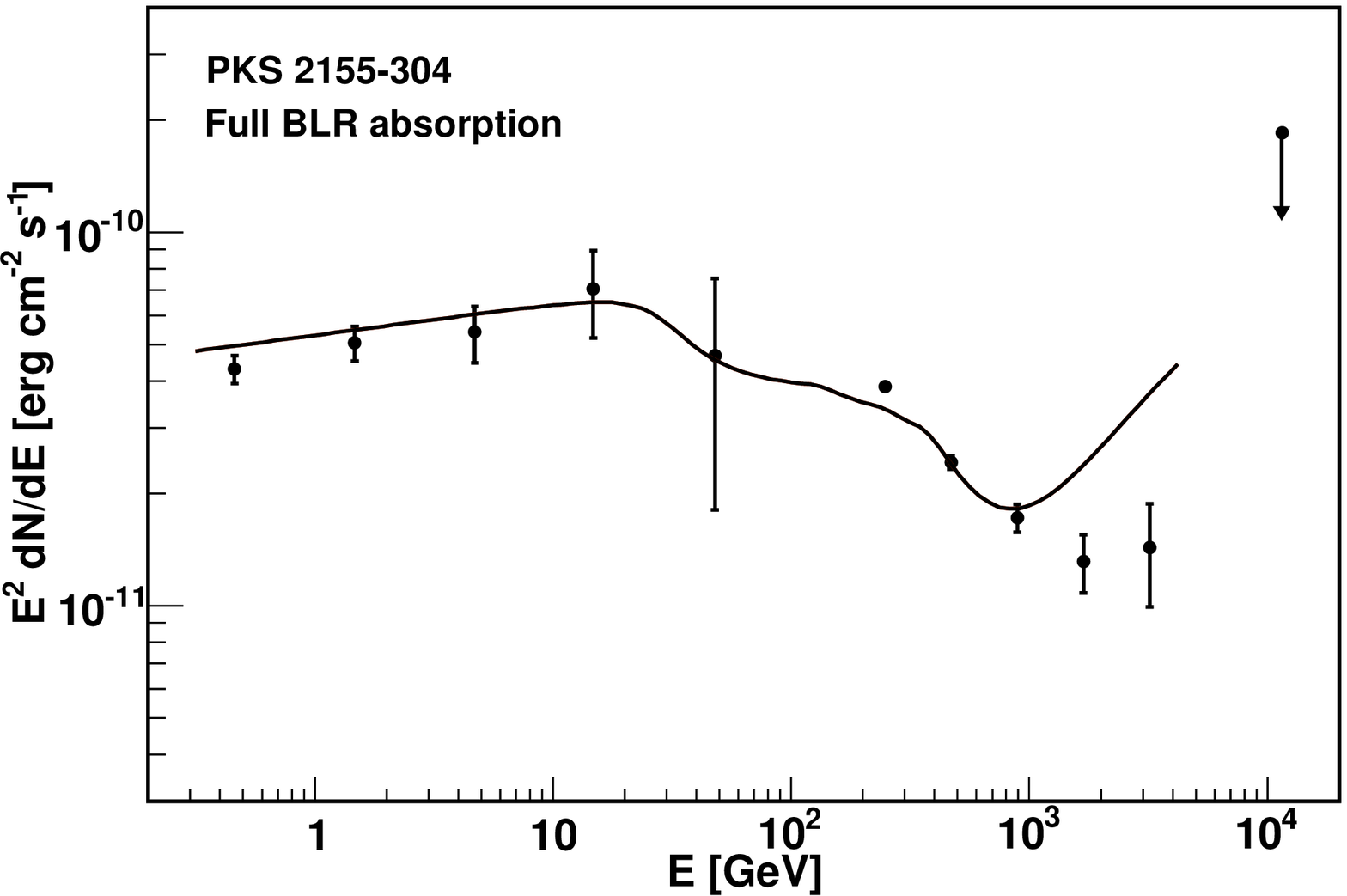}
\caption{ Contemporaneous GeV-TeV spectra with power-law (dashed lines), broken power-law (solid lines) and power-law
with full-BLR-absorption fits. BPL and full BLR absorption scenarios seem to fit well the combined spectra of the
blazars PKS~1424+240 and PKS~2155-304. A BPL (full BLR absorption) function is preferred over the PL for the blazars
PKS~1424+240 and PKS~2155-304 with a significance of $\sim5\sigma$ ($\sim4.8\sigma$) and $\sim12\sigma$
($\sim8.5\sigma$) respectively.}
\label{fig:gevTevFits}
\end{figure*}

\subsection{Quasi-simultaneous GeV-TeV spectra}\label{sim}
Seven of the TeV spectra in our sample are contemporaneous with \fermi observations and therefore merit a deeper analysis. 
We extended the work described in section \ref{spectralFeatures} to this subsample. 
This time, in addition to PL and BPL fits, we tested four different scenarios of absorption due to photons emitted
from the BLR: $\mathrm{H}_\mathrm{I}$ line (13.6 eV), $\mathrm{He}_\mathrm{II}$ line (54.4 eV), $\mathrm{H}_\mathrm{I}$ \& $\mathrm{He}_\mathrm{II}$ combined, and full BLR
spectrum taken from~\citep{poutanen}.
\par
For single- and double-line absorption scenarios, $\mathrm{H}_\mathrm{I}$ and $\mathrm{He}_\mathrm{II}$ recombination continua are the most plausible cases given that they are the most dominant ones in the BLR spectrum, and that the breaks we see in the \fermi spectra are located around a few GeV. 
As for the full BLR spectrum, taken from~\citep{poutanen}, it was modeled assuming a photoionized gas with the ionization parameter and the cloud density changing with the distance to the central ionizing source.
See~\citep{poutanen} for a detailed discussion on the $\gamma$-ray absorption within the BLR in the \fermi spectra. 
No general trend can be seen in the contemporaneous data sample. 
BPL and full BLR absorption scenarios seem to fit well the combined spectra of the blazars PKS 1424+240 and PKS 2155-304. 
A BPL (full BLR absorption) function is preferred over the PL for the blazars PKS 1424+240 and PKS 2155-304 with a significance of $\sim5\sigma$ ($\sim4.8\sigma$) and $\sim12\sigma$ ($\sim8.5\sigma$), respectively (see Figure~\ref{fig:gevTevFits}). 
The $\chi^{2}$/dof values of PL, BPL and APL fits are 32/9, 3.5/7, 5.3/7 for PKS~1424+240 and 148/8, 5.8/6, and 71/6 for PKS~2155-304, respectively. 
BPL fits yield $\Delta\Gamma=1.4$ (PKS 1424+240) and $\Delta\Gamma=0.7$ (PKS 2155-304), both larger than what electron cooling would predict, which might indicate that an additional mechanism is at work. 
Both BPL and full BLR absorption scenarios provide a slight improvement in the MAGIC and VERITAS spectra of 3C 66A, albeit not significant. 
Similarly, for PKS 1222+21, BPL, $\mathrm{H}_\mathrm{I}$ single line and $\mathrm{H}_\mathrm{I}$ + $\mathrm{He}_\mathrm{II}$ double line absorptions slightly improve the fit over PL.  
In the case of RGB J0710+091 and 1ES 1218+304, we don't observe any preference over the power-law fit.
In case a $\gamma$-$\gamma$ absorption from BLR is at work, the cascades initiated in this process might produce
observable GeV $\gamma$-ray emission~\citep{parisa_radio}, and their synchrotron
emission could contribute to the big blue bump seen in several blazars~\citep{parisa_bbb}.
\clearpage
\section{Summary}\label{summary}
We study blazar spectral properties with a focus on the GeV-TeV energy range for a sample of VHE blazars. 
In order to obtain a set of joint GeV-TeV blazar spectra, we analyze the first 27 month \fermi data for VHE blazars and combine our results with archival VHE data. 
In cases where the \fermi data set does not overlap with the TeV observations but has enough statistics, we split the data into high and low flux states and join the best-matching subset with the corresponding TeV spectrum.
The peak frequency band of the inverse Compton component increases following the order FSRQ -$>$ LBL\&IBL -$>$ HBL. 
Thus, our results confirm the positive correlation between $\nu_\mathrm{syn}$ and $\nu_\mathrm{IC}$. 
We note that \fermi spectra from different flux states for a given TeV blazar do not undergo a significant change in photon index. 
The variability amplitudes within our \fermi data set do not show an immediate correlation with the reported TeV variabilities for individual blazars.
We find that in many cases a power law is insufficient to describe the GeV-TeV spectra and a broken power law improves the fits, especially for non-HBL blazars, where the BLR emission may have an effect on the observed spectral shape. 
In some blazars we observe absorption-like spectral features. 
We present seven quasi-simultaneous joint spectra, on which we test possible absorption scenarios from the BLR. 
Even though the absorption seems to describe well some of the observed spectra, no general pattern can be identified.

\acknowledgements
This work was supported in part by the NSF grant Phy-0855627 and NASA grant NNX09AU14G. We acknowledge Paolo Coppi for helpful discussions and Juri Poutanen for providing the BLR emission templates.



\end{document}